\RequirePackage{fix-cm}
\documentclass[smallcondensed]{svjour3}     % onecolumn (ditto)
\smartqed  % flush right qed marks, e.g. at end of proof
\usepackage{graphicx}
\usepackage{lineno,hyperref}
\modulolinenumbers[10]

\usepackage{ifpdf}
\usepackage{ifthen}

\usepackage{array}
\usepackage{tabularx}
\usepackage{multirow}
\usepackage{booktabs}

\usepackage{graphicx}
\usepackage{grffile}
\usepackage{color}
\usepackage[dvipsnames]{xcolor}
\usepackage{float}
\usepackage{caption}
\usepackage{subcaption}
\usepackage{rotating}

\usepackage{tablefootnote}

\usepackage{pgf,pgfplots,pgfplotstable}
\pgfplotsset{compat=newest}
\pgfplotsset{
xlabel near ticks,
ylabel near ticks,
label style={font=\footnotesize},
tick label style={font=\footnotesize},
legend style={font=\scriptsize},
xticklabel style={/pgf/number format/set thousands separator={\,}},
yticklabel style={/pgf/number format/set thousands separator={\,}},
tick scale binop=\times,
try min ticks=6,
legend pos=outer north east
}
\usepgfplotslibrary{external} 
\usepgfplotslibrary{units}
\usepgfplotslibrary{patchplots}
\usepgfplotslibrary{groupplots}
\usepgfplotslibrary{fillbetween}

\usepackage{tikz,tkz-base,xstring}
\usetikzlibrary{shapes,decorations,shadows}
\usetikzlibrary{decorations.pathreplacing}
\usetikzlibrary{decorations.pathmorphing}
\usetikzlibrary{decorations.shapes}
\usetikzlibrary{decorations.text}
\usetikzlibrary{decorations.footprints}
\usetikzlibrary{decorations.fractals}
\usetikzlibrary{fadings}
\usetikzlibrary{patterns}
\usetikzlibrary{calc}
\usetikzlibrary{shapes.geometric}
\usetikzlibrary{shapes.gates.logic.IEC}
\usetikzlibrary{shapes.gates.logic.US}
\usetikzlibrary{fit,chains}
\usetikzlibrary{positioning}
\usepgflibrary{shapes}
\usetikzlibrary{scopes}
\usetikzlibrary{arrows}
\usetikzlibrary{arrows.meta}
\usetikzlibrary{backgrounds}
\usetikzlibrary{intersections}
\usetikzlibrary{matrix}
\usetikzlibrary{plotmarks}
\usetikzlibrary{pgfplots.units}
\usetikzlibrary{trees}
\usetikzlibrary{hobby}

\newlength\figureheight
\newlength\figurewidth
\usepgfplotslibrary{units}

\usepackage[all]{xy}

\usepackage{amsmath,amssymb,amsfonts,%amsthm,
bm}
\usepackage{textcomp,wasysym,eurosym,stmaryrd}
\usepackage{mathrsfs}
\usepackage{mathtools}
\usepackage{pifont}
\usepackage{upgreek}
\usepackage{centernot}
\usepackage{algorithm,algorithmic}

\usepackage{xspace}
\usepackage{enumitem}
\setlist[itemize]{label={--}}
\newlength{\hatchspread}
\newlength{\hatchthickness}
\newlength{\hatchshift}
\newcommand{\hatchcolor}{}
\tikzset{hatchspread/.code={\setlength{\hatchspread}{#1}},
         hatchthickness/.code={\setlength{\hatchthickness}{#1}},
         hatchshift/.code={\setlength{\hatchshift}{#1}},% must be >= 0
         hatchcolor/.code={\renewcommand{\hatchcolor}{#1}}}
\tikzset{hatchspread=3pt,
         hatchthickness=0.4pt,
         hatchshift=0pt,% must be >= 0
         hatchcolor=black}
\pgfdeclarepatternformonly[\hatchspread,\hatchthickness,\hatchshift,\hatchcolor]% variables
   {custom north west lines}% name
   {\pgfqpoint{\dimexpr-2\hatchthickness}{\dimexpr-2\hatchthickness}}% lower left corner
   {\pgfqpoint{\dimexpr\hatchspread+2\hatchthickness}{\dimexpr\hatchspread+2\hatchthickness}}% upper right corner
   {\pgfqpoint{\dimexpr\hatchspread}{\dimexpr\hatchspread}}% tile size
   {% shape description
    \pgfsetlinewidth{\hatchthickness}
    \pgfpathmoveto{\pgfqpoint{0pt}{\dimexpr\hatchspread+\hatchshift}}
    \pgfpathlineto{\pgfqpoint{\dimexpr\hatchspread+0.15pt+\hatchshift}{-0.15pt}}
    \ifdim \hatchshift > 0pt
      \pgfpathmoveto{\pgfqpoint{0pt}{\hatchshift}}
      \pgfpathlineto{\pgfqpoint{\dimexpr0.15pt+\hatchshift}{-0.15pt}}
    \fi
    \pgfsetstrokecolor{\hatchcolor}
    \pgfusepath{stroke}
   }
\pgfdeclarepatternformonly[\hatchspread,\hatchthickness,\hatchshift,\hatchcolor]% variables
   {custom north east lines}% name
   {\pgfqpoint{\dimexpr-2\hatchthickness}{\dimexpr-2\hatchthickness}}% lower left corner
   {\pgfqpoint{\dimexpr\hatchspread+2\hatchthickness}{\dimexpr\hatchspread+2\hatchthickness}}% upper right corner
   {\pgfqpoint{\dimexpr\hatchspread}{\dimexpr\hatchspread}}% tile size
   {% shape description
    \pgfsetlinewidth{\hatchthickness}
    \pgfpathmoveto{\pgfqpoint{\dimexpr\hatchshift-0.15pt}{-0.15pt}}
    \pgfpathlineto{\pgfqpoint{\dimexpr\hatchspread+0.15pt}{\dimexpr\hatchspread-\hatchshift+0.15pt}}
    \ifdim \hatchshift > 0pt
      \pgfpathmoveto{\pgfqpoint{-0.15pt}{\dimexpr\hatchspread-\hatchshift-0.15pt}}
      \pgfpathlineto{\pgfqpoint{\dimexpr\hatchshift+0.15pt}{\dimexpr\hatchspread+0.15pt}}
    \fi
    \pgfsetstrokecolor{\hatchcolor}
    \pgfusepath{stroke}
   }

\newcommand{\eb}{\boldsymbol{e}}
\newcommand{\fb}{\boldsymbol{f}}

\newcommand{\kb}{\boldsymbol{k}}

\newcommand{\nb}{\boldsymbol{n}}

\newcommand{\pb}{\boldsymbol{p}}

\newcommand{\ub}{\boldsymbol{u}}
\newcommand{\vb}{\boldsymbol{v}}

\newcommand{\xb}{\boldsymbol{x}}

\newcommand{\zerob}{\boldsymbol{0}}

\newcommand{\alphab}{\boldsymbol{\alpha}}

\newcommand{\deltab}{\boldsymbol{\delta}}
\newcommand{\epsilonb}{\boldsymbol{\varepsilon}}
\newcommand{\zetab}{\boldsymbol{\zeta}}

\newcommand{\sigmab}{\boldsymbol{\sigma}}

\newcommand{\ebf}{\mathbf{e}}

\newcommand{\Ab}{\boldsymbol{A}}

\newcommand{\Cb}{\boldsymbol{C}}
\newcommand{\Db}{\boldsymbol{D}}
\newcommand{\Eb}{\boldsymbol{E}}

\newcommand{\Ib}{\boldsymbol{I}}
\newcommand{\Jb}{\boldsymbol{J}}

\newcommand{\Lb}{\boldsymbol{L}}

\newcommand{\Sb}{\boldsymbol{S}}

\newcommand{\Ub}{\boldsymbol{U}}

\newcommand{\varPib}{\boldsymbol{\varPi}}
\newcommand{\varSigmab}{\boldsymbol{\varSigma}}

\newcommand{\Abf}{\mathbf{A}}

\newcommand{\Cbf}{\mathbf{C}}

\newcommand{\Ebf}{\mathbf{E}}
\newcommand{\Fbf}{\mathbf{F}}
\newcommand{\Gbf}{\mathbf{G}}
\newcommand{\Hbf}{\mathbf{H}}

\newcommand{\Kbf}{\mathbf{K}}

\newcommand{\Mbf}{\mathbf{M}}

\newcommand{\Ubf}{\mathbf{U}}
\newcommand{\Vbf}{\mathbf{V}}

\newcommand{\Cc}{\mathcal{C}}

\newcommand{\Fc}{\mathcal{F}}

\newcommand{\Cbb}{\mathbb{C}}

\newcommand{\Fbb}{\mathbb{F}}
\newcommand{\Gbb}{\mathbb{G}}
\newcommand{\Hbb}{\mathbb{H}}
\newcommand{\Ibb}{\mathbb{I}}
\newcommand{\Jbb}{\mathbb{J}}
\newcommand{\Kbb}{\mathbb{K}}

\newcommand{\Mbb}{\mathbb{M}}

\newcommand{\Rbb}{\mathbb{R}}

\newcommand{\Ubb}{\mathbb{U}}

\newcommand{\abs}[1]{\lvert#1\rvert}

\newcommand{\norm}[1]{\lVert#1\rVert}

\newcommand{\set}[1]{\{#1\}}

\newcommand{\setst}[2]{\{#1\mathrel{;}#2\}}

\newcommand{\scalprod}[2]{#1\cdot#2}

\DeclareMathOperator{\diver}{div}

\DeclareMathOperator{\divb}{\textbf{div}}
\DeclareMathOperator{\gradd}{grad\kern-.5em{grad}}

\DeclareMathOperator{\nablab}{\boldsymbol{\nabla}}
\DeclareMathOperator{\tr}{tr}

\DeclareMathOperator{\adj}{adj}
\DeclareMathOperator{\re}{Re}
\DeclareMathOperator{\im}{Im}

\newcommand{\interval}[4]{\mathopen{#1}#2 \mathclose{}\mathpunct{},#3 \mathclose{#4}}
\newcommand{\intervalcc}[2]{\interval{[}{#1}{#2}{]}}
\newcommand{\intervaloc}[2]{\interval{]}{#1}{#2}{]}}

\newcommand{\intervaloo}[2]{\interval{]}{#1}{#2}{[}}

\renewcommand{\(}{\left(}
\renewcommand{\)}{\right)}
\renewcommand{\[}{\left[}
\renewcommand{\]}{\right]}

\renewcommand{\geq}{\geqslant}
\renewcommand{\leq}{\leqslant}
\let\oldtimes\times
\renewcommand{\times}{\!\oldtimes\!}

\newcommand{\ie}{\textit{i.e.}\xspace}
\newcommand{\eg}{\textit{e.g.}\xspace}

\newcommand{\etc}{etc.\xspace}

\journalname{Archives of Computational Methods in Engineering}

\begin{document}

\title{Review and recent developments on the perfectly matched layer (PML) method for the numerical modeling and simulation of elastic wave propagation in unbounded domains
}

\titlerunning{Review on the PML method for elastic wave propagation in unbounded domains}        % if too long for running head

\author{Florent Pled         \and
        Christophe Desceliers %etc.
}


\institute{F. Pled \and C. Desceliers \at
              Univ Gustave Eiffel, MSME UMR 8208, F-77454, Marne-la-Vall\'ee, France \\
              Tel.: +33 (0)1 60 95 77 82, +33 (0)1 60 95 77 79 \\
              \email{\{florent.pled,christophe.desceliers\}@univ-eiffel.fr}           %  \\
           \and
%           C. Desceliers \at
%              Univ Gustave Eiffel, MSME UMR 8208, F-77454, Marne-la-Vall\'ee, France \\
%              Tel.: +33 (0)1 60 95 77 79\\
%              %Fax: +123-45-678910\\
%              \email{christophe.desceliers@univ-eiffel.fr}
}

\date{Received: 18 November 2020 / Accepted: 24 March 2021}

\maketitle

\begin{abstract}
This review article revisits and outlines the perfectly matched layer (PML) method and its various formulations developed over the past 25 years for the numerical modeling and simulation of wave propagation in unbounded media. Based on the concept of complex coordinate stretching, an efficient mixed displacement-strain unsplit-field PML formulation for second-order (displacement-based) linear elastodynamic equations is then proposed for simulating the propagation and absorption of elastic waves in unbounded (infinite or semi-infinite) domains. Both time-harmonic (frequency-domain) and time-dependent (time-domain) PML formulations are derived for two- and three-dimensional linear elastodynamic problems. Through the introduction of only a few additional variables governed by low-order auxiliary differential equations, the resulting mixed time-domain PML formulation is second-order in time, thereby allowing the use of standard time integration schemes commonly employed in computational structural dynamics and thus facilitating the incorporation into existing displacement-based finite element codes. For computational efficiency, the proposed time-domain PML formulation is implemented using a hybrid approach that couples a mixed (displacement-strain) formulation for the PML region with a classical (displacement-based) formulation for the physical domain of interest, using a standard Galerkin finite element method (FEM) for spatial discretization and a Newmark time scheme coupled with a finite difference (Crank-Nicolson) time scheme for time sampling. Numerical experiments show the performances of the PML method in terms of accuracy, efficiency and stability for two-dimensional linear elastodynamic problems in single- and multi-layer isotropic homogeneous elastic media.
\keywords{Elastic wave propagation \and Elastodynamics \and Unbounded domain \and Perfectly matched layers (PMLs) \and Absorbing boundary conditions (ABCs) \and Absorbing layers (ALs) \and Finite Element Method (FEM) \and Time-domain analysis}
\subclass{35L05 \and 35L20 \and 35B35 \and 65M60 \and 65M12 \and 74J05 \and 74J10 \and 74S05 \and 74H15}
\end{abstract}

\section{Introduction}\label{sec:intro}

Large-scale complex wave propagation, radiation or scattering problems are often set on unbounded (or very large) domains. When using classical numerical approximation methods such as finite difference (FD), finite volume (FV), finite element (FE), spectral element (SE) or discontinuous Galerkin (DG) methods, their numerical solutions can be computed in using computational models that are constructed for bounded subdomains (corresponding to the physical domains of interest), thereby requiring the truncation of the unbounded domain with \textit{ad hoc} modeling artifacts, \ie{} appropriate conditions at the artificial truncation boundaries, to properly simulate the unimpeded outward propagation of waves without any spurious reflections coming from the truncation boundaries. Among the numerous methods dealing with the construction of such bounded (or finite) computational domains, such as artificial transparent and absorbing boundary conditions (ABCs), absorbing layers (ALs) methods and alternative techniques (see \cite{Giv92,Wolf96,Tsy98,Giv99a,Hag99,Hag03a,Col04,Hag07a,Giv08,Ant08,Han13,Gao17b} and the references therein for a comprehensive overview of the existing approaches), the high-order local ABCs (involving only low-order %normal 
derivatives, \ie{} no high-order spatial and temporal derivatives, having recourse to special auxiliary variables on the artificial boundary) \cite{%Col93,
Gro95,Gro96,Hag98,Ant99,%Huan00,
Gud00,%Giv03a,Giv03b,Giv03c,
Giv04,Hag04,%Joo05,
Giv06,Gud06,%Zah06,
Hag07a,Hag07b,%Hag08,Hag09,
Hag10,Beca10,Baf12,Hag14,Sch15} and the perfectly matched layers (PMLs) \cite{Bere94,Kat94,%Bere96a,
Bere96b,Hu96,Has96,Tur98,Tei01b,Col01,App06a,Bere07,Hu08a,Berm10,Kuc11,Fat15a} have been extensively studied during the last decades, since they are neither computationally expensive, nor difficult to implement, nor limited to domains with simple geometric shapes (\eg{} rectangular, cuboidal, circular, spherical, ellipsoidal, \etc), and allow the outgoing %propagative
 waves to be efficiently absorbed with arbitrarily-high controlled accuracy for a broad class of wave propagation problems.

The PML method has been introduced in the mid 1990s for simulating the propagation of electromagnetic waves over time in two-dimensional (2D) \cite{Bere94,Bere96a} and three-dimensional (3D) \cite{Kat94,Bere96b} unbounded spatial domains. %by considering a physical bounded domain (resulting from the truncation of the unbounded domain) to which an artificial (non-physical) absorbing region is attached with a specific treatment allowing the outgoing waves to leave the physical domain without any spurious reflections. 
It basically consists in surrounding the bounded physical domain of interest (resulting from the truncation of the unbounded domain) by artificial (non-physical) absorbing non-reflecting layers of finite thickness especially designed to absorb the outgoing waves (leaving the physical domain) without any spurious (unwanted) reflections. In theory, within the continuous framework, a PML model is then constructed such that the outgoing waves are perfectly transmitted at the interface between the physical domain and the absorbing layers whatever their non-zero angles of incidence and their non-zero angular frequencies, and then they decrease exponentially in the direction normal to the interface %between the physical domain and the absorbing layers 
when propagating inside the absorbing layers. Nevertheless, in practice, within the discrete framework, the numerical approximations introduced in the PML computational model do not allow such a perfect matching (reflectionless) property to be still satisfied due to the numerical dispersion inherent to any discretization method. Furthermore, since the absorbing layers (hereinafter referred to as the PML region) have a limited (finite) thickness, the outgoing waves that are not completely damped while traveling inside the PML region, being reflected at the exterior boundary and again propagating back through the PML region (hence traveling twice through it) re-enter the physical domain and therefore generate spurious waves with a non-zero amplitude, which is relatively small in practical applications since it is damped by an exponential factor that depends on twice the PML thickness. In numerical practice, the layers are then designed to have minimal reflection and strong absorption properties. Besides, due to its simplicity of implementation and versatility (broad applicability), the PML method can be easily implemented in conjunction with any classical numerical approximation method available in standard computer codes and often performs very well compared to other existing ABCs for artificially handling problems defined on unbounded domains \cite{Kan97,Col98a,Har00,Col01,Kom03,Giv08,Rab10,Zhao13,Zaf16,Gao17b}, especially for arbitrarily heterogeneous materials/media.

\subsection{Overview of existing PML formulations}\label{sec:overview}

The classical PML formulation, henceforth known as split-field PML (sometimes referred to as S-PML), is built by artificially splitting the physical fields solution of the time-domain wave equations (written in Cartesian coordinates) into unphysical components (according to their spatial derivatives) in the PML region %(hence introducing additional degrees of freedom) 
and considering specific dissipative anisotropic material properties. These dissipative material properties are introduced through the use of artificial damping terms satisfying a matching impedance condition to achieve a perfect transmission of outgoing propagative waves (without any reflections) at the interface between the physical domain and the absorbing layers, with an exponential decay in the PML region along the direction normal to the interface %between the physical domain and the absorbing layers
 \cite{Bere94,Bere96a,Bere96b}. Such a field-splitting PML formulation alters the original formulation of second-order time-domain wave equations and introduces two distinct systems of equations set on the physical domain and the PML region, requiring a special treatment at the interface between the two and making the implementation into existing codes rather cumbersome. In the frequency domain, the PML method can be interpreted as a complex space coordinate stretching approach \cite{Chew94,Rap95,Rap96,Chew96,Chew97} through an analytic continuation of the real spatial coordinates into the complex space via complex stretching functions, corresponding to a change of metric of the physical coordinate space \cite{Tei99a,Tei99b,Las01a,Las01b}. As already mentioned in \cite{Berm10}, the complex coordinate stretching strategy is closely related to the so-called complex scaling technique (also referred to as analytic dilatation technique) originally introduced in the mid 1970s for the numerical simulation and analysis of the dynamical properties (eigenstates) of atomic systems in quantum mechanics \cite{Agui71,Sim78,His96}. 
The resulting PML formulations based on complex stretched coordinates then involve complex operators, rendering the PML method difficult to implement along with any conventional numerical approximation method. Alternative unsplit-field (or nonsplitting) PML formulations, frequently called uniaxial PMLs (sometimes referred to as U-PML or N-PML), preserve the original (unstretched) formulation of second-order time-harmonic wave equations (involving physical fields and classical real operators), but with specific frequency-dependent complex anisotropic material properties, thereby allowing the PML region to be interpreted as an artificial anisotropic absorbing medium (sometimes called anisotropic material absorber) \cite{Sac95,Ged96a,Ged96b,Zhao96a,Zhao96b,Zhao98,Zio97,Rod97,Wu97,Tei97b,Tei98a,Tei98b,Tei99b,Aba98}. The unsplit-field PML formulations can then be straightforwardly incorporated into existing codes (with minor modifications) and implemented using any classical numerical method unlike their split-field counterparts. Time-dependent PML formulations are usually derived from their time-harmonic counterparts by taking the inverse Fourier transform in time of time-harmonic PML equations, which can be performed by directly using convolution products that are usually computed by means of a recursive convolution update technique \cite{Lue92} to avoid expensive temporal convolution operations \cite{Rod00a,Rod00b,Wang03,Ryl04,Wang06a,App06b,Dro07b,Kom07,Mar08a,Mar08b,Mar09,Guan09,Li10,Mat11,Gir12,Xie14,Ping16,Per16,Li17,Li18,Ma18} (henceforth known as the convolutional PML or C-PML), or by using a recursive time integration approach \cite{Zeng04a,Dro07a,Gia08,Gia12,Gia18} (henceforth known as the recursive integration PML or RI-PML), or by using digital signal processing (DSP) techniques such as the Z-transform and digital filtering methods \cite{Ram02,Ram03b,Ram03c,%Ram03d,Dong04,
Ram06,%Li06a,Li06b,
Li06c,Li07b,Li08a,Shi12,Feng12,Feng13b,Feng15a,Feng15b} (henceforth known as the matched Z-transform PML or MZT-PML), or by introducing additional fields and auxiliary differential equations \cite{Tur98,Hu01,Zhe02,Kom03,Ram03a,Basu04,Sjo05,Wang06c,Li07a,%Li08b,
Basu09,Kri09,Qin09,Ged10,Mar10,Zha10,Gro10,Duru12a,Duru12b,Feng13a,Kal13,Ben13,Duru14a,Duru14b,Duru14c,Duru15,Sag14,Xie14,Zha14,Ma14,Gao15,Wei16a,Wei16b,Zhou16,Mod17} (henceforth known as the auxiliary-differential-equation PML or ADE-PML), and thus require additional resources in terms of computational cost and memory storage capacity. Note that ADE-PML formulations provide more flexibility allowing for an easy extension to high-order time integration schemes and are also easier to implement with classical numerical methods as compared to C-PML formulations \cite{Ged10,Mar10,Zha10,Sag14,Xie14,Zha14}. In addition, such ADE-PML formulations can be seen as perturbations of the original (unstretched) formulation of time-domain equations since they reduce to solving the original system of time-domain governing equations over the physical domain by considering all additional fields (auxiliary variables) set to zero and adding the non-zero source terms, boundary and initial conditions imposed over the physical domain of interest.

Some mathematical analysis results on the existence and convergence properties of the PML method can be found in \cite{Col98a,Las98,Las01b,Hoh03,Bao06,Bra07}. The well-posedness and stability properties\footnote{The interested reader can refer to \cite{Kre89,%Beca02,
Beca03,Kre04b,App06a,App06b,Beca15} for the definition of well-posedness, hyperbolicity and stability of initial boundary value problems for hyperbolic systems.} of the PML method have also been studied in numerous research works \cite{Neh96,Aba97,Aba98,Aba99,Pet00,Tei01a,Aba02,Beca02,Diaz03,Beca03,Beca04a,Beca04b,Rah04,App06a,App06b,App09a,Diaz06,Meza08,Meza10,Meza12,Dmi12,Loh09,Sav10a,Sav10b,Hal11,Chen12,Beca12,Joly12,Duru12a,Duru12b,Kal13,Kre13,Dem13,Duru14a,Duru15,Duru16,Hal16,Beca15,Beca17a,Beca17b,Kac17,Ben15,Sal17,Gao17a,Beca18}. On the one hand, for wave propagation problems governed by symmetric hyperbolic systems (\eg{} in electromagnetics, acoustics and elastodynamics), classical split-field PML formulations %(based on an unphysical (or artificial) splitting of the physical fields within the PML region) 
result in weakly hyperbolic %(but neither strictly nor strongly hyperbolic)
systems 
and lead to only weakly well-posed %(but not strongly well-posed) 
problems, which may become ill-posed under some low-order (small) perturbations on the split fields \cite{Aba97,Aba98,Hes98,Beca02,Joly12,Kre13}, thus making numerical computations impracticable. On the other hand, unsplit-field PML formulations result in symmetric strongly hyperbolic systems and lead to strongly well-posed problems \cite{Aba98,Aba99,Pet00,Rah04,Hal11,Duru12a}. Furthermore, standard PML systems derived from both split- and unsplit-field PML formulations have been proved to be only weakly stable under some suitable conditions (depending on the symbol of the %linear 
hyperbolic partial differential operator) \cite{Aba02,Beca02,Beca03,Beca04a,Kom03,Chen12}, which means exponentially growing waves are not supported by the time-domain PML formulation, but linear (or polynomial) growing waves may potentially exist and be triggered for instance by numerical integration (quadrature) or round-off (machine precision) errors \cite{Xie14}. Such growing waves can then lead to spurious instabilities that spread and adversely pollute the solution within the physical domain in long-time simulations. Despite its general success and satisfactory performances in most practical applications, the PML method has some apparent drawbacks and may fail to absorb the outgoing waves in some specific situations \cite{Hu96,Tam98,Hes98,Aba99,Hu01,Beca03,Diaz03,Nav04,Cor04,Cum04,Dong04,Shi06,Diaz06,App06b,Ske07,Kom07,Mar08b,Meza08,Osk08,Loh09,App09b,Li10,Dmi12,Duru12a,Kre13,Bon14,Ping14a,Ping14b,Ping16,Assi15,Beca15,Gao17a,Li17,Li18,Kac17,Beca17a,Beca17b,Beca18}. In particular, numerical instabilities may occur in time-domain simulations for some wave propagation problems in anisotropic and/or dispersive media (such as negative refractive index metamaterials, also called left-handed or double negative metamaterials, and cold plasmas, also called non-thermal plasmas), \eg{} in (visco)elastodynamics \cite{Beca03,App06b,Ske07,Kom07,Ske07,Mar08b,Meza08,App09b,Li10,Dmi12,Duru12a,Bon14,Ping14a,Ping14b,Ping16,Assi15,Beca15,Gao17a,Li17,Li18}, electromagnetics \cite{%Neh96,
Beca03,Cor04,Cum04,Dong04,Shi06,Osk08,Loh09,Beca15,Kac17,Beca17a,Beca17b,Beca18}, aeroacoustics \cite{Hu96,Tam98,Hes98,Aba99,Hu01,Beca03,Diaz03,Diaz06,Dem13}, geophysical fluid dynamics \cite{Nav04} and even in anisotropic scalar wave propagation problems \cite{Kre13}. Such instabilities are due to the presence of backward propagating waves (for which the phase and group velocities point in opposite directions with respect to the interface, \ie{} they are not oriented in the same way and then have differing signs with respect to the direction normal to the interface) exponentially growing within the PML region and producing strong spurious reflections within the physical domain. 
Even for isotropic media, the PML method may not only suffer from long-time instabilities but also exhibit instabilities for near-grazing incident propagating waves or evanescent waves \cite{DeMoe95,Bere98,Bere99,Col98a,Liu99,Win00,Fes03,Mar03,Kom03,Fes05a,Fes05b,Dro07a,Dro07b,Kom07,Mar08a,Mar08b,Mar09,Mar10,Li10,Mat11,Meza08,Meza10,Meza12,Zha10,Zha14}. Besides, the PML method also performs very poorly in the presence of low-frequency propagating waves or long waves (for which the wavenumber is almost zero) arising near cut-off frequencies and decaying slowly within the PML region \cite{Ske07,Bon14}. 

Several methods have been proposed to avoid and remove growing waves and improve the stability and accuracy of PML formulations %(in order to derive stable and accurate PML formulations)
 in the presence of evanescent, near-grazing incident propagating or long waves, \eg{} by applying a numerical filtering (low-pass filter% used to maintain stability of the numerical scheme
 ) \cite{Hu96,Nav04}, by introducing a complex frequency shift %in the complex stretching function 
 (known as the complex-frequency-shifted PML or CFS-PML) \cite{Kuz96,Ged96b,Liu99,Tong99,Rod00a,Rod00b,Bere02a,Bere02b,Beca04a,Fes05a,Fes05b,Cor05a,%Li06a,
 Li06c,%Li08b,
App06a,App06b,Wang06a,Wang06c,Dro07a,Dro07b,Kom07,Gia08,Mar08a,Mar08b,Mar09,Qin09,Ged10,Mar10,Zha10,Li10,Mat11,Gir12,Shi12,Pro08,Pro13,Feng15a,Xie14,Duru12a,Duru12b,Duru14a,Duru14c,Duru15,Gao15,Xie16,Ma18}, by adding artificial stabilizing parameters %in the PML formulation 
 (with loss of the perfect matching property) and/or making use of \textit{ad hoc} space-time coordinate transformations \cite{Tam98,Hes98,Tur98,Aba99,Hu01,Aba02,Aba03,Hag03b,%Hag03c,
 Diaz03,Beca04b,Hu05,Beca06,Diaz06,Parr09}, or by considering different multi-dimensional damping functions (known as the multiaxial PML or M-PML) \cite{Meza08,Meza10,Meza12,%Mar08b,Li10,
 Zeng11,Dmi11,Dmi12,Fat15a,Ping14a,Ping14b,Ping16,Gao17a}. To overcome (or at least alleviate) the stability issues, the complex-frequency-shifted PML (CFS-PML) method consists in using a more sophisticated (frequency-shifted) complex coordinate stretching with a frequency-dependent real part (in addition to the standard frequency-dependent imaginary part) by simply shifting the frequency-dependent pole (zero-frequency singularity) of the complex stretching function off the real axis and onto the negative imaginary axis of the complex plane, so that the PML region can act as a low-pass (%first-order 
Butterworth-type \cite{But30}) filter with a cut-off angular frequency $\omega_c$ to efficiently improve the absorption of near-grazing incident propagating waves \cite{Bere02b,Fes05a,Dro07a,Kom07,Mar08a,Mar08b,Mar09,Mar10,Li10,Zha10} and also the attenuation of evanescent waves \cite{Liu99,Rod00a,Rod00b,Bere02a,Bere02b,Fes05b,Cor06,Dro07b}. The PML region then depends on the angular frequency $\omega$ since it behaves like the physical medium (losing its absorption properties) at low frequencies ($\omega\ll \omega_c$) and like a dissipative layer (losing its stabilizing properties) at high frequencies ($\omega\gg \omega_c$), thus switching from a transparent behaviour (without attenuation) to a dissipative behaviour (with an exponential amplitude decay) in a frequency range around the cut-off frequency $\omega_c$ \cite{Fes05a,Meza08,Mat11}. The CFS-PML method can then be interpreted as a low-pass filter allowing to efficiently absorb high-frequency propagating waves (as it reduces asymptotically to the classical PML method at high frequencies) and improve the absorption of near-grazing incident propagating waves as well as evanescent waves, while slightly degrading the absorption of low-frequency propagating waves \cite{Rod00b,Bere02a,Bere02b,Beca04a,Fes05a,Cor06,Meza08,Mat11}. To overcome this potential limitation, a second-order PML method has been proposed in \cite{Cor05b,Cor06,Lou07,Ged10,Mar10,Feng12,Feng13a,Feng13b,Feng14,Feng17,Wei16a,Wei16b} and extended to a general higher-order PML method in \cite{Gia12,Feng15b}. It allows combining the advantages of both classical PML and CFS-PML methods (in terms of absorption performance) by simply considering a more complicated (multipole) stretching function defined as the product of individual CFS stretching functions. Despite a rather complex implementation in the time domain %(due to the complexity for inverting the stretched time-harmonic (frequency-domain) PML equations back into the time domain by applying an inverse Fourier transform in time) 
and more computational resource requirements (in terms of computational cost and memory storage capacity), the resulting higher-order PML method is then highly effective in absorbing both low-frequency propagating waves and strong evanescent waves, especially for open-region and periodic problems as well as for waveguide problems defined over elongated domains. More recently, another multiple-pole (or multipole) PML method has been introduced in \cite{Gia18} and is based on an \textit{ad hoc} multipole complex stretching function simply defined as the sum of individual CFS stretching functions in order to facilitate the implementation and parameter optimization of the multipole PML method when compared to a higher-order PML method. Note that other generalized (multipole) complex stretching functions have also been considered in \cite{Beca18} to derive stable PML formulations for dispersive %electromagnetic 
media (\eg{} negative index metamaterials). As another attempt to prevent late-time instabilities occurring in some kinds of anisotropic elastic media, the multiaxial PML (M-PML) method extends the classical PML method by considering different multi-dimensional damping functions (that are proportional to each other in orthogonal directions and in which the ratios of damping functions are additional user-tunable correction parameters for improving stability), leading to a more general complex coordinate stretching involving extra stabilizing correction parameters, so that the outgoing waves are absorbed along all coordinate directions (\ie{} not only along the direction normal to the interface% between the physical domain and the absorbing layers
) with different amplitude attenuation factors (decay rates). Nevertheless, it has been shown that such a M-PML method results in an improved sponge layer with better absorption and stability properties in long-time simulations for some particular 2D orthotropic (visco)elastic media \cite{Meza08,Meza10,Meza12,Mar08b,Li10,Ping14a,Ping14b,Ping16,Gao17a} or 2D isotropic (visco)elastic media with high Poisson's ratios \cite{Zeng11,Ping14a,Ping16} or even 2D and 3D general anisotropic elastic media \cite{Gao17a}, but it is not perfectly matched at the interface with the physical domain even within the continuous framework %(in that the theoretical reflection coefficient for an infinite PML region is not exactly zero anymore within the continuous framework) 
\cite{Dmi11,Dmi12,Wang13,Ping14a,Ping14b,Ping16,Gao17a}. In addition, let us mention that the CFS-PML and M-PML methods have been recently combined in \cite{Zha14,Li17,Li18} (hence termed the hybrid PML or H-PML) to further improve the stability in some strongly anisotropic (2D orthotropic) elastic media without degrading the accuracy (absorption capacity), thus combining the advantages of both CFS-PML and M-PML methods for maximizing both accuracy and stability through the optimization (fine tuning) of the PML parameters involved in the H-PML (\ie{} the combined CFS-M-PML) stretching function. Lastly, we refer to \cite{Joh10,Beca15} for further discussions about the limitations and failure cases of the PML method as well as some workarounds.

During the last 25 years, the PML method has been widely used for (transient) electromagnetic problems \cite{Bere94,Chew94,Kat94,Nav94,Sac95,Rap95,Rap96,Bere96a,Bere96b,Kuz96,Ged96a,Ged96b,Fang96,Zhao96a,Zhao96b,Zhao98,Rod97,Kuz97,Bere97,Bere98,Kan97,Zio97,Wu97,Mal97,Yang97,Chew97,Tei97a,Tei97b,Tei98a,Tei98c,Aba97,Aba98,Xu98,Hwa99,Tei99a,Tei99b,Tei99c,Win00,Tei00a,Tei00b,Tei01a,Tei01b,Col98a,Col98b,Tur98,Liu98,He99,Tong99,Bere99,Pet98,Pet00,Rod00a,Rod00b,Fan00,Liu01,Liu04,Aba02,Bere02a,Bere02b,Jiao02a,%Jiao02b,
Jiao03,Ram03a,Ram03b,Ram03c,%Ram03d,
Ram06,Pet03,Ryl04,Beca02,Beca04a,Cor04,Cum04,Dong04,Gond04,Sjo05,Taf05,Cor05a,Cor05b,Cor06,Lou07,Shi06,Aba06,Wang06a,%Wang06b,
Wang06c,Ozg06,Ozg07,%Li06a,Li06b,
Li06c,Li07a,Li07b,Li08a,%Li08b,
Bere07,Don08a,Don08b,deOli07,Gia08,Kau08,Osk08,Loh09,Pro08,Pro13,Aba09,Dos10,Ged10,Gia12,Alv13,Feng12,Feng13a,Feng13b,Feng14,Feng15a,Feng15b,Sun15,Duru12b,Duru14b,Duru16,Wei16a,Wei16b,Beca15,Beca17a,Beca17b,Kac17,Beca18,Gia18}, then quickly extended and successfully applied to various wave propagation problems \cite{Mic07,Liu09,Berm07a,Berm10,Mat13a} in diverse scientific and engineering fields including hydrodynamics and aeroacoustics (%such as linearized and nonlinear Euler and Navier-Stokes equations
\eg{} arising in radiation and scattering problems)
\cite{Hu96,Yuan97,Liu97,Hes98,Qi98,Tam98,Tur98,Hay99,Aba99,Har00,Hu01,%Hu02,
Kat02,%Hag02,
Hag03b,%Hag03c,
Diaz03,Rah04,Beca03,Beca04b,Zeng04b,Hu04a,%Hu04b,
Hu05,%Hu06a,Hu06b,Hag05,
Nat05,Nat06,Har06b,Beca06,Diaz06,App06a,Shir06,Zam07,%Hag07c,
Berm07b,Hu08a,Hu08b,Kor08,Parr09,App09a,Zhou10,%Li10conf,
Isa11,Zam13,Rao13,Kal13,Dem13,Ma14,Velu14,Gao15}, geophysical fluid dynamics (%such as linearized and nonlinear shallow water wave equations
\eg{} arising in atmospheric and oceanic sciences) 
\cite{Dar97,Nav04,Aba03,Lav08,Bar10,Mod10}, elastodynamics (%such as first- and second-order linear elastic wave equations 
\eg{} arising in soil dynamics, soil-structure interaction, earth seismology, geotechnical site characterization and earthquake engineering, geophysical subsurface sensing/probing)
\cite{Chew96,Has96,Zha97,Liu99,Col01,Beca01,Zhe02,Beca03,Kom03,Fes03,Mar03,Wang03,Liu03,Basu03,Zeng04a,Coh05,Fes05a,Fes05b,Bin05,Ma06,Har06b,App06b,Ske07,Dro07a,Dro07b,Kom07,Mar08b,Mar10,Qin09,Gao08,Zha10,Zha11,Basu09,Li10,Meza08,Meza10,Meza12,Kang10,Kuc10,Kuc11,Kuc13,Mat11,Zeng11,Duru12a,Shi12,Kau12,Zhao13,Ben13,Assi15,Xie14,Ping14a,Zha14,Fat15a,Fat15b,Sag14,Duru14c,Duru14d,Duru15,Brun16,Per16,Zhou16,Ben15,Sal17,Ben17,Gao17a,Li17,Li18}, viscoelastodynamics \cite{Basu04,Beca05,Mar09,Ping14b,Ping16}, poroelastodynamics %(such as first- and second-order Biot's equations)
\cite{Zeng01a,Zeng01b,Song05,Mar08a,Guan09,He19}, poroviscoelastodynamics \cite{Gir12}, optics and quantum mechanics (%such as Schr\"{o}dinger, Klein-Gordon and Dirac equations 
\eg{} arising in atomic, molecular and laser physics) \cite{Col97,Ahl99,Pra99,Levy01,Hag03a,Far04,Cheng07,Zhe07,Doh07,Ant08,Doh09,Ant11,Nis11,Men14,Pin15,Ant17}, fluid-porous medium coupled problems or fluid-solid interaction problems (\eg{} arising in marine seismology, in non-destructive testing techniques such as ultrasound evaluation techniques for the characterization of complex biological systems) \cite{Berm07a,Berm10,Calvo10,Mat12,Mat13a,Assi16,Xie16,Shi16}, as well as other general wave-like hyperbolic problems \cite{Lio02,App06a,App07,App09a}. Initially derived in Cartesian coordinates for simply-shaped domains with straight (planar) artificial boundaries (\eg{} squared, rectangular or cuboidal domains), some PML formulations have been extended to other (non-Cartesian) coordinate systems such as polar, cylindrical and spherical coordinates \cite{Mal97,Yang97,Chew97,Tei97a,Tei97b,Yang98,Col98a,Col98b,Tei98c,%Qi98,%Liu98,
He99,Liu99,Tei99c,Zhao00,Tei00a,Pet00,Zhe02,Pet03,Liu03,Berm07a,%Guan09,
Kuc10,Sag14}, %providing absorbing boundary conditions in these coordinate systems, 
and applied to generally-shaped convex domains using local polygonal or (non-)orthogonal curvilinear coordinates \cite{Tei98a,Tei99a,Hwa99,Tei00b,Tei01a,Las01a,Las01b,Liu01,Liu04,Fes05a,Zsc06,Ozg06,Ozg07,Zam07,Don08b,Gao08,Dos10,Zha11,Zha14,Alv13,Mat13a,Dem13,Velu14,Mod17}. Such improvements provide more flexibility and choice in the shape and geometry of the underlying physical domain of interest. As already pointed out in \cite{Zhe02}, recall that the first attempts to extend the classical PML formulation (written in Cartesian coordinates) to non-orthogonal curvilinear coordinates (such as proposed in \cite{Nav94,Wu95,Rod97,Kuz97%,deOli07,Rao13
} for instance) were based on an approximate impedance matching condition derived under the assumption that the complex coordinate stretching function is independent of the spatial coordinates, hence resulting in an approximate PML formulation. Furthermore, it should be noticed that a straightforward extension of the classical PML formulation to cylindrical coordinates (such as proposed in \cite{Kuz97} for instance) leads a loss of the perfect matching (reflectionless) property for curvilinear (non-planar) interfaces even within the continuous framework, hence resulting in an approximate PML formulation (henceforth known as the quasi-PML) \cite{Liu98,He99}. In the present work, for the sake of simplicity, we restrict to rectangular or cuboidal domains with straight (planar) boundaries in Cartesian coordinates, but the proposed PML formulation could be easily extended and applied to more complex domains in other coordinate systems.

\subsection{Overview of existing PML formulations in linear elastodynamics}\label{sec:elastodynamics}

Following this general overview of the PML method and its variants addressed in the literature, we now focus on the different PML formulations as well as their implementations in linear (visco)elastodynamics. Classical split-field PML formulations for the linear elastodynamic equations have been initially derived for the first-order mixed (velocity-stress) formulation \cite{Chew96,Has96,Zha97,Liu99,Col01,Beca01,%Zeng01b,
Beca03,Fes03,Mar03,Liu03,Coh05,Ma06,Gao08,Zha11}, then extended to the second-order (displacement-based) formulation (with a velocity field as additional variable to render the resulting mixed formulation second-order in time) \cite{%Zeng01a,
Kom03,Zhao13} and to the second-order mixed (displacement-stress) formulation of the viscoelastodynamic equations \cite{Beca05}. Classical unsplit-field PML formulations have also been obtained for the linear elastic wave equations formulated as a first-order mixed (velocity-stress) system \cite{Wang03,%Song05,Guan09,He19
Ben17} and as a second-order system in displacement or velocity (with potentially a stress, strain and/or other auxiliary field(s) as additional unknown(s), such as time-integral or memory-like terms, thus resulting in a second-order mixed formulation) \cite{Zhe02,Basu03,Basu04,Basu09,Bin05,Har06b,Ske07,Kang10,Kuc10,Kuc11,Kuc13,Ben13,Ben15,Sal17,Ben17,%Sag14%(first-order velocity-stress system from second-order equations in Bin05),
Fat15a,Fat15b,Assi15,Brun16,Zhou16}. Later, split-field CFS-PML (resp. M-PML) formulations have been proposed for the linear elastic wave equations written as a first-order system in velocity and stress \cite{Fes05a,Fes05b,Mar08b} (resp. \cite{Meza08,Meza10,Meza12,Zeng11,Gao17a} and as a second-order system in displacement \cite{Ping14a,Ping14b,Ping16}). Lastly, unsplit-field CFS-PML (resp. M-PML) formulations have been developed based on either the first-order mixed (velocity-stress) formulation \cite{App06b,Dro07a,Dro07b,Kom07,%Mar08a,
Mar09,Mar10,Qin09,Zha10,Zha11,%Gir12,
Duru15,Per16} (with an unsplit-field H-PML, \ie{} combined CFS-M-PML, formulation proposed in \cite{Zha14,Li17} and an unsplit-field combined CFS-MZT-PML formulation developed in \cite{Shi12}) or the second-order (displacement-only or mixed displacement-stress) formulation (with potentially additional auxiliary fields) \cite{Li10,Mat11,Duru12a,Duru14c,Duru14d,Xie14} (resp. \cite{Fat15a}, with an unsplit-field H-PML formulation developed in \cite{Li18}) of the linear elastodynamic equations. A brief overview of the relations between different split-field and unsplit-field (CFS-)PML formulations and their classification for the linear elastic wave equations formulated as a first-order mixed (velocity-stress) system can be found in \cite{Kri09}. All the aforementioned PML formulations require either the use of convolution products (and related recursive convolution update techniques) or specific (specialized or recursive) time integration schemes with sometimes a large number of additional fields and auxiliary differential equations as well as additional assumptions on the solution fields to derive the PML formulations in the time domain. By and large, the PML formulations differ not only by the use of split or unsplit primary fields (and auxiliary fields if should be the case) and the definition of the complex coordinate stretching function, but also by the choice of numerical approximation methods for space and time discretizations. The interested reader can refer to \cite{Kuc10,Kuc11,Kuc13,Fat15a,Xie14} for an overview of the various PML formulations and implementations (based on FD, FE or SE methods) developed for time-domain elastodynamic problems. As already pointed out in \cite{Kom03,Kom07,Sag14}, most existing PML formulations based on the linear elastic wave equations written as a first-order system in velocity and stress have been developed for FD methods and cannot be straightforwardly used in classical numerical approximation methods that are based on the linear elastic wave equations written as a second-order system in displacement, such as most FE, SE, DG methods and some FD methods. Conversely, PML formulations based on the second-order (displacement-based) linear elastodynamic equations can be more readily implemented in such numerical schemes \cite{Kom07,Li10,Kang10,Kuc10,Kuc11,Kuc13,Duru12a,Xie14,Sag14,Assi15,Zhou16,Ben17} and are more robust (\ie{} have much better discrete stability properties) than the ones based on first-order (velocity-stress) linear elastodynamic equations \cite{Duru12a,Kre13,Assi15,Ben17}.

\subsection{Introduction of an efficient PML formulation in linear elastodynamics}\label{sec:novelty}

The main objective of this paper is to derive both time-harmonic (frequency-domain) and time-dependent (time-domain) PML formulations as well as their numerical implementations for simulating the propagation of elastic waves in both two- and three-dimensional unbounded media. %In this paper, we consider the linear elastodynamic equations in two- and three-dimensional multi-layer isotropic homogeneous media. 
For this purpose, we develop an unsplit-field PML formulation for second-order (displacement-based) linear elastodynamic equations, which is well-suited for numerical implementation using standard FE, SE or DG methods on unstructured meshes that are well adapted to handle computational domains with arbitrarily complex geometries and curved boundaries and allow for a natural application of boundary conditions, as already mentioned in \cite{Kuc10,Sag14}. Despite the zero-frequency singularity, we adopt a standard stretching function with careful parameterization, since it leads to a simple and straightforward implementation of the resulting PML formulation and exhibits better performance in the presence of low-frequency propagating waves (and in the absence of strong evanescent waves) when compared to CFS-PML formulations, as clearly explained in \cite{Fes05a,Mat11,Kuc11,Xie14} and numerically observed in \cite{Cor06} for instance. 
In order to relax or at least alleviate the temporal complexity arising from the PML-transformed equations and to derive an unsplit-field non-convolutional PML formulation in the time domain for both two- and three-dimensional transient elastodynamic analyses, we first introduce an auxiliary (symmetric) tensor-valued strain field (defined as the difference between the complex stretched tensor-valued strain field and the real classical one) treated as an additional unknown variable (only in the PML region) and governed by a second-order ordinary differential equation in time derived from the strain-displacement kinematic relation. We then propose to construct a space weak formulation of both the equilibrium equations and the strain-displacement kinematic relation, resulting in a mixed (displacement-strain) unsplit-field PML formulation that is second-order in time and can be discretized in space using a standard Galerkin finite element method. We finally employ a classical implicit second-order time integration scheme commonly used in computational structural dynamics, that is a Newmark time scheme coupled with a finite difference (Crank-Nicolson) time scheme, in order to solve the time-dependent PML equations while preserving second-order accuracy. The resulting mixed PML formulation can be easily implemented using classical numerical tools incorporated in existing (displacement-based) finite element codes. Such codes originally developed for bounded computational domains can then be easily modified to accommodate both time-harmonic (frequency-domain) and time-dependent (time-domain) PML formulations for the propagation of elastic waves in unbounded domains. In addition, it requires no field splitting, no nonlinear solvers (such as Newton-type iterative solvers), no mass lumping and no critical time step size (such as the stability criterion required in explicit time integration schemes) as well as no complicated or costly convolution operations in time and no high-order spatial and temporal derivatives, so that the derivation, resolution and implementation of the proposed PML formulation are made easier compared to most other existing PML formulations. As in \cite{Kang10,Kuc10,Kuc11,Kuc13,Fat15a,Fat15b,Assi15}, the proposed PML formulation preserves the second-order form of the original (unstretched) time-domain linear elastic wave equations while using a very small number of additional fields and auxiliary differential equations as compared to most existing PML formulations based on either first- or second-order equations for time-domain elastic wave propagation in both two- and three-dimensional unbounded domains. For further computational savings, we finally consider a hybrid approach, originally developed for two-dimensional elastodynamic problems in \cite{Kuc13} and later extended to three-dimensional elastodynamic problems in \cite{Fat15a}, that couples a mixed displacement-strain unsplit-field PML formulation in the PML region with a purely (non-mixed) displacement-based formulation (\ie{} a standard displacement-only formulation) in the physical domain.

\subsection{Outline of the paper}\label{sec:outline}

The paper is organized as follows. In Section~\ref{sec:formulation}, we introduce the linear elastodynamic formulation in both time and frequency domains. In Section~\ref{sec:PML}, we revisit the time-harmonic (frequency-domain) and time-dependent (time-domain) PML formulations based on complex coordinate stretching by introducing an auxiliary tensor-valued strain field and using a space weak formulation of the resulting PML equations. We also discuss the proposed hybrid approach, whereby a mixed displacement-strain unsplit-field PML formulation in the PML region is coupled with a standard (non-mixed) displacement-only formulation in the physical domain of interest, and provide details on its finite element implementation in both the frequency domain and the time domain. In Section~\ref{sec:results}, we present numerical experiments carried out on two-dimensional linear elastodynamic problems in single- and multi-layer isotropic homogeneous media to numerically simulate the propagation of elastic waves in semi-infinite (unbounded) domains. The %efficiency 
capability of the proposed time-dependent PML formulation to efficiently absorb outgoing propagative waves is shown by computing relevant local and global quantities of interest such as the components of the displacement field at some selected receiving points (called receivers) and the kinetic, internal and total energies stored in the physical domain of interest. The performances of the PML formulation in terms of accuracy (absorption capability) are compared to the ones of a classical absorbing layer (CAL) formulation. Finally in Section~\ref{sec:concl}, we conclude with further discussions and give possible perspectives.

\section{Problem statement}\label{sec:formulation}

\subsection{Time-domain formulation of linear elastodynamics}\label{sec:timeformulation}

Within the framework of linear elasticity theory, we consider a two- or three-dimensional linear elastodynamic problem. The material is assumed to be linear and elastic, characterized by the fourth-order Hooke's elasticity tensor $\Cb$ and the mass density $\rho$. The medium is subjected to given external forces represented by a body force field $\fb$. In the time domain, the original problem, initially defined on an unbounded domain of the $d$-dimensional Euclidean physical space $\Rbb^d$ ($d=2,3$ being the spatial dimension), consists in finding the vector-valued displacement field $\ub(\xb,t)$ and its associated tensor-valued Cauchy stress field $\sigmab(\xb,t)$ satisfying the second-order partial differential equation (derived from linear momentum conservation and hereinafter referred to as the equilibrium equations)
\begin{subequations}\label{systemtime}
\begin{equation}\label{equilibriumtime}
\rho \ddot{\ub} - \divb(\sigmab) = \fb,
\end{equation}
for spatial position vector $\xb \in \Rbb^d$ and time $t>0$, with given initial conditions at time $t=0$
\begin{equation}\label{initialconditions}
\ub(\xb,0) = \ub_0(\xb) \quad \text{and} \quad \dot{\ub}(\xb,0) = \vb_0(\xb),
\end{equation}
where $\divb$ denotes the divergence operator of a tensor-valued field with respect to $\xb$, while the dot %$\dot{}$ 
and the double (or superposed) dots %$\ddot{}$ 
over a variable denote respectively the first- and second-order partial derivatives with respect to time $t$, \ie{} $\dot{\ub}(\xb,t) = \dfrac{\partial \ub}{\partial t}(\xb,t)$ and $\ddot{\ub}(\xb,t) = \dfrac{\partial^2 \ub}{\partial t^2}(\xb,t)$. Body force field $\fb(\cdot,t)$ is %assumed to be 
a given %(square integrable) 
function from $\Rbb^d$ into $\Rbb^d$ for all $t\geq 0$. Both fields $\ub$ and $\sigmab$ are linked through the linear stress-strain constitutive relation (known as the generalized Hooke's law)
\begin{equation}\label{constitutiverelationtime}
\sigmab = \Cb : \epsilonb(\ub),
\end{equation}
in which the colon symbol $:$ denotes the twice contracted tensor product, and where $\epsilonb(\ub(\xb,t))$ is the classical linearized tensor-valued strain field associated to displacement field $\ub(\xb,t)$, that is
\begin{equation}\label{kinematicrelationtime}
\epsilonb(\ub) = \frac{1}{2}\(\nablab \ub + (\nablab \ub)^T\),
\end{equation}
where the superscript ${}^T$ denotes the transpose operator and $\nablab$ (resp. $\nabla$) denotes the gradient operator of a vector-valued (resp. scalar-valued) field with respect to $\xb$.
\end{subequations}
Recall that both second-order tensor-valued stress field $\sigmab% = \sigma_{ij} \ebf_i \otimes \ebf_j
$ and strain field $\epsilonb% = \varepsilon_{ij} \ebf_i \otimes \ebf_j
$ are symmetric, \ie{} $\sigma_{ij}=\sigma_{ji}$ and $\varepsilon_{ij} = \varepsilon_{ji}$. Also, the fourth-order elasticity tensor $\Cb% = C_{ijkl} \ebf_i \otimes \ebf_j \otimes \ebf_k \otimes \ebf_l
$ satisfies the classical major and minor symmetry properties $C_{ijkl} = C_{klij} = C_{jikl} = C_{ijlk}$ (without summation) and the following ellipticity (hence, positive-definiteness) and boundedness properties on $\Rbb^d$ (see \eg{} \cite{Hug87}) : there exist constants $0<\alpha\leq\beta<+\infty$ such that $0< \alpha \norm{\zetab}^2 \leq \zetab : \Cb : \zetab \leq \beta \norm{\zetab}^2 < +\infty$ for all non-zero second-order symmetric real tensors $\zetab% = \zeta_{ij} \ebf_i \otimes \ebf_j
=(\zeta_{ij})_{1\leq i,j\leq d}$ of dimension $d(d+1)/2$, where $\zetab : \Cb : \zetab = C_{ijkl} \zeta_{ij}\zeta_{kl}$ and $\norm{\zetab}^2 = \zeta_{ij}^2$ (with summation over all indices). Besides, the mass density $\rho$ is a strictly positive-valued field.

In the case of isotropic homogeneous linear elastic materials, the constitutive relation between $\sigmab$ and $\epsilonb(\ub)$ restricts to
\begin{equation}\label{Hookelaw}
\begin{aligned}
\sigmab &= \lambda \tr(\epsilonb(\ub)) \Ib + 2 \mu \epsilonb(\ub)\\
&= \lambda \diver(\ub) \Ib + \mu (\nablab \ub + (\nablab \ub)^T),
\end{aligned}
\end{equation}
where $\tr$ denotes the trace operator, $\diver$ denotes the divergence operator of a vector-valued field with respect to $\xb$, $\Ib$ is the second-order symmetric identity (or unit) tensor such that $I_{ij} = \delta_{ij}$, with the Kronecker delta $\delta_{ij}=1$ if $i=j$ and $\delta_{ij}=0$ if $i\neq j$, $\lambda>0$ and $\mu>0$ are the Lam\'e's coefficients (that must be strictly positive according to the ellipticity properties of elasticity tensor $\Cb$ \cite{Wang95,Ernst98}) defined by
\begin{equation*}
\lambda = \begin{dcases}
\dfrac{E \nu}{(1+\nu) (1-2\nu)} & \text{in 2D under plane strain assumption and in 3D} \\
\dfrac{E \nu}{1-\nu^2} & \text{in 2D under plane stress assumption}
\end{dcases}
\end{equation*}
and
\begin{equation*}
\mu = \dfrac{E}{2 (1+\nu)} \quad \text{(also called shear modulus)}
\end{equation*}
with $E$ and $\nu$ respectively the Young's modulus and the Poisson's ratio of the isotropic linear elastic material. The fourth-order elasticity tensor $\Cb$ is then constituted of $2$ algebraically independent coefficients $(\lambda,\mu)$ (or equivalently, $(E,\nu)$) independent of $\xb$ such that $C_{ijkl} = \lambda \delta_{ij} \delta_{kl} + \mu (\delta_{ik} \delta_{jl} + \delta_{il} \delta_{jk})$, leading to $\sigma_{ij} = C_{ijkl} \varepsilon_{kl} = \lambda \varepsilon_{kk} \delta_{ij} + 2 \mu \varepsilon_{ij}$ (with summation over indices $k$ and $l$ only). In such an isotropic homogeneous linear elastic medium, the time-dependent elastic wave system \eqref{systemtime} of homogeneous equations (under no external body forces) set on an unbounded domain of $\Rbb^d$ admits bulk (or body) plane wave solutions, namely longitudinal pressure (or compressional) waves (often referred to as $P$-waves) with characteristic velocity $c_p = %\sqrt{\dfrac{\lambda+2\mu}{\rho}}
\sqrt{(\lambda+2\mu)/\rho}$ and whose polarization vector is parallel to the wavefront propagation direction, and transverse shear waves (often referred to as $S$-waves) with characteristic velocity $c_s = %\sqrt{\dfrac{\mu}{\rho}}
\sqrt{\mu/\rho} < \sqrt{2} c_p$ and whose polarization vector is perpendicular to the wavefront propagation direction. It also admits surface- or interface-guided wave solutions, such as Rayleigh waves (propagating along a free surface) and Stoneley waves (propagating along the interface between two semi-infinite elastic media), whose amplitudes decay exponentially in the direction normal to and away from the surface (or the interface if should be the case) \cite{Basu03}. In the more general case of anisotropic and/or heterogeneous linear elastic media, the fourth-order elasticity tensor $\Cb$ is constituted of $21$ algebraically independent coefficients possibly dependent on $\xb$. The time-dependent elastic wave system \eqref{systemtime} admits further plane wave solutions with different characteristic velocities (depending on the material symmetry properties and on the wavefront propagation direction), such as quasi-longitudinal (or quasi-compressional) waves (often referred to as $qP$-waves) and quasi-shear (or quasi-transverse) waves (often referred to as $qS$-waves) for orthotropic elastic media \cite{Beca03,App06b,Meza08}, as well as other wave types in the presence of heterogeneities and interfaces.

For computational purposes, we introduce a bounded domain $\Omega_{\text{BD}}\subset \Rbb^d$ with sufficiently smooth boundary $\partial \Omega_{\text{BD}}$, where the subscript ${}_{\text{BD}}$ refers to ``Bounded Domain'', which corresponds to the physical domain of interest over which the numerical computations are performed. We assume that the external body force field $\fb$ and the initial displacement and velocity fields $\ub_0$ and $\vb_0$ have bounded supports in $\Omega_{\text{BD}}$ and therefore remain confined within the bounded (or finite) computational domain. The problem of interest then consists in finding the solution fields $\ub$ and $\sigmab$ satisfying the time-dependent elastic wave system \eqref{systemtime} %, that is the equilibrium equations \eqref{equilibriumtime}, the initial conditions \eqref{initialconditions}, the stress-strain constitutive relation \eqref{constitutiverelationtime} and the strain-displacement kinematic relation \eqref{kinematicrelationtime}, 
only in the physical bounded domain $\Omega_{\text{BD}}$ (and not in the entire unbounded domain of $\Rbb^d$).
 
 \subsection{Frequency-domain formulation of linear elastodynamics}\label{sec:freqformulation}
 
In the frequency domain, the time-harmonic elastic wave system is obtained by taking the Fourier transform in time of \eqref{systemtime} and reads
\begin{subequations}\label{systemfreq}
\begin{align}
-\rho \omega^2 \hat{\ub} - \divb(\hat{\sigmab}) &= \hat{\fb},\label{equilibriumfreq}\\
\hat{\sigmab} &= \Cb : \epsilonb(\hat{\ub}),\label{constitutiverelationfreq}\\
\epsilonb(\hat{\ub}) &= \frac{1}{2}\(\nablab \hat{\ub} + (\nablab \hat{\ub})^T\),\label{kinematicrelationfreq}
\end{align}
\end{subequations}
where $\omega>0$ is the angular frequency and the hat (or caret) %$\hat{}$ 
over a variable denotes the Fourier transform in time, \ie{} $\hat{\ub}(\cdot,\omega) = \displaystyle\int_{\Rbb} \ub(\cdot,t) e^{-\mathrm{i}\omega t} \, dt$, in which $\mathrm{i}$ denotes the unit imaginary number satisfying $\mathrm{i}^2 = -1$, and where the spatial and frequency dependencies of variables are dropped for the sake of readability and notational brevity. The time-harmonic dependence of the displacement, stress and strain fields is chosen as the multiplicative factor $e^{\mathrm{i}\omega t}$ by convention. %A time-harmonic regime in $e^{\mathrm{i}\omega t}$ is chosen by convention.
Furthermore, all the vector-valued and tensor-valued fields are expressed throughout the paper in a Cartesian coordinate system with canonical basis $\set{\ebf_i}_{1 \leq i\leq d}$.

\section{Perfectly matched layers}\label{sec:PML}

We now revisit the PML method that aims at constructing artificial %absorbing non-reflecting 
layers (of finite thickness) surrounding the physical bounded domain $\Omega_{\text{BD}}$ and designed in such a way that the incoming waves (entering into the layers) are not reflected at the interface with the physical domain and decay exponentially into them. In the following, we implicitly assume that all the outgoing waves (leaving the physical domain of interest) correspond to purely forward propagating waves in the direction normal to the interface, \ie{} there is no backward propagating waves or evanescent waves.

\begin{figure}
\centering
\tikzsetnextfilename{PML_infinite_domain}
\begin{tikzpicture}[scale=0.75]

\def \lx {4}
\def \ly {2}
\def \hx {1}
\def \hy {1}
%\draw[help lines,step=.5] (-\lx-\hx-1,-\ly-\hy-1) grid (\lx+\hx+1,\ly+\hy+1);

% Origin
\coordinate (O) at (0,0);

% Domain coordinates
\coordinate (A) at (-\lx,-\ly);
\coordinate (B) at (\lx,\ly);
\coordinate (C) at (-\lx,\ly);
\coordinate (D) at (\lx,-\ly);

% PML coordinates
\coordinate (E) at (-\lx-\hx,-\ly-\hy);
\coordinate (F) at (\lx+\hx,\ly+\hy);
\coordinate (G) at (-\lx-\hx,\ly+\hy);
\coordinate (H) at (\lx+\hx,-\ly-\hy);
\coordinate (I) at (-\lx,-\ly-\hy);
\coordinate (J) at (\lx,-\ly-\hy);
\coordinate (K) at (\lx+\hx,-\ly);
\coordinate (L) at (\lx+\hx,\ly);
\coordinate (M) at (\lx,\ly+\hy);
\coordinate (N) at (-\lx,\ly+\hy);
\coordinate (P) at (-\lx-\hx,\ly);
\coordinate (Q) at (-\lx-\hx,-\ly);
\coordinate (P1) at (-\lx-\hx/2,0);
\coordinate (P2) at (0,-\ly-\hy/2);
\coordinate (Interface) at (\lx,\ly/2);
\coordinate (Boundary) at (\lx+\hx,\ly/2);

% Domain
\draw[black,thick,fill=cyan!30] (A) rectangle (B);
\draw (O) node[below left] {$O$} node{$\bullet$};
\node[above right=0.5] at (A) {$\Omega_{\text{BD}}$};
\draw[solid,black,stealth-stealth] (0,-0.5) -- (\lx,-0.5) node[midway,below]{$\ell_1$};
\draw[solid,black,stealth-stealth] (-0.5,0) -- (-0.5,\ly) node[midway,left]{$\ell_2$};

% PML
\fill[preaction={fill=magenta!30}, pattern=custom north east lines, hatchspread=3pt, hatchthickness=1pt, hatchcolor=orange!30] (A) rectangle (E);
\fill[preaction={fill=magenta!30}, pattern=custom north east lines, hatchspread=3pt, hatchthickness=1pt, hatchcolor=orange!30] (B) rectangle (F);
\fill[preaction={fill=magenta!30}, pattern=custom north west lines, hatchspread=3pt, hatchthickness=1pt, hatchcolor=orange!30] (C) rectangle (G);
\fill[preaction={fill=magenta!30}, pattern=custom north west lines, hatchspread=3pt, hatchthickness=1pt, hatchcolor=orange!30] (D) rectangle (H);
\fill[fill=orange!30] (A) rectangle (J);
\fill[fill=orange!30] (C) rectangle (M);
\fill[fill=magenta!30] (Q) rectangle (C);
\fill[fill=magenta!30] (D) rectangle (L);
\draw[black,thick] (E) rectangle (F);
\draw[dashed,black,thin] (A) -- (I);
\draw[dashed,black,thin] (A) -- (Q);
\draw[dashed,black,thin] (D) -- (J);
\draw[dashed,black,thin] (D) -- (K);
\draw[dashed,black,thin] (B) -- (L);
\draw[dashed,black,thin] (B) -- (M);
\draw[dashed,black,thin] (C) -- (N);
\draw[dashed,black,thin] (C) -- (P);
\node[above right=0.2] at (E) {$\Omega_{\text{PML}}$};
\node at (P1) {$\Omega_1$};
\node at (P2) {$\Omega_2$};
\draw[solid,black,stealth-stealth] (\lx,-0.5) -- (\lx+\hx,-0.5) node[midway,below]{$h_1$};
\draw[solid,black,stealth-stealth] (-0.5,\ly) -- (-0.5,\ly+\hy) node[midway,left]{$h_2$};

% Interface
\node[above right=0.1] at (Interface) {$\Gamma$};
\node[above right=0.1] at (Boundary) {$\Gamma_{\text{ext}}$};

% Axis
\draw[solid,black,-stealth] (0,0) -- (\lx+\hx+1,0) node[below]{$x_1$};
\draw[solid,black,-stealth] (0,0) -- (0,\ly+\hy+1) node[left]{$x_2$};

\end{tikzpicture}
\caption{Physical bounded domain $\Omega_{\text{BD}}$ surrounded by an artificial PML region $\Omega_{\text{PML}} = \bigcup_{i=1}^d \Omega_i$ divided into $d$ overlapping subregions $\Omega_i$ with interface $\Gamma = \partial\Omega_{\text{BD}} \cap \partial\Omega_{\text{PML}}$ and exterior boundary $\Gamma_{\text{ext}} = \partial\Omega_{\text{PML}}\setminus \Gamma$}\label{fig:PML}
\end{figure}
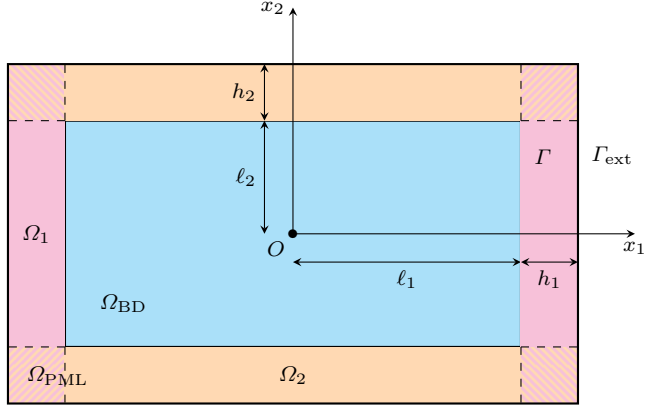

We consider a bounded Cartesian domain $\Omega_{\text{BD}} = \bigtimes\limits_{i=1}^{d} \intervalcc{-\ell_i}{\ell_i} \subset \Rbb^d$ surrounded by an artificial absorbing layer (the PML region) $\Omega_{\text{PML}} = \bigtimes\limits_{i=1}^{d} \intervalcc{-\ell_i-h_i}{\ell_i+h_i} \setminus \Omega_{\text{BD}}$ with a constant thickness $h_i$ in each $x_i$-direction. %, for $i=1,\dots,d$ 
The PML region $\Omega_{\text{PML}}$ is divided into $d$ overlapping subregions $\set{\Omega_i}_{i=1}^{d}$ such that $\Omega_i = \setst{\xb\in \Omega_{\text{PML}}}{\abs{x_i} > \ell_i}$ and the outgoing propagative waves are attenuated in the $x_i$-direction within the PML subregion $\Omega_i$ (see Figure~\ref{fig:PML}). We denote by $\Gamma = \partial\Omega_{\text{BD}} \cap \partial\Omega_{\text{PML}}$ (resp. $\Gamma_i = \partial\Omega_{\text{BD}} \cap \partial\Omega_i$) the interface between the physical domain $\Omega_{\text{BD}}$ and the PML region $\Omega_{\text{PML}}$ (resp. the PML subregion $\Omega_i$), and by $\nb = n_i \ebf_i$ the unit normal vector to $\Gamma$ pointing outward $\Omega_{\text{BD}}$. Furthermore, we define the entire domain $\Omega = \Omega_{\text{BD}} \cup \Omega_{\text{PML}}$ as the disjoint union between $\Omega_{\text{BD}}$ and $\Omega_{\text{PML}}$, and we denote by $\Gamma_{\text{ext}}$ the part of its boundary $\partial\Omega$ that coincides with the exterior boundary of $\Omega_{\text{PML}}$, that is $\Gamma_{\text{ext}} = \partial\Omega_{\text{PML}}\setminus (\Gamma \cup \Gamma_{\text{free}})$, where $\Gamma_{\text{free}}$ stands for the possible free surfaces in the case where a semi-infinite medium with free-surface boundary conditions is considered.

\subsection{Complex coordinate stretching}\label{sec:complexcoordinatestretching}

In the frequency domain, the PML method consists in analytically stretching one or more real spatial coordinate(s) of the physical wave equations into the complex plane $\Cbb$, the ones corresponding to the direction(s) along which the outwardly propagating waves have to be attenuated in the absorbing layers. %Such an analytic continuation of the physical wave equations and solutions into the complex coordinate plane implicitly assumes that the medium properties in the PML region are analytic functions along the direction(s) of the complex stretched coordinate(s).
In the following, for any complex number $z\in \Cbb$, we denote by $\re(z)$ and $\im(z)$ the real and imaginary parts of $z$, respectively.

Following the complex coordinate stretching approach \cite{Chew94,Rap95,Rap96,Chew96,Chew97}, we introduce a complex change of spatial variable $\Sb\colon \xb \in \Omega\subset \Rbb^d \mapsto \Sb(\xb) = \tilde{\xb} \in \Cc = \Sb(\Omega) \subset \Cbb^d$ defining a complex stretching (analytic continuation) $\tilde{x}_i = S_i(x_i) \in \Cc_i$ of the real spatial coordinate $x_i$ along a curve $\Cc_i$ in the complex plane $\Cbb$ parameterized by 
\begin{equation}\label{coordstretch}
\tilde{x}_i = S_i(x_i) = \int_0^{x_i} s_i(\xi) \, d\xi,
\end{equation} 
where the tilde %\textasciitilde{} 
over a variable or an operator denotes its stretched version in the complex space, and $s_i(x_i)$ is a complex coordinate stretching function defining the absorption profile along the $x_i$-direction, which is a nowhere zero, continuous, complex-valued function such that
\begin{equation}\label{stretchingcondition}
\begin{cases}
\im\(s_i(x_i)\) < 0 & \text{in } \Omega_i, \\
s_i(x_i) = 1 & \text{elsewhere}.
\end{cases}
\end{equation}
Condition \eqref{stretchingcondition} ensures that $\tilde{x}_i = S_i(x_i) \in \Cbb\setminus\Rbb$ in the PML subregion $\Omega_i$, whereas $\tilde{x}_i = S_i(x_i) = x_i \in \Rbb$ elsewhere and in particular in the physical domain $\Omega_{\text{BD}}$, so that the outgoing propagative waves are perfectly transmitted at the interface $\Gamma$ and attenuated (or damped) in the $x_i$-direction inside the PML subregion $\Omega_i$.
As an inverse Fourier transform in time may be required to come back into the time domain by inverting the frequency-domain PML equations, the stretching function $s_i(x_i)$ is usually chosen frequency-dependent \cite{Chew94,Rap95,Rap96,Col01,Kom03,Beca03,Berm07b,Ske07,Kom07,Dro07a,Berm10,Gro10,Kre13,Mod14,Beca15,Zhou16,Mod17} and defined as
\begin{equation}\label{stretchingfunction}
s_i(x_i) = 1 + \frac{d_i(x_i)}{\mathrm{i}\omega}% = 1 - \mathrm{i} \frac{d_i(x_i)}{\omega}
,
\end{equation}
where $d_i(x_i)$ is the so-called damping (or absorption or attenuation) function that is a continuous and monotonically increasing (\ie{} non-decreasing) positive-valued function of $x_i$ such that 
\begin{equation}\label{dampingfunction}
\begin{cases}
d_i(x_i) > 0 & \text{in } \Omega_i, \\
d_i(x_i) = 0 & \text{elsewhere}.
\end{cases}
\end{equation}
The dependence of the stretching function $s_i(x_i)$ on the factor $\mathrm{i}\omega$ in \eqref{stretchingfunction} allows for an easy application of the inverse Fourier transform in time to the time-harmonic (frequency-domain) PML equations, and thus a simple derivation of the time-dependent (time-domain) PML equations. It should be noted that the particular form \eqref{stretchingfunction} of stretching function $s_i(x_i)$ exhibits a zero-frequency singularity (zero-frequency pole). Therefore, the derivation of the PML formulation considered in this work naturally excludes the static case corresponding to $\omega = 0$, that means it does not allow static solutions (displacement and stress fields associated to a given static loading) to be recovered \cite{Kuc11}. In theory, any choice of damping function $d_i(x_i)$ satisfying \eqref{dampingfunction} ensure both continuity at the interface $\Gamma_i$ between $\Omega_i$ and $\Omega_{\text{BD}}$ and exponential amplitude decay along the $x_i$-direction in the PML subregion $\Omega_i$ for the outgoing propagative waves. Since the damping function $d_i(x_i)$ does not depend on angular frequency $\omega$, it further allows obtaining an exponential amplitude decay of the propagating waves in the PML subregion $\Omega_i$ that is independent of $\omega$ for non-dispersive PML media (see Section~\ref{sec:absorptionproperties} for a discussion). In practice, typical choices for $d_i(x_i)$ in the PML subregion $\Omega_i$ are low-degree polynomial functions \cite{Bere94,Chew96,Col98a,Col98b,Col01,Zhe02,Mar03,Wang03,Kom03,Fes03,Basu03,Beca03,Sin04,Taf05,Coh05,Fes05a,Har06b,App06b,Ske07,Bere07,Zhe07,Kom07,Dro07a,Mar08a,Mar08b,Mar09,Qin09,Zha10,Li10,Mat11,Zeng11,Kang10,Kuc10,Kuc11,Kuc13,Xie14,Ping14a,Ping14b,Ping16,Assi15,Brun16,Per16,Zhou16} or singular (non-integrable) hyperbolic or shifted hyperbolic functions \cite{Berm04,Berm07a,Berm07b,Berm08,Berm10,Kal13,Mod10,Mod13,Mod14,Cim15,Mod17} according to the underlying problem. In the numerical experiments shown in Section~\ref{sec:results}, we adopt the most widely used damping profile (due to its straightforward implementation) by choosing a polynomial function $d_i(x_i)$ with a power law dependence on the distance $\abs{x_i}-\ell_i$ to ensure a progressive (gradual) and smoothly increasing damping %(with a smoothly varying profile) 
of the outgoing propagative waves across the PML subregion $\Omega_i$, that is
\begin{equation}\label{powerlawfunction}
d_i(x_i) = d_i^{\mathrm{max}} \(\frac{\abs{x_i}-\ell_i}{h_i}\)^p \quad \text{in } \Omega_i,
\end{equation}
where $p$ is the polynomial degree (that is a nonnegative integer) and $d_i^{\mathrm{max}}>0$ is the damping coefficient (that is a strictly positive parameter) having the same unit as the angular frequency $\omega$ and corresponding to the maximum value of damping function $d_i(x_i)$. Both $p$ and $d_i^{\mathrm{max}}$ are user-chosen scalar parameters allowing the shape (sharpness) and the intensity (strength) of the imposed damping profile of the propagating waves to be controlled within the PML subregion $\Omega_i$ \cite{Kuc11,Kuc13}. From a computational viewpoint, choosing a smoothly varying damping profile is of particular interest for the attenuation of outgoing propagative waves within the PML region to be adequately resolved by the spatial discretization in order to minimize the spurious reflections. Furthermore, for any classical numerical method that implicitly assumes continuity of the spatial derivatives of the solution field, choosing a constant or linear function $d_i(x_i)$ in the PML subregion $\Omega_i$ (\ie{} taking $p=0$ or $p=1$ in \eqref{powerlawfunction}) can generate spurious numerical reflections at the interface $\Gamma_i$. By contrast, for given values of the damping coefficient $d_i^{\mathrm{max}}$ and the PML thickness $h_i$, choosing a quadratic or higher-degree polynomial function $d_i(x_i)$ in the PML subregion $\Omega_i$ (\ie{} taking $p\geq 2$ in \eqref{powerlawfunction}) avoids those numerical issues and also minimizes the spurious reflections produced at the exterior boundary of the PML subregion $\Omega_i$ \cite{Col98b,Sin04,Bin05,Ske07}. In the numerical examples presented in Section~\ref{sec:results}, we choose a quadratic form for $d_i(x_i)$ (with a fixed polynomial degree $p=2$) so as to enforce a smooth and gradual spatial attenuation of the outgoing propagating waves within the PML subregion $\Omega_i$.

In the general case of heterogeneous materials, we further assume that the mass density $\rho$ and the components $C_{ijkl}$ of the fourth-order elasticity tensor $\Cb$ depend analytically on $\xb$ in the $d$-dimensional complex manifold $\Cc = \Sb(\Omega)$, so that the solution fields $\ub$ and $\sigmab$ of problem \eqref{systemtime} (or equivalently, their Fourier transforms in time $\hat{\ub}$ and $\hat{\sigmab}$ solution of problem \eqref{systemfreq}) are also analytic with respect to $\xb$ and can be evaluated along the complex path $\Cc$ \cite{Osk08,Berm10,Joh10}.

\subsection{Reflectionless and absorption properties}\label{sec:absorptionproperties}

The wave propagation properties induced by the complex coordinate stretching in the PML region $\Omega_{\text{PML}}$ can be merely analyzed by means of a plane wave analysis \cite{%Bere94,
Col01,Basu03,Basu04,Beca03,Fes03,Fes05a,App06b,Bere07,Berm10,Meza08,Meza10,Meza12,Duru12a,Beca15}. We consider harmonic propagative plane waves corresponding to particular solutions of the time-dependent elastic wave system \eqref{systemtime} (or equivalently, of its time-harmonic counterpart \eqref{systemfreq}) of homogeneous equations (under no external body force field, \ie{} with $\fb = \zerob$) of the form
\begin{equation}\label{propagativewave}
\ub(\xb,t) = \Ab e^{\mathrm{i}(\omega t - \scalprod{\kb}{\xb})},
\end{equation}
where $\Ab$ is the polarization vector, $\kb = k_i \ebf_i = k \pb$ is the wave vector, with $k=\norm{\kb}$ the wavenumber, and $\pb = p_i \ebf_i$ is the unit vector denoting the wavefront propagation direction, with $p_i=k_i/k$ the direction cosinus of $\kb$ along the $x_i$-direction, and $\scalprod{\kb}{\xb} = k_i x_i$ denotes the usual Euclidean inner product between $\kb$ and $\xb$ in $\Rbb^d$ and $\norm{\kb}^2 = \scalprod{\kb}{\kb} = k_i^2$ the associated Euclidean norm of $\kb$ in $\Rbb^d$, where the Einstein notation convention is used for summation over repeated indices. The %complex-valued coordinate transformation 
complex-valued change of coordinates $S_i\colon x_i \mapsto S_i(x_i) = \tilde{x}_i$ introduced in \eqref{coordstretch} allows a propagative plane wave of the form \eqref{propagativewave} to be transformed into an attenuated wave
\begin{equation}\label{attenuatedwave}
\tilde{\ub}(\xb,t) = \ub(\tilde{\xb},t) = \Ab e^{\mathrm{i}(\omega t - \scalprod{\kb}{\tilde{\xb}})} %= \Ab e^{\mathrm{i}(\omega t - \scalprod{\kb}{(\xb+\alphab(\xb)/(\mathrm{i} k))}} 
= \Ab e^{\mathrm{i}(\omega t - \scalprod{\kb}{\xb})} e^{- \scalprod{\alphab(\xb)}{\pb}}
%= \ub(\xb,t) \, e^{- \scalprod{\alphab(\xb)}{\pb}} 
= \ub(\xb,t) \, e^{-\alpha(\xb)}
\end{equation}
exponentially decreasing as $\abs{x_i}$ increases in each PML subregion $\Omega_i$, where $\alphab(\xb) = \alpha_i(x_i) \ebf_i$ is the amplitude attenuation vector whose components $\alpha_i(x_i)$ are such that
\begin{equation}\label{attenuationfactor}
\alpha_i(x_i) = \begin{cases}
\dfrac{k}{\omega} \int_0^{x_i} d_i(\xi) \, d\xi >0 & \text{in } \Omega_i, \\
0 & \text{elsewhere},
\end{cases}
\end{equation}
and $\alpha(\xb) = \scalprod{\alphab(\xb)}{\pb} = \alpha_i(x_i) p_i$ (with summation over index $i$) is the so-called amplitude attenuation factor (or decay rate). The phase of the outgoing propagative plane waves is maintained as they penetrate the PML subregion $\Omega_i$, whereas the amplitude is modulated by an exponential factor $e^{-\alpha(\xb)}$ that introduces an exponential decay of the outgoing propagative plane waves along the $x_i$-direction once they enter the PML subregion $\Omega_i$. As the attenuation factor $\alpha(\xb)$ depends on the direction of propagation $\pb$ (characterizing the angle of incidence) and on the phase velocity $v_p = \omega/k$ but not directly on the angular frequency $\omega$ itself, it turns out to be frequency-independent (\ie{} independent of $\omega$) for non-dispersive PML media, thus leading to a uniform exponential decay of the outgoing propagative plane waves with respect to the angular frequency $\omega$. Also note that, using \eqref{attenuationfactor}, the stretching function $s_i(x_i)$ and damping function $d_i(x_i)$ introduced in \eqref{stretchingfunction} can be respectively expressed as
\begin{equation}\label{stretchingfunctionbis}
s_i(x_i) = 1 + \frac{\alpha_i'(x_i)}{\mathrm{i}k}% = 1 - \mathrm{i} \frac{\alpha_i'(x_i)}{k}
\quad \text{and} \quad d_i(x_i) = \alpha_i'(x_i) \frac{\omega}{k}% = \alpha_i'(x_i) \frac{\mathrm{i}\omega}{\mathrm{i}k}
,
\end{equation}
where the prime %${}'$ 
over a variable denotes the first derivative with respect to spatial coordinate $x_i$, \ie{} $\alpha_i'(x_i) = \dfrac{d\alpha_i}{d x_i}(x_i)$. Also, the complex stretched coordinate $\tilde{x}_i = S_i(x_i)$ defined in \eqref{coordstretch} can be rewritten as \cite{Chew94,Rap96,Col01,Kom03,Beca03,Kom07}
\begin{equation}\label{coordstretchbis}
\tilde{x}_i = S_i(x_i) = x_i + \frac{1}{\mathrm{i}\omega} \int_0^{x_i} d_i(\xi) \, d\xi = x_i + \frac{\alpha_i(x_i)}{\mathrm{i} k}.
\end{equation}
Except for near-grazing incident propagative waves (for which $\scalprod{\kb}{\nb}\approx 0$), all the outgoing propagative waves (for which $\scalprod{\kb}{\nb}>0$) are exponentially damped in the PML region $\Omega_{\text{PML}}$ since the amplitude attenuation factor $\alpha(\xb)$ is positive and increases as $\abs{x_i}$ increases in each PML subregion $\Omega_i$, while all the ingoing ones (for which $\scalprod{\kb}{\nb}<0$) are exponentially amplified in $\Omega_{\text{PML}}$ since the amplitude attenuation factor $\alpha(\xb)$ is negative and increases as $\abs{x_i}$ decreases in $\Omega_i$.

As all outgoing propagative waves are expected to be damped in the PML region $\Omega_{\text{PML}}$, an \textit{ad hoc} (Dirichlet, Neumann, characteristic or mixed) boundary condition can be applied on the exterior boundary $\Gamma_{\text{ext}}$ of the PML region. Let us mention that, provided that the PML model is carefully parameterized, the choice of boundary conditions specified on the exterior boundary of $\Omega_{\text{PML}}$ does not seem to have a significant impact on the PML absorption performance \cite{Col98b,Pet98}. In the numerical experiments presented in Section~\ref{sec:results} (as in most practical applications), for the sake of convenience and simplicity, we choose to impose homogeneous Dirichlet boundary conditions %$\ub=\zerob$ 
on the exterior boundary %$\Gamma_{\text{ext}}$ 
of $\Omega_{\text{PML}}$, that is for all $t\geq 0$,
\begin{equation}\label{boundaryconditions}
\ub(\cdot,t)=\zerob \quad \text{on } \Gamma_{\text{ext}}% \quad (\text{\ie{} at } \abs{x_i}=\ell_i+h_i)
.
\end{equation}

The absorptive and attenuative properties in the PML region $\Omega_{\text{PML}}$ can be analyzed by studying the reflection of an incident plane wave at the fixed exterior boundary $\Gamma_{\text{ext}}$ of $\Omega_{\text{PML}}$. A propagative plane wave traveling outward from the physical domain $\Omega_{\text{BD}}$ is absorbed into the PML region $\Omega_{\text{PML}}$ without any reflections at the interface $\Gamma$ between $\Omega_{\text{BD}}$ and $\Omega_{\text{PML}}$. At the continuous level, the PML region $\Omega_{\text{PML}}$ is thus said perfectly matched to the physical domain $\Omega_{\text{BD}}$. A propagative plane wave entering and traversing the PML subregion $\Omega_i$, being reflected at the exterior boundary $\Gamma_{\text{ext}}$ of $\Omega_{\text{PML}}$ and again propagating back through $\Omega_i$ is attenuated by an exponential factor \cite{Col98a,Col98b,Col01,Ske07,Kal13}
\begin{equation}\label{reflectioncoefficient}
R_i = \exp(-2 \alpha_i(\ell_i+h_i) p_i) = \exp\(-2 \dfrac{p_i}{v_p} \int_{\ell_i}^{\ell_i+h_i} d_i(\xi) \, d\xi\)
\end{equation}
before re-entering the physical domain $\Omega_{\text{BD}}$, as it travels twice through the PML subregion $\Omega_i$ (of finite thickness $h_i$). The theoretical reflection coefficient $R_i$ \textit{a priori} depends on the choice of three user-defined and user-tunable PML parameters, namely the position $\ell_i$ of the interface $\Gamma_i$, the thickness $h_i$ of the PML subregion $\Omega_i$ and the damping function $d_i(x_i)$, but it is also influenced by the phase velocity $v_p = \omega/k$ and the direction of propagation $p_i$ of the plane waves along the $x_i$-direction. For the quadratic damping function $d_i(x_i)$ previously defined in \eqref{powerlawfunction} (with a polynomial degree $p=2$), the reflection coefficient $R_i$ can be rewritten as \cite{Bere94,Bere96b,Col01,Ske07,Mat11}
\begin{equation}\label{reflectioncoefficientparabolic}
R_i = \exp\(-\frac{2 d_i^{\mathrm{max}} h_i p_i}{3 v_p}\).
\end{equation}
As a result, the reflected-wave amplitude $R_i$ defined in \eqref{reflectioncoefficientparabolic} decreases as the PML thickness $h_i$, the wavefront propagation direction $p_i$ (along the $x_i$-direction) or the damping coefficient $d_i^{\mathrm{max}}$ increases, whereas it increases as the phase velocity $v_p$ increases. Note that it does not depend of the size $\ell_i$ of the physical domain $\Omega_{\text{BD}}$. Also, for non-dispersive media, the reflection coefficient $R_i$ is independent of the angular frequency $\omega$ since the phase velocity $v_p = \omega/k$ is equal to the wave velocity \cite{Fes05a}. In practical computations, increasing the PML thickness $h_i$ allows the outgoing propagative waves to be more attenuated but may induce substantial computational costs. Besides, for a given PML thickness $h_i$, the damping coefficient $d_i^{\mathrm{max}}$ cannot be chosen arbitrarily large, since large values of $d_i^{\mathrm{max}}$ would require fine spatial discretizations to well reproduce and approximate the fast attenuation of outgoing waves within the PML region with sufficiently good numerical accuracy, thus leading to high computational costs. Indeed, disregarding any numerical integration or rounding error, the discretization error arising from the numerical approximation method can cause spurious reflections at the interface $\Gamma$ between $\Omega_{\text{BD}}$ and $\Omega_{\text{PML}}$ \cite{Bin05}. Also, it becomes predominant compared to the truncation error related to the spurious reflections at the exterior boundary of the PML subregion $\Omega_i$ (of finite thickness $h_i$) as the damping coefficient $d_i^{\mathrm{max}}$ increases \cite{Berm07b,Ske07}. Therefore, for fixed values of PML thickness $h_i$ and discretization parameters, choosing an optimal damping coefficient $d_i^{\mathrm{max}}$ is essential for producing minimal reflections from the exterior boundary of the PML region and thus maximizing the attenuation of outgoing waves while maintaining a desired accuracy level. Such an optimal value results from a trade-off between absorption strength (requiring large enough values of $d_i^{\mathrm{max}}$) and numerical accuracy (requiring small enough values of $d_i^{\mathrm{max}}$) \cite{Col98a,Col98b,Col01,Asv03,Har06b,Sag14}. It \textit{a priori} depends on the PML thickness $h_i$, the wavefront propagation direction $p_i$, the phase velocity $v_p$ and the discretization error introduced by the numerical scheme. In practice, the damping coefficient $d_i^{\mathrm{max}}$ is usually chosen such that the reflection coefficient $R_i$ at normal incidence (\ie{} for $p_i=1$) is equal to an arbitrarily small value $R_0<1$ (typically $10^{-3}$ to $10^{-8}$) and defined as \cite{Col98a,Col98b,Col01,Beca03,Beca04a,Kom03,Fes05a,Ma06,Ske07,Kom07,Dro07a,Mar08a,Mar09,Qin09,Li10,Kang10,Kuc10,Kuc11,Kuc13,Fat15a,Mat11,Zeng11,Xie14,Ping14a,Ping14b,Ping16,Assi15,Per16,Zhou16}
\begin{equation}\label{dampingcoefficient}
d_i^{\mathrm{max}} = \frac{3 v_p}{2 h_i} \log\(\dfrac{1}{R_0}\),
\end{equation}
or more generally as \cite{Fes03,Fes05a,Meza08,Meza10,Meza12}
\begin{equation}\label{dampingcoefficientgeneral}
d_i^{\mathrm{max}} = A_i \frac{v_p}{h_i},
\end{equation}
with $A_i$ a given dimensionless constant adapted to the problem being considered. Note that there are only few guidelines reported in the literature and based on either simple heuristics (\textit{ad hoc} rules of thumb), \textit{a priori} error estimates or costly general nonlinear optimization algorithms (see \cite{Col98b,Dat02,Asv03,Sin04,Sjo05,Bin05,Har06b,Ske07,Mic07,Lis08,Nis11,Sag14} for instance) to properly select appropriate values for the PML model and discretization parameters. Hence, as already pointed out in \cite{Kuc10,Kuc11,Kuc13}, owing to a critical lack of rigorous methodology and simple process for providing proper guidance on the choice of damping functions and on the selection of optimal PML parameters, the derivation of an optimal %bounded 
damping function $d_i$ independently of the underlying problem and the discretization method is still an open question. A possible alternative would consist in using the singular (non-integrable) damping functions proposed in \cite{Berm04,Berm07a,Berm07b,Berm08,Berm10} that exhibit a power-type singularity at the exterior boundary $\Gamma_{\text{ext}}$ of the PML region and allows recovering a reflection coefficient $R_i = 0$ for any angle of incidence and for any angular frequency, since $d_i$ is a non-integrable function such that $\int_{\ell_i}^{\ell_i+h_i} d_i(\xi) \, d\xi = +\infty$ \cite{Berm04,Berm07b,Berm08}. Numerical computations \cite{Berm07b,Berm08,Rab10,Mod10,Mod13,Mod14,Mod17} have shown that such (shifted) hyperbolic damping functions do not require any tuning of the PML parameters (whatever the numerical method employed) and provide good results that may outperform (or at least be comparable with) those obtained with optimized polynomial damping functions.

\subsection{Frequency-domain PML formulation}\label{sec:PMLformulationfreq}

The derivation of the time-harmonic (frequency-domain) PML equations classically results from the application of the complex coordinate stretching \eqref{coordstretch} to the original time-harmonic governing equations \eqref{systemfreq}% so that the resulting system of time-harmonic PML equations govern the wave motion over the entire computational domain $\Omega$ consisting of both the (physical) bounded domain $\Omega_{\text{BD}}$ and the (non-physical) PML region $\Omega_{\text{PML}}$ \cite{Kang10,Kuc10,Kuc11,Sag14}
.

\subsubsection{Classical frequency-domain PML formulation based on complex coordinate stretching}\label{sec:classicalPMLformulationfreq}

%We classically obtain the time-harmonic PML formulation by applying the complex change of spatial variable $S_i\colon x_i \mapsto S_i(x_i)=\tilde{x}_i$ defined in \eqref{coordstretch} to the time-harmonic elastic wave system \eqref{systemfreq}. 
From the complex coordinate transformation \eqref{coordstretch} together with \eqref{stretchingfunction} and making use of the fundamental theorem of calculus, we have the following relation and derivative rule between the complex stretched coordinate system and the real physical coordinate system
\begin{equation*}
\frac{\partial \tilde{x}_i}{\partial x_i} = s_i(x_i) = 1 + \frac{d_i(x_i)}{\mathrm{i}\omega} %= 1 + \frac{\alpha_i'(x_i)}{\mathrm{i}k} 
\quad \text{and} \quad \frac{\partial(\cdot)}{\partial \tilde{x}_i} = \frac{1}{s_i(x_i)} \frac{\partial(\cdot)}{\partial x_i} = \frac{\mathrm{i}\omega}{\mathrm{i}\omega+d_i(x_i)} \frac{\partial(\cdot)}{\partial x_i},
\end{equation*}
where the partial derivatives $\dfrac{\partial(\cdot)}{\partial \tilde{x}_i}$ with respect to the complex %spatial 
coordinates $\tilde{x}_i$ are interpreted in the usual Cauchy-Riemann sense \cite{Rud74,Berm10}, \ie{} $\dfrac{\partial(\cdot)}{\partial \tilde{x}_i} = \dfrac{\partial \re(\cdot)}{\partial \re(\tilde{x}_i)} - \mathrm{i} \dfrac{\partial \im(\cdot)}{\partial \im(\tilde{x}_i)}$. The jacobian matrix $\Jb = \nablab \Sb %= \dfrac{\partial S_i}{\partial x_j}(x_i) \ebf_i \otimes \ebf_j 
= \dfrac{\partial \tilde{x}_i}{\partial x_j} \ebf_i \otimes \ebf_j$ of the complex coordinate transformation \eqref{coordstretch} is then a complex-valued diagonal matrix that is everywhere non-singular and defined by
\begin{equation}\label{defjacobianmatrix}
\Jb = s_i(x_i) \ebf_i \otimes \ebf_i,
\end{equation}
in which the symbol $\otimes$ denotes the usual tensor product. In the following, the functional (spatial) dependence of stretching function $s_i$ and damping function $d_i$ will be often omitted for the sake of readability and notational brevity.
%For any scalar-valued field $\psi$, the gradient $\widetilde{\nabla} \psi = \dfrac{\partial \psi}{\partial \tilde{x}_i} \ebf_i$ can be expressed in terms of the original gradient $\nabla \psi = \dfrac{\partial \psi}{\partial x_i} \ebf_i$ as
%\begin{subequations}
%\begin{equation*}
%\widetilde{\nabla} \psi = \Jb^{-T} \nabla \psi %= \Jb^{-1} \nabla \psi 
%= \frac{1}{s_i} \frac{\partial \psi}{\partial x_i} \ebf_i.
%\end{equation*}
%Similarly, for any vector-valued field $\psib = \psi_i \ebf_i$, the gradient $\widetilde{\nablab} \psib = \dfrac{\partial \psi_i}{\partial \tilde{x}_j} \ebf_i \otimes \ebf_j$ can be expressed in terms of the original gradient $\nablab \psib = \dfrac{\partial \psi_i}{\partial x_j} \ebf_i \otimes \ebf_j$ as
%\begin{equation*}
%\widetilde{\nablab} \psib = (\nablab \psib) \Jb^{-1} = \frac{1}{s_j} \frac{\partial \psi_i}{\partial x_j} \ebf_i \otimes \ebf_j.
%\end{equation*}
%\end{subequations}

Therewith, the analytic continuation of the time-harmonic elastic wave system \eqref{systemfreq} of homogeneous equations (in the absence of body force field, \ie{} with $\hat{\fb} = \zerob$) in the PML region $\Omega_{\text{PML}}$ is basically obtained by replacing the partial derivatives $\dfrac{\partial(\cdot)}{\partial x_i}$ with respect to $x_i$ by the partial derivatives $\dfrac{\partial(\cdot)}{\partial \tilde{x}_i}$ with respect to $\tilde{x}_i$ in \eqref{systemfreq} and reads \cite{Beca03,Basu03,Basu04,Bin05,Kuc11,Kuc13,Xie14}
\begin{subequations}\label{systemfreqPML}
\begin{align}
-\rho \omega^2 \hat{\ub} - \widetilde{\divb}(\hat{\sigmab}) &= \zerob,\label{equilibriumfreqPML}\\
\hat{\sigmab} &= \Cb : \tilde{\epsilonb}(\hat{\ub}),\label{constitutiverelationfreqPML}\\
\tilde{\epsilonb}(\hat{\ub}) &= \frac{1}{2}\(\widetilde{\nablab} \hat{\ub} + (\widetilde{\nablab} \hat{\ub})^T\),\label{kinematicrelationfreqPML}
\end{align}
\end{subequations}
which is similar to the original time-harmonic elastic wave system \eqref{systemfreq} defined over the physical domain $\Omega_{\text{BD}}$, except that the external body force field $\hat{\fb}$ has been set to zero (since $\hat{\fb}$ is non-vanishing only within $\Omega_{\text{BD}}$) while the partial differential operators $\nabla$, $\nablab$, $\divb$ and $\epsilonb$ in the real physical $\xb$-coordinate system have been replaced with the frequency-dependent complex stretched operators $\widetilde{\nabla}$, $\widetilde{\nablab}$ and $\widetilde{\divb}$ and $\tilde{\epsilonb}$ in the complex stretched $\tilde{\xb}$-coordinate system, respectively defined by
%\begin{alignat*}{2}
%\widetilde{\nabla} \hat{u} &= \Jb^{-T} \nabla \hat{u} %= \Jb^{-1} \nabla \hat{u}
%= \frac{1}{s_i} \frac{\partial \hat{u}}{\partial x_i} \ebf_i, \quad &&\widetilde{\divb}(\hat{\sigmab}) = \widetilde{\nabla} \cdot \hat{\sigmab} = (\Jb^{-T} \nabla) \cdot \hat{\sigmab} = \frac{1}{s_j} \frac{\partial \hat{\sigma}_{ij}}{\partial x_j} \ebf_i,\\
%\widetilde{\nablab} \hat{\ub} &= (\nablab \hat{\ub}) \Jb^{-1} = \frac{1}{s_j} \frac{\partial \hat{u}_i}{\partial x_j} \ebf_i \otimes \ebf_j, \quad &&\tilde{\epsilonb}(\hat{\ub}) %= \frac{1}{2}\(\widetilde{\nablab} \hat{\ub} + (\widetilde{\nablab} \hat{\ub})^T\) 
%= \frac{1}{2}\((\nablab \hat{\ub}) \Jb^{-1} + \Jb^{-T} (\nablab \hat{\ub})^T\) %= \frac{1}{2}\((\nablab \hat{\ub}) \Jb^{-1} + ((\nablab \hat{\ub}) \Jb^{-1})^T\) 
%= \frac{1}{2}\(\frac{1}{s_j} \frac{\partial \hat{u}_i}{\partial x_j} + \frac{1}{s_i} \frac{\partial \hat{u}_j}{\partial x_i}\) \ebf_i \otimes \ebf_j.
%\end{alignat*}
\begin{align*}
\widetilde{\nabla} \hat{u} &= \Jb^{-T} \nabla \hat{u} %= \Jb^{-1} \nabla \hat{u}
= \frac{1}{s_i} \frac{\partial \hat{u}}{\partial x_i} \ebf_i, \\
\widetilde{\nablab} \hat{\ub} &= (\nablab \hat{\ub}) \Jb^{-1} = \frac{1}{s_j} \frac{\partial \hat{u}_i}{\partial x_j} \ebf_i \otimes \ebf_j, \\
\widetilde{\divb}(\hat{\sigmab}) &= \widetilde{\nabla} \cdot \hat{\sigmab} = (\Jb^{-T} \nabla) \cdot \hat{\sigmab} = \frac{1}{s_j} \frac{\partial \hat{\sigma}_{ij}}{\partial x_j} \ebf_i,\\
\tilde{\epsilonb}(\hat{\ub}) %= \frac{1}{2}\(\widetilde{\nablab} \hat{\ub} + (\widetilde{\nablab} \hat{\ub})^T\) 
&= \frac{1}{2}\((\nablab \hat{\ub}) \Jb^{-1} + \Jb^{-T} (\nablab \hat{\ub})^T\) %= \frac{1}{2}\((\nablab \hat{\ub}) \Jb^{-1} + ((\nablab \hat{\ub}) \Jb^{-1})^T\) 
= \frac{1}{2}\(\frac{1}{s_j} \frac{\partial \hat{u}_i}{\partial x_j} + \frac{1}{s_i} \frac{\partial \hat{u}_j}{\partial x_i}\) \ebf_i \otimes \ebf_j.
\end{align*}
Note that the complex coordinate stretching \eqref{coordstretch} allows the symmetry properties of the second-order tensor-valued strain field $\epsilonb$ to be preserved, so that the second-order complex stretched tensor-valued strain field $\tilde{\epsilonb}% = \tilde{\varepsilon}_{ij} \ebf_i \otimes \ebf_j
$ is symmetric, \ie{} $\tilde{\varepsilon}_{ij} = \tilde{\varepsilon}_{ji}$. Besides, although by construction the complex coordinate stretching \eqref{coordstretch} implicitly excludes the static case for which $\omega = 0$ \cite{Kuc10,Kuc11,Kuc13,Fat15a,Assi15}, we consider continuous extensions of the complex stretched strain tensor $\tilde{\epsilonb}(\hat{\ub}(\xb,\omega))$ and stress tensor $\hat{\sigmab}(\xb,\omega)$ so that in the static case, \ie{} for $\omega=0$, we have for all $\xb \in \Omega_{\text{PML}}$,
\begin{equation}\label{zerofreq}
\tilde{\epsilonb}(\hat{\ub}(\xb,0)) = \zerob \quad \text{and} \quad \hat{\sigmab}(\xb,0) = \zerob.
\end{equation}
Note that condition \eqref{zerofreq} is (often implicitly) stated as an assumption in other PML formulations previously derived in the literature, such as in \cite{Basu04,Basu09,Kang10,Kuc10,Kuc11,Kuc13,Fat15a,Assi15,Brun16}.

Multiplying \eqref{equilibriumfreqPML} by $\det(\Jb)$, where $\det$ denotes the determinant operator, and noticing that $\det(\Jb)/s_j = \prod_{\substack{i=1, i \neq j}}^{d} s_i$ is independent of $x_j$, the complex stretched (PML-transformed) time-harmonic elastic wave system \eqref{systemfreqPML} can be rewritten as a system similar to \eqref{systemfreq} with classical real partial differential operators but for an anisotropic medium made up with a specific frequency-dependent complex anisotropic heterogeneous material. It reads \cite{Zhe02,Har06b,Li10,Mat11,Mat13a,Xie14,Shi16}
\begin{subequations}\label{systemfreqPMLaniso}
\begin{align}
-\tilde{\rho} \omega^2 \hat{\ub} - \divb(\hat{\tilde{\sigmab}}) &= \zerob,\label{equilibriumfreqPMLaniso}\\
\hat{\tilde{\sigmab}} &= \widetilde{\Cb} : \nablab \hat{\ub},\label{constitutiverelationfreqPMLaniso}
\end{align}
\end{subequations}
where $\tilde{\rho} = \rho \det(\Jb) = \rho \prod_{i=1}^d s_i$ is a frequency-dependent complex scalar-valued field, $\hat{\tilde{\sigmab}} = \hat{\sigmab} \det(\Jb) \Jb^{-1}$ is a second-order non-symmetric complex tensor-valued field such that $\hat{\tilde{\sigma}}_{ij} = \hat{\sigma}_{ij} \det(\Jb) / s_j$, and $\widetilde{\Cb}$ is a fourth-order frequency-dependent complex anisotropic %heterogeneous 
tensor-valued field such that $\widetilde{C}_{ijkl} = C_{ijkl} \det(\Jb)/(s_j s_l)$ with only the major symmetry property preserved, but not the minor ones (due to the non-symmetry of both fields $\hat{\tilde{\sigmab}}$ and $\nablab \hat{\ub}$)\footnote{As already mentioned in \cite{Har06b,Berm10,Mat13a} and unlike stated in \cite{Li10,Mat11}, the fourth-order tensor-valued field $\widetilde{\Cb}$ does have the major symmetry property, \ie{} $\widetilde{C}_{ijkl} = \widetilde{C}_{klij}$, but both minor ones are lost \ie{} $\widetilde{C}_{ijkl} \neq \widetilde{C}_{jikl}$ and $\widetilde{C}_{ijkl} \neq \widetilde{C}_{ijlk}$.}. In view of the time-harmonic PML formulation \eqref{systemfreqPMLaniso}, the PML region can be interpreted as an artificial anisotropic heterogeneous absorbing medium with specific complex material properties \cite{Sac95,Ged96a,Ged96b,Zhao96a,Zhao96b,Zhao98,Zio97,Wu97,Tei97b,Tei98a,Tei98b,Tei99b,Aba98}.

\begin{remark}
Alternatively, left multiplying (or pre-multiplying) \eqref{constitutiverelationfreqPMLaniso} by $\Jb^{-T}$, the constitutive relation \eqref{constitutiverelationfreqPMLaniso} can be rewritten as \cite{Fes05a}
\begin{equation*}
\hat{\tilde{\sigmab}}' = \widetilde{\Cb}' : \nablab \hat{\ub},
\end{equation*}
where $\hat{\tilde{\sigmab}}' = \Jb^{-T} \hat{\tilde{\sigmab}} = \det(\Jb) \Jb^{-T} \hat{\sigmab} \Jb^{-1}$ is a second-order symmetric complex tensor-valued field such that $\hat{\tilde{\sigma}}'_{ij} = \hat{\tilde{\sigma}}_{ij} / s_i = \hat{\sigma}_{ij} \det(\Jb) / (s_i s_j)$, and $\widetilde{\Cb}'$ is a fourth-order frequency-dependent complex anisotropic %heterogeneous 
tensor-valued field such that $\widetilde{C}'_{ijkl} = \widetilde{C}_{ijkl}/s_i = C_{ijkl} \det(\Jb)/(s_i s_j s_l)$ with only the left minor symmetry property preserved (resulting from the symmetry of $\hat{\tilde{\sigmab}}'$), but neither the right minor one, nor the major one (due to the non-symmetry of $\nablab \hat{\ub}$).
\end{remark}

\begin{remark}
Another anisotropic medium re-interpretation of the time-harmonic PML formulation \eqref{systemfreqPML} is proposed in \cite{Bin05}. It allows for a simple implementation of the time-harmonic (frequency-domain) PML equations within the context of finite element methods. Rewriting the kinematic relation \eqref{kinematicrelationfreq} as $\epsilonb(\hat{\ub}) = \Lb : \nablab \hat{\ub}$, where $\Lb$ is the fourth-order symmetric identity (or unit) tensor such that $L_{ijpq} = (\delta_{ip} \delta_{jq} + \delta_{iq} \delta_{jp})/2$, the complex stretched kinematic relation \eqref{kinematicrelationfreqPML} then becomes $\tilde{\epsilonb}(\hat{\ub}) = \widetilde{\Lb} : \nablab \hat{\ub}$, where $\widetilde{\Lb}$ is a fourth-order complex tensor-valued field such that $\widetilde{L}_{ijpq} = (\delta_{ip} (\Jb^{-1})_{qj} + (\Jb^{-1})_{qi} \delta_{jp})/2 = (\delta_{ip} \delta_{jq}/s_j + \delta_{iq} \delta_{jp}/s_i)/2$. Note that both tensors $\Lb$ and $\widetilde{\Lb}$ satisfy only the left minor symmetry property (due to the symmetry of strain tensors $\epsilonb(\hat{\ub})$ and $\tilde{\epsilonb}(\hat{\ub})$), but not the major and right minor ones (due to the non-symmetry of $\nablab \hat{\ub}$). The time-harmonic elastic wave system \eqref{systemfreqPMLaniso} can then be recast in the same form as the original one \eqref{systemfreq} and reads \cite{Bin05}
\begin{subequations}\label{systemfreqPMLanisosym}
\begin{align}
-\tilde{\rho} \omega^2 \hat{\ub} - \divb(\hat{\tilde{\sigmab}}) &= \zerob,\label{equilibriumfreqPMLanisosym}\\
\hat{\tilde{\sigmab}} &= \widetilde{\Cb}'' : \epsilonb(\hat{\ub}),\label{constitutiverelationfreqPMLanisosym}
\end{align}
\end{subequations}
where $\widetilde{\Cb}''$ is a fourth-order frequency-dependent complex anisotropic %heterogeneous 
tensor-valued field such that $\widetilde{C}''_{pqrs} = \widetilde{L}_{ijpq} C_{ijkl} \widetilde{L}_{klrs} \det(\Jb)$ which inherits the major and minor symmetry properties of $\Cb$ (thanks to the left minor symmetry of $\widetilde{\Lb}$).
\end{remark}

\subsubsection{Mixed frequency-domain PML formulation with additional auxiliary field}\label{sec:mixedPMLformulationfreq}

For the stretching function $s_i$ chosen as in \eqref{stretchingfunction}, the jacobian matrix $\Jb$ defined in \eqref{defjacobianmatrix} can be rewritten as
\begin{equation}\label{jacobianmatrix}
\Jb = \Ib + \frac{1}{\mathrm{i}\omega} \Db,
\end{equation}
where $\Db = d_i(x_i) \ebf_i \otimes \ebf_i$ is a diagonal real matrix with positive diagonal terms. The stress field $\hat{\tilde{\sigmab}}$ can then be expressed in terms of the original one $\hat{\sigmab}$ as
\begin{align}\label{stretchedstress}
\hat{\tilde{\sigmab}} = \begin{cases}
\hat{\sigmab} \(\Ib + \dfrac{1}{\mathrm{i}\omega} \Db_2^{\Sigma}\) & \text{for } d=2, \\
\hat{\sigmab} \(\Ib + \dfrac{1}{\mathrm{i}\omega} \Db_3^{\Sigma} -\dfrac{1}{\omega^2} \Db_3^{\Pi}\) & \text{for } d=3,
\end{cases}
\end{align}
where $\Db_2^{\Sigma}$, $\Db_3^{\Sigma}$ and $\Db_3^{\Pi}$ are diagonal real matrices expressed in the canonical basis $\set{\ebf_i}_{1\leq i \leq d}$ of $\Rbb^d$ such that
\begin{align*}
\Db_2^{\Sigma} &= \begin{bmatrix}
d_2 & 0\\
0 & d_1
\end{bmatrix}
= \(\sum_{\substack{j=1, j \neq i}}^{2} d_j\) \ebf_i \otimes \ebf_i, \\
\Db_3^{\Sigma} &= \begin{bmatrix}
d_2+d_3 & 0 & 0\\
0 & d_1+d_3 & 0\\
0 & 0 & d_1+d_2\\
\end{bmatrix} 
= \(\sum_{\substack{j=1, j \neq i}}^{3} d_j\) \ebf_i \otimes \ebf_i, \\
\Db_3^{\Pi} &= \begin{bmatrix}
d_2 d_3 & 0 & 0\\
0 & d_1 d_3 & 0\\
0 & 0 & d_1 d_2\\
\end{bmatrix} 
= \(\prod_{\substack{j=1, j \neq i}}^{3} d_j\) \ebf_i \otimes \ebf_i.
\end{align*}
Note that, since $\Db$ is a diagonal matrix, $\Db_2^{\Sigma} = \tr(\Db) \Ib - \Db = \adj(\Db)$ in the two-dimensional case (\ie{} for $d=2$), while $\Db_3^{\Sigma} = \tr(\Db) \Ib - \Db$ and $\Db_3^{\Pi} = \adj(\Db)$ in the three-dimensional case (\ie{} for $d=3$), where $\adj(\Db)$ denotes the adjugate of square matrix $\Db$ (that is the transpose of its cofactor matrix). Accordingly, $\Db_2^{\Sigma}$ (resp. $\Db_3^{\Pi}$) is such that $\Db\Db_2^{\Sigma} = \Db_2^{\Sigma}\Db = \det(\Db)\Ib = d_1 d_2 \Ib$ (resp. $\Db\Db_3^{\Pi} = \Db_3^{\Pi}\Db = \det(\Db)\Ib = d_1 d_2 d_3 \Ib$) in the two-dimensional (resp. three-dimensional) case. Besides, $\Db_3^{\Sigma}$ and $\Db_3^{\Pi}$ are linked through the relation $\Db \Db_3^{\Sigma} = \tr(\Db_3^{\Pi}) \Ib - \Db_3^{\Pi}$.

Even though the system of time-harmonic PML equations \eqref{systemfreqPML} (or equivalently \eqref{systemfreqPMLaniso}) could be recast as a single vector equation involving only a single (displacement) field, the resulting system of time-dependent PML equations would involve convolution products or would require specific (specialized or recursive) time integration schemes due to the temporal complexity. To further maintain the second-order (in time) character of the original linear elastodynamic equations \eqref{systemtime} while avoiding convolution operations in time, thereby greatly facilitating time integration, we now introduce an auxiliary tensor-valued strain field $\eb(\xb,t)$ defined by $\eb(\xb,t) = \tilde{\epsilonb}(\ub(\xb,t)) - \epsilonb(\ub(\xb,t))$ for any time $t\geq 0$ and spatial position vector $\xb \in \Omega_{\text{PML}}$. Its Fourier transform in time $\hat{\eb}$ is then such that
\begin{equation}\label{auxiliarystrain}
\tilde{\epsilonb}(\hat{\ub}) = \epsilonb(\hat{\ub}) + \hat{\eb}.
\end{equation}
%Note that for all $\xb \in \Omega_{\text{BD}}$, $\eb(\xb,t) = \zerob$ for all $t\geq 0$ and $\hat{\eb}(\xb,\omega) = \zerob$ for all $\omega> 0$.

Introducing \eqref{stretchedstress} into \eqref{equilibriumfreqPMLaniso}, substituting \eqref{auxiliarystrain} into \eqref{constitutiverelationfreqPML} and \eqref{kinematicrelationfreqPML}, left multiplying (or pre-multiplying) \eqref{kinematicrelationfreqPML} by $\mathrm{i}\omega \Jb^T$ and right multiplying (or post-multiplying) it by $\mathrm{i}\omega \Jb$ with \eqref{jacobianmatrix}, rearranging and grouping similar terms, we obtain the following time-harmonic PML formulation that consists in finding the time-harmonic vector-valued displacement field $\hat{\ub}(\xb,\omega)$, tensor-valued stress field $\hat{\sigmab}(\xb,\omega)$ and auxiliary tensor-valued strain field $\hat{\eb}(\xb,\omega)$ satisfying
\begin{subequations}\label{newsystemfreqPML}
\begin{equation}
\rho \(-\omega^2 + \mathrm{i}\omega \tr(\Db) + \det(\Db)\) \hat{\ub} = \divb\(\hat{\sigmab} + \dfrac{1}{\mathrm{i}\omega} \hat{\sigmab} \Db_2^{\Sigma}\) \label{newequilibriumfreqPML2D}
\end{equation}
in the two-dimensional case (\ie{} for $d=2$),
\begin{equation}
\rho \(-\omega^2 + \mathrm{i}\omega \tr(\Db) + \tr(\Db_3^{\Pi}) + \dfrac{1}{\mathrm{i}\omega}\det(\Db)\) \hat{\ub} = \divb\(\hat{\sigmab} + \dfrac{1}{\mathrm{i}\omega}\hat{\sigmab} \Db_3^{\Sigma} - \dfrac{1}{\omega^2} \hat{\sigmab} \Db_3^{\Pi}\) \label{newequilibriumfreqPML3D}
\end{equation}
in the three-dimensional case (\ie{} for $d=3$),
\begin{equation}
\hat{\sigmab} = \Cb : \epsilonb(\hat{\ub}) + \Cb : \hat{\eb} \label{newconstitutiverelationfreqPML}
\end{equation}
and
\begin{equation}
-\omega^2 \hat{\eb} + \mathrm{i}\omega \(\Db^T \hat{\eb} + \hat{\eb} \Db\) + \Db^T \hat{\eb} \Db + \mathrm{i}\omega \epsilonb^D(\hat{\ub}) + \Db^T \epsilonb(\hat{\ub}) \Db = \zerob,\label{newkinematicrelationfreqPML}
\end{equation}
\end{subequations}
for any spatial position vector $\xb \in \Omega_{\text{PML}}$ and angular frequency $\omega>0$, where $\epsilonb^D(\hat{\ub})$ is a second-order symmetric tensor-valued field defined by
\begin{equation*}
\epsilonb^D(\hat{\ub}) = \frac{1}{2} \((\nablab \hat{\ub}) \Db + \Db^T (\nablab \hat{\ub})^T\) %= \frac{1}{2} \((\nablab \hat{\ub}) \Db + ((\nablab \hat{\ub}) \Db)^T\)
 = \frac{1}{2}\(d_j \frac{\partial \hat{u}_i}{\partial x_j} + d_i \frac{\partial \hat{u}_j}{\partial x_i}\) \ebf_i \otimes \ebf_j.
\end{equation*}
Multiplying \eqref{newequilibriumfreqPML3D} by $\mathrm{i}\omega$, the PML equilibrium equations \eqref{newequilibriumfreqPML3D} in the three-dimensional case  (\ie{} for $d=3$) can be recast as
\begin{equation}\label{newequilibriumfreqPML3Dthirdorder}
\rho \(-\mathrm{i}\omega^3 - \omega^2 \tr(\Db) + \mathrm{i}\omega \tr(\Db_3^{\Pi}) + \det(\Db)\) \hat{\ub} = \divb\(\mathrm{i}\omega \hat{\sigmab} + \hat{\sigmab} \Db_3^{\Sigma} + \dfrac{1}{\mathrm{i}\omega} \hat{\sigmab} \Db_3^{\Pi}\)% \quad \text{for } d=3
.
\end{equation}

%The time-harmonic kinematic relation \eqref{kinematicrelationfreqPML} can be recast as
%\begin{equation}\label{kinematicrelationfreqPMLbis}
%-\omega^2 \tilde{\epsilonb}(\hat{\ub}) + \mathrm{i}\omega \(\Db^T \tilde{\epsilonb}(\hat{\ub}) + \tilde{\epsilonb}(\hat{\ub}) \Db\) + \Db^T \tilde{\epsilonb}(\hat{\ub}) \Db = -\omega^2 \epsilonb(\hat{\ub}) + \mathrm{i}\omega \(\Db^T \epsilonb(\hat{\ub}) + \epsilonb(\hat{\ub}) \Db - \epsilonb^D(\hat{\ub})\),
%\end{equation}

%To avoid convolution operations in time, we introduce an auxiliary tensor-valued strain field $\eb(\xb,t)$ such that $\tilde{\epsilonb}(\ub(\xb,t)) = \epsilonb(\ub(\xb,t)) + \eb(\xb,t)$. Substituting $\tilde{\epsilonb}(\hat{\ub})$ by $\epsilonb(\hat{\ub}) + \hat{\eb}$ into \eqref{kinematicrelationfreqPMLbis}, the time-harmonic kinematic relation \eqref{kinematicrelationfreqPMLbis} becomes
%\begin{equation}\label{newkinematicrelationfreqPML}
%-\omega^2 \hat{\eb} + \mathrm{i}\omega \(\Db^T \hat{\eb} + \hat{\eb} \Db\) + \Db^T \hat{\eb} \Db + \mathrm{i}\omega \epsilonb^D(\hat{\ub}) + \Db^T \epsilonb(\hat{\ub}) \Db = \zerob.
%\end{equation}

\begin{remark}
Alternatively, multiplying \eqref{kinematicrelationfreqPML} by $\det(\Jb)$ similarly to \cite{Kuc13,Fat15a}, using \eqref{auxiliarystrain} and \eqref{jacobianmatrix}, the time-harmonic kinematic relation \eqref{newkinematicrelationfreqPML} can be replaced with 
\begin{subequations}\label{newkinematicrelationfreqPMLsym}
\begin{equation}
\(-\omega^2 + \mathrm{i}\omega \tr(\Db) + \det(\Db)\) \hat{\eb} = -\mathrm{i}\omega \epsilonb^D(\hat{\ub}) - \det(\Db) \epsilonb(\hat{\ub}) \label{newkinematicrelationfreqPML2D}
\end{equation}
in the two-dimensional case (\ie{} for $d=2$), and
\begin{equation}
\(-\omega^2 + \mathrm{i}\omega \tr(\Db) + \tr(\Db_3^{\Pi}) + \dfrac{1}{\mathrm{i}\omega} \det(\Db)\) \hat{\eb} = -\mathrm{i}\omega \epsilonb^D(\hat{\ub}) - \epsilonb_3^D(\hat{\ub}) - \dfrac{1}{\mathrm{i}\omega} \det(\Db) \epsilonb(\hat{\ub}) \label{newkinematicrelationfreq3D}
\end{equation}
\end{subequations}
in the three-dimensional case (\ie{} for $d=3$), where $\epsilonb_3^D(\hat{\ub})$ is a second-order symmetric tensor-valued field defined by
\begin{align*}
\epsilonb_3^D(\hat{\ub}) &= \frac{1}{2} \((\nablab \hat{\ub}) \Db\Db_3^{\Sigma} + (\Db\Db_3^{\Sigma})^T (\nablab \hat{\ub})^T\) %= \frac{1}{2} \((\nablab \hat{\ub}) \Db\Db_3^{\Sigma} + ((\nablab \hat{\ub}) \Db\Db_3^{\Sigma})^T\)
\\
 &= \frac{1}{2}\(d_j \(\sum_{\substack{k=1, k \neq j}}^{3} d_k\) \frac{\partial \hat{u}_i}{\partial x_j} + d_i \(\sum_{\substack{k=1, k \neq i}}^{3} d_k\) \frac{\partial \hat{u}_j}{\partial x_i}\) \ebf_i \otimes \ebf_j.
\end{align*}
Since both time-harmonic equilibrium equations and kinematic relation are stretched in the same manner (both multiplied by $\det(\Jb)$), the left-hand sides of the PML equilibrium equations \eqref{newequilibriumfreqPML2D}-\eqref{newequilibriumfreqPML3D} and the PML kinematic relations \eqref{newkinematicrelationfreqPMLsym} have a similar form, thus leading to a fully-symmetric mixed displacement-strain unsplit-field PML formulation %(that is similar to the symmetric mixed displacement-stress unsplit-field formulations developed in \cite{Kuc13,Fat15a}) 
unlike the non-symmetric one \eqref{newsystemfreqPML} considered here% (that is similar to the non-symmetric mixed displacement-stress unsplit-field formulation proposed in \cite{Kuc11})
. Similar mixed unsplit-field PML formulations have already been proposed in the literature, such as the fully-symmetric ones developed in \cite{Kuc13,Fat15a} or the non-symmetric ones proposed in \cite{Kuc11,Fat15a} for instance, but they retain the stress field (or, more precisely, the time-history of stress field) as unknown variable (instead of the auxiliary strain field considered here) in addition to the displacement field.
\end{remark}

\subsection{Time-domain PML formulation}\label{sec:PMLformulationtime}

The time-harmonic PML equations can then be easily transformed into the corresponding time-dependent PML equations for transient analysis due to their simple frequency-dependence resulting from the specific choice of stretching function \eqref{stretchingfunction}. For the subsequent derivations, we introduce the auxiliary fields $\Ub(\xb,t)$, $\varSigmab(\xb,t)$ and $\varPib(\xb,t)$, corresponding to the time-integrals (or time-histories) of displacement field $\ub$, stress field $\sigmab$ and stress time-history field $\varSigmab$, respectively, defined for any spatial position vector $\xb \in \Omega_{\text{PML}}$ and time $t\geq 0$ by 
\begin{equation*}
\Ub(\xb,t) = \int_{0}^t \ub(\xb,\tau) \, d\tau, \quad \varSigmab(\xb,t) = \int_{0}^t \sigmab(\xb,\tau) \, d\tau
\end{equation*}
and
\begin{equation*}
\varPib(\xb,t) = \int_{0}^t \varSigmab(\xb,\tau) \, d\tau = \int_{0}^t \int_{0}^{\tau} \sigmab(\xb,\theta) \, d\theta \, d\tau.
\end{equation*}
For any spatial position vector $\xb \in \Omega_{\text{PML}}$, we make use of the Fourier transform in time $\hat{\varSigmab}(\xb,\omega)$ of $\varSigmab(\xb,t)$ defined by
\begin{equation}\label{Fouriertransform}
\hat{\varSigmab}(\xb,\omega) = \dfrac{1}{\mathrm{i}\omega} \hat{\sigmab}(\xb,\omega) + \pi \delta_0(\omega) \hat{\sigmab}(\xb,\omega),
\end{equation}
where $\delta_0(\omega)$ is the Dirac distribution on $\Rbb$ at the origin (\ie{} at point $\omega=0$), in order to derive the following inverse Fourier transform in time
\begin{align}\label{inverseFouriertransform}
\Fc^{-1}\[\dfrac{1}{\mathrm{i}\omega} \hat{\sigmab}(\xb,\omega)\] &= \varSigmab(\xb,t) - \dfrac{1}{2} \hat{\sigmab}(\xb,0) \notag\\
&= \int_{0}^t \sigmab(\xb,\tau) \, d\tau - \dfrac{1}{2} \int_{\Rbb^{+}} \sigmab(\xb,\tau) \, d\tau,
\end{align}
where $\Fc^{-1}$ denotes the inverse Fourier transform operator. Then, using relation \eqref{zerofreq} at angular frequency $\omega=0$, that is $\hat{\sigmab}(\xb,0) = \zerob$, the inverse Fourier transform in time \eqref{inverseFouriertransform} reduces to
\begin{equation}\label{newinverseFouriertransform}
\Fc^{-1}\[\dfrac{1}{\mathrm{i}\omega} \hat{\sigmab}(\xb,\omega)\] = \varSigmab(\xb,t).
\end{equation}
By applying the inverse Fourier transform in time to the time-harmonic PML equations \eqref{newsystemfreqPML} in which \eqref{newequilibriumfreqPML3D} is replaced with \eqref{newequilibriumfreqPML3Dthirdorder} and with the aid of \eqref{newinverseFouriertransform}, we derive the corresponding time-dependent PML equations. The time-dependent PML formulation then consists in finding the time-dependent vector-valued displacement field $\ub(\xb,t)$, tensor-valued stress field $\sigmab(\xb,t)$ and auxiliary tensor-valued strain field $\eb(\xb,t)$ satisfying
\begin{subequations}\label{systemtimePML}
\begin{equation}
\rho \(\ddot{\ub} + \tr(\Db) \dot{\ub} + \det(\Db) \ub\) = \divb\(\sigmab + \varSigmab \Db_2^{\Sigma}\) \label{equilibriumtimePML2D}
\end{equation}
in the two-dimensional case (\ie{} for $d=2$),
\begin{equation}
\rho \(\dddot{\ub} + \tr(\Db) \ddot{\ub} + \tr(\Db_3^{\Pi}) \dot{\ub} + \det(\Db) \ub\) = \divb\(\dot{\sigmab} + \sigmab \Db_3^{\Sigma} + \varSigmab \Db_3^{\Pi}\) \label{equilibriumtimePML3D}
\end{equation}
in the three-dimensional case (\ie{} for $d=3$),
\begin{equation}
\sigmab = \Cb : \epsilonb(\ub) + \Cb : \eb,\label{constitutiverelationtimePML}
\end{equation}
and
\begin{equation}
\ddot{\eb} + \(\Db^T \dot{\eb} + \dot{\eb} \Db\) + \Db^T \eb \Db + \epsilonb^D(\dot{\ub}) + \Db^T \epsilonb(\ub) \Db = \zerob,\label{kinematicrelationtimePML}
\end{equation}
for any spatial position vector $\xb \in \Omega_{\text{PML}}$ and time $t>0$, where the space and time dependencies of variables are implicit and the triple dots %$\dddot{}$ 
over a variable denotes the third-order partial derivative with respect to time $t$, \ie{} $\dddot{\ub}(\xb,t) = \dfrac{\partial^3 \ub}{\partial t^3}(\xb,t)$. The set of time-dependent PML equations \eqref{systemtimePML} is complemented with the boundary conditions on the exterior boundary $\Gamma_{\text{ext}}$ given by \eqref{boundaryconditions}, the usual continuity conditions on the displacement and normal stress fields at the interface $\Gamma$ (which are not recalled here for the sake of brevity) and the following homogeneous initial conditions at time $t=0$ given for any $\xb \in \Omega_{\text{PML}}$ by
\begin{equation}\label{initialconditionsPML}
\ub(\xb,0) = \dot{\ub}(\xb,0) = \zerob \quad \text{and} \quad \eb(\xb,0) = \dot{\eb}(\xb,0) = \zerob,
\end{equation}
with
\begin{equation}\label{initialconditionsPML3D}
\ddot{\ub}(\xb,0) = \zerob \quad \text{for } d=3.
\end{equation}
\end{subequations}
Note that the PML equilibrium equations \eqref{equilibriumtimePML2D} derived in the two-dimensional case are second-order in time, while the PML equilibrium equations \eqref{equilibriumtimePML3D} derived in the three-dimensional case are third-order in time. As proposed in \cite{Fat15a}, in order to preserve a fully second-order in time PML formulation in the three-dimensional case, one may analytically integrate \eqref{equilibriumtimePML3D} in time over $\intervalcc{0}{t}$ for any time $t>0$, using the initial conditions \eqref{initialconditionsPML} and \eqref{initialconditionsPML3D}. We then obtain the following PML equilibrium equations in the three-dimensional case (\ie{} for $d=3$)
\begin{equation}\label{newequilibriumtimePML3D}
\rho \(\ddot{\ub} + \tr(\Db) \dot{\ub} + \tr(\Db_3^{\Pi}) \ub + \det(\Db) \Ub\) = \divb\(\sigmab + \varSigmab \Db_3^{\Sigma} + \varPib \Db_3^{\Pi}\)% \quad \text{for } d=3
.
\end{equation}
The time-dependent PML formulation is then second-order in time for both displacement field $\ub$ and auxiliary strain field $\eb$ in both two- and three-dimensional cases.

\begin{remark}
Alternatively, taking the inverse Fourier transform in time of \eqref{newequilibriumfreqPML3D} in the three-dimensional case, as it was done in \cite{Basu09} for instance, requires the use of the Fourier transform in time $\hat{\Ub}(\xb,\omega)$ of $\Ub(\xb,t)$ and the one $\hat{\varPib}(\xb,\omega)$ of $\varPib(\xb,t)$, respectively defined for any spatial position vector $\xb \in \Omega_{\text{PML}}$ by
\begin{subequations}\label{Fouriertransform3D}
\begin{align}
\hat{\Ub}(\xb,\omega) &= \dfrac{1}{\mathrm{i}\omega} \hat{\ub}(\xb,\omega) + \pi \delta_0(\omega) \hat{\ub}(\xb,\omega),\\
\hat{\varPib}(\xb,\omega) &= -\dfrac{1}{\omega^2} \hat{\sigmab}(\xb,\omega) + \mathrm{i} \pi \delta_0'(\omega) \hat{\sigmab}(\xb,\omega),
\end{align}
\end{subequations}
where $\delta_0'(\omega)$ is the first derivative of $\delta_0(\omega)$ with respect to $\omega$, in order to derive the following inverse Fourier transforms in time
\begin{subequations}\label{inverseFouriertransform3D}
\begin{align}
\Fc^{-1}\[\dfrac{1}{\mathrm{i}\omega} \hat{\ub}(\xb,\omega)\] &= \Ub(\xb,t) - \dfrac{1}{2} \hat{\ub}(\xb,0) \notag\\
&= \int_{0}^t \ub(\xb,\tau) \, d\tau - \dfrac{1}{2} \int_{\Rbb^{+}} \ub(\xb,\tau) \, d\tau,\\
\Fc^{-1}\[-\dfrac{1}{\omega^2} \hat{\sigmab}(\xb,\omega)\] &= \varPib(\xb,t) + \dfrac{\mathrm{i}}{2} \(\dfrac{\partial\hat{\sigmab}}{\partial\omega}(\xb,0) + \mathrm{i} \, t \, \hat{\sigmab}(\xb,0)\) \notag\\
&= \int_{0}^t \int_{0}^{\tau} \sigmab(\xb,\theta) \, d\theta \, d\tau - \dfrac{1}{2} \int_{\Rbb^{+}} (t-\tau) \sigmab(\xb,\tau) \, d\tau.
\end{align}
\end{subequations}
By applying the inverse Fourier transform in time to the time-harmonic PML equilibrium equations \eqref{newequilibriumfreqPML3D} with the aid of \eqref{newinverseFouriertransform} and \eqref{inverseFouriertransform3D}, we may recover \eqref{newequilibriumtimePML3D} subject to silent initial conditions \eqref{initialconditionsPML} at time $t=0$ under the additional assumptions $\hat{\ub}(\xb,0) = \zerob$ and $\dfrac{\partial\hat{\sigmab}}{\partial\omega}(\xb,0) = \zerob$ for all $\xb \in \Omega_{\text{PML}}$.
\end{remark}

\subsection{Space weak formulation}\label{sec:spaceweakformulation}

Similarly to \cite{Kuc13,Fat15a}, we adopt a hybrid approach for solving the overall mixed problem defined over the entire domain $\Omega = \Omega_{\text{BD}}\cup \Omega_{\text{PML}}$ by considering a mixed displacement-strain formulation in the PML region $\Omega_{\text{PML}}$ coupled with a purely (non-mixed) displacement-based formulation (\ie{} a standard displacement-only formulation) in the physical domain $\Omega_{\text{BD}}$ for computational efficiency.

Introducing the constitutive relation \eqref{newconstitutiverelationfreqPML} into \eqref{newequilibriumfreqPML2D} and \eqref{newequilibriumfreqPML3D}, the weak formulation of the system of time-harmonic PML equations \eqref{newsystemfreqPML} consists in finding $\hat{\ub}(\xb,\omega)$ over the entire domain $\Omega = \Omega_{\text{BD}} \cup \Omega_{\text{PML}}$ and $\hat{\eb}(\xb,\omega)$ over the PML region $\Omega_{\text{PML}}$ such that for all vector-valued test functions $\deltab\ub(\xb)$ in an appropriate admissible vector space such that $\deltab\ub(\xb)=\zerob$ for $\xb\in \partial\Omega\setminus \Gamma_{\text{free}}$, and for all second-order symmetric tensor-valued test functions $\deltab\eb(\xb)$ in an appropriate admissible vector space,
\begin{subequations}\label{weakformfreqPML}
\begin{align}
-\omega^2 m(\hat{\ub},\deltab\ub) + \mathrm{i}\omega c(\hat{\ub},\deltab\ub) + k(\hat{\ub},\deltab\ub) + k_2(\hat{\ub},\deltab\ub) + \frac{1}{\mathrm{i}\omega} g(\hat{\ub},\deltab\ub)&\notag\\
+ \tilde{k}_1(\hat{\eb},\deltab\ub) + \frac{1}{\mathrm{i}\omega} \tilde{g}_1(\hat{\eb},\deltab\ub) = \ell(\deltab\ub) &\quad \text{for } d=2,\label{weakequilibriumfreqPML2D}\\
-\omega^2 m(\hat{\ub},\deltab\ub) + \mathrm{i}\omega c(\hat{\ub},\deltab\ub) + k(\hat{\ub},\deltab\ub) + k_3(\hat{\ub},\deltab\ub) + \frac{1}{\mathrm{i}\omega} (g(\hat{\ub},\deltab\ub) &+ g_1(\hat{\ub},\deltab\ub)) \notag\\
- \frac{1}{\omega^2} h(\hat{\ub},\deltab\ub) + \tilde{k}_1(\hat{\eb},\deltab\ub) + \frac{1}{\mathrm{i}\omega} \tilde{g}_1(\hat{\eb},\deltab\ub) - \frac{1}{\omega^2} \tilde{h}_1(\hat{\eb},\deltab\ub) = \ell(\deltab\ub) &\quad \text{for } d=3,\label{weakequilibriumfreqPML3D}\\
-\omega^2 \tilde{m}(\hat{\eb},\deltab\eb) + \mathrm{i}\omega \tilde{c}(\hat{\eb},\deltab\eb) + \tilde{k}(\hat{\eb},\deltab\eb) + \mathrm{i}\omega c_1(\hat{\ub},\deltab\eb) + k_1(\hat{\ub},\deltab\eb) &= 0,\label{weakkinematicrelationfreqPML}
\end{align}
\end{subequations}
for any angular frequency $\omega>0$, where $m$, $c$, $k$, $\tilde{k}_1$, $k_2$, $k_3$, $g$, $\tilde{g}_1$, $h$, $\tilde{h}_1$ and $g_1 = k_2$ are sesquilinear forms and $\ell$ is an antilinear (\ie{} conjugate-linear) form, respectively defined by
%\begin{alignat*}{3}
%&m(\hat{\ub},\deltab\ub) = \int_{\Omega} \rho \scalprod{\hat{\ub}}{\overline{\deltab\ub}} \, d\Omega, \quad &&c(\hat{\ub},\deltab\ub) = \int_{\Omega_{\text{PML}}} \rho \tr(\Db) \scalprod{\hat{\ub}}{\overline{\deltab\ub}} \, d\Omega, \quad &&k(\hat{\ub},\deltab\ub) = \int_{\Omega} (\Cb : \epsilonb(\hat{\ub})) : \epsilonb(\overline{\deltab\ub}) \, d\Omega,\\
%&k_2(\hat{\ub},\deltab\ub) = \int_{\Omega_{\text{PML}}} \rho \det(\Db) \scalprod{\hat{\ub}}{\overline{\deltab\ub}} \, d\Omega, \quad &&k_3(\hat{\ub},\deltab\ub) = \int_{\Omega_{\text{PML}}} \rho \tr(\Db_3^{\Pi}) \scalprod{\hat{\ub}}{\overline{\deltab\ub}} \, d\Omega, \quad &&\tilde{k}_1(\hat{\eb},\deltab\ub) = \int_{\Omega_{\text{PML}}} (\Cb : \hat{\eb}) : \epsilonb(\overline{\deltab\ub}) \, d\Omega,\\
%&g(\hat{\ub},\deltab\ub) = \int_{\Omega_{\text{PML}}} (\Cb : \epsilonb(\hat{\ub})) : \epsilonb_d^{\Sigma}(\overline{\deltab\ub}) \, d\Omega, \quad &&g_1(\hat{\ub},\deltab\ub) = k_2(\hat{\ub},\deltab\ub), \quad &&\tilde{g}_1(\hat{\eb},\deltab\ub) = \int_{\Omega_{\text{PML}}} (\Cb : \hat{\eb}) : \epsilonb_d^{\Sigma}(\overline{\deltab\ub}) \, d\Omega,\\
%&h(\hat{\ub},\deltab\ub) = \int_{\Omega_{\text{PML}}} (\Cb : \epsilonb(\hat{\ub})) : \epsilonb_3^{\Pi}(\overline{\deltab\ub}) \, d\Omega, \quad &&\tilde{h}_1(\hat{\eb},\deltab\ub) = \int_{\Omega_{\text{PML}}} (\Cb : \hat{\eb}) : \epsilonb_3^{\Pi}(\overline{\deltab\ub}) \, d\Omega, \quad &&\ell(\deltab\ub) = \int_{\Omega_{\text{BD}}} \scalprod{\hat{\fb}}{\overline{\deltab\ub}} \, d\Omega,
%\end{alignat*}
 \begin{alignat*}{2}
&m(\hat{\ub},\deltab\ub) = \int_{\Omega} \rho \scalprod{\hat{\ub}}{\overline{\deltab\ub}} \, d\Omega, \quad &&c(\hat{\ub},\deltab\ub) = \int_{\Omega_{\text{PML}}} \rho \tr(\Db) \scalprod{\hat{\ub}}{\overline{\deltab\ub}} \, d\Omega,\\
&k(\hat{\ub},\deltab\ub) = \int_{\Omega} (\Cb : \epsilonb(\hat{\ub})) : \epsilonb(\overline{\deltab\ub}) \, d\Omega, \quad &&\tilde{k}_1(\hat{\eb},\deltab\ub) = \int_{\Omega_{\text{PML}}} (\Cb : \hat{\eb}) : \epsilonb(\overline{\deltab\ub}) \, d\Omega,\\
&k_2(\hat{\ub},\deltab\ub) = \int_{\Omega_{\text{PML}}} \rho \det(\Db) \scalprod{\hat{\ub}}{\overline{\deltab\ub}} \, d\Omega, \quad &&k_3(\hat{\ub},\deltab\ub) = \int_{\Omega_{\text{PML}}} \rho \tr(\Db_3^{\Pi}) \scalprod{\hat{\ub}}{\overline{\deltab\ub}} \, d\Omega,\\
&g(\hat{\ub},\deltab\ub) = \int_{\Omega_{\text{PML}}} (\Cb : \epsilonb(\hat{\ub})) : \epsilonb_d^{\Sigma}(\overline{\deltab\ub}) \, d\Omega, \quad &&\tilde{g}_1(\hat{\eb},\deltab\ub) = \int_{\Omega_{\text{PML}}} (\Cb : \hat{\eb}) : \epsilonb_d^{\Sigma}(\overline{\deltab\ub}) \, d\Omega,\\
&h(\hat{\ub},\deltab\ub) = \int_{\Omega_{\text{PML}}} (\Cb : \epsilonb(\hat{\ub})) : \epsilonb_3^{\Pi}(\overline{\deltab\ub}) \, d\Omega, \quad &&\tilde{h}_1(\hat{\eb},\deltab\ub) = \int_{\Omega_{\text{PML}}} (\Cb : \hat{\eb}) : \epsilonb_3^{\Pi}(\overline{\deltab\ub}) \, d\Omega,\\
&g_1(\hat{\ub},\deltab\ub) = k_2(\hat{\ub},\deltab\ub), \quad &&\ell(\deltab\ub) = \int_{\Omega_{\text{BD}}} \scalprod{\hat{\fb}}{\overline{\deltab\ub}} \, d\Omega,
\end{alignat*}
where the overline denotes the complex conjugate transpose operator and
\begin{align*}
%\epsilonb(\deltab\ub) &= \frac{1}{2}(\nablab \deltab\ub + (\nablab \deltab\ub)^T),
\epsilonb_d^{\Sigma}(\deltab\ub) &= \dfrac{1}{2} \((\nablab \deltab\ub) \Db_d^{\Sigma} + {\Db_d^{\Sigma}}^T (\nablab \deltab\ub)^T\) = \tr(\Db) \epsilonb(\deltab\ub) - \epsilonb^D(\deltab\ub),\\
\epsilonb_3^{\Pi}(\deltab\ub) &= \dfrac{1}{2} \((\nablab \deltab\ub) \Db_3^{\Pi} + {\Db_3^{\Pi}}^T (\nablab \deltab\ub)^T\) = \tr(\Db_3^{\Pi}) \epsilonb(\deltab\ub) - \epsilonb_3^D(\deltab\ub),
\end{align*}
and where $\tilde{m}$, $\tilde{c}$, $\tilde{k}$, $c_1$ and $k_1$ are sesquilinear forms respectively defined by
\begin{alignat*}{2}
&\tilde{m}(\hat{\eb},\deltab\eb) = \int_{\Omega_{\text{PML}}} \hat{\eb} : \overline{\deltab\eb} \, d\Omega, \quad &&\tilde{c}(\hat{\eb},\deltab\eb) = \int_{\Omega_{\text{PML}}} \(\Db^T \hat{\eb} + \hat{\eb} \Db\) : \overline{\deltab\eb} \, d\Omega, \\
&\tilde{k}(\hat{\eb},\deltab\eb) = \int_{\Omega_{\text{PML}}} \Db^T \hat{\eb} \Db : \overline{\deltab\eb} \, d\Omega,\quad &&c_1(\hat{\ub},\deltab\eb) = \int_{\Omega_{\text{PML}}} \epsilonb^D(\hat{\ub}) : \overline{\deltab\eb} \, d\Omega,\\
&k_1(\hat{\ub},\deltab\eb) = \int_{\Omega_{\text{PML}}} \Db^T \epsilonb(\hat{\ub}) \Db : \overline{\deltab\eb} \, d\Omega.
\end{alignat*}

Since the auxiliary strain field $\hat{\eb}$ needs to be introduced only in the PML region $\Omega_{\text{PML}}$ and the diagonal real matrices $\Db$ (gathering the damping functions $d_i$), $\Db_2^{\Sigma}$, $\Db_3^{\Sigma}$ and $\Db_3^{\Pi}$ are non-zero only in $\Omega_{\text{PML}}$, all the space-integrals involved in the sesquilinear forms $\tilde{m}$, $c$, $c_1$, $\tilde{c}$, $k_1$, $k_2$, $k_3$, $\tilde{k}$, $\tilde{k}_1$, $g$, $g_1$, $\tilde{g}_1$, $h$ and $\tilde{h}_1$ of \eqref{weakformfreqPML} restrict to $\Omega_{\text{PML}}$. Besides, since the physical domain $\Omega_{\text{BD}}$ contains the support of the complex vector $\hat{\fb}$ associated to external body forces, the space-integral involved in the antilinear form $\ell$ restricts to $\Omega_{\text{BD}}$.

In the following, we will consider vectorial representations of second-order symmetric tensor-valued fields $\eb$ and $\hat{\eb}$. Also, for the sake of readability and notational convenience, we will use the same notation for both second-order symmetric tensors (or tensor-valued fields) of dimension $d(d+1)/2$ and their vectorial representations.

\subsection{Hybrid finite element approximation}\label{sec:FEapproximation}

For the mixed finite element implementation of space weak formulation \eqref{weakformfreqPML}, both displacement field $\ub$ and auxiliary strain field $\eb$ are retained as unknowns and treated as independent variables that can be approximated separately. We then use a classical (displacement-based) finite element method (FEM) \cite{Hug87,Zie05} by introducing finite-dimensional approximation spaces spanned by piecewise polynomial basis functions for the spatial discretization of both displacement field $\ub$ (resp. its Fourier transform in time $\hat{\ub}$) over $\Omega = \Omega_{\text{BD}}\cup \Omega_{\text{PML}}$ and auxiliary strain field $\eb$ (resp. its Fourier transform in time $\hat{\eb}$) over $\Omega_{\text{PML}}$ in the time domain (resp. the frequency domain). For any time $t\geq 0$, the finite element approximations of displacement field $\ub(\cdot,t)$ and auxiliary strain field $\eb(\cdot,t)$ can then be identified with the real vectors $\Ubf(t)$ and $\Ebf(t)$ gathering the nodal values of $\ub(\cdot,t)$ in the entire domain $\Omega = \Omega_{\text{BD}}\cup \Omega_{\text{PML}}$ and the ones of $\eb(\cdot,t)$ in the PML region $\Omega_{\text{PML}}$, respectively. Similarly, for any angular frequency $\omega> 0$, the finite element approximations of Fourier transforms in time $\hat{\ub}(\cdot,\omega)$ and $\hat{\eb}(\cdot,\omega)$ can be identified with the complex vectors $\hat{\Ubf}(\omega)$ and $\hat{\Ebf}(\omega)$ gathering the nodal values of $\hat{\ub}(\cdot,\omega)$ in the entire domain $\Omega = \Omega_{\text{BD}}\cup \Omega_{\text{PML}}$ and the ones of $\hat{\eb}(\cdot,\omega)$ in the PML region $\Omega_{\text{PML}}$, respectively.

\subsubsection{Frequency-domain implementation}\label{sec:implementationfreq}

For the finite element discretization of sesquilinear forms $m$, $c$, $k$, $k_2$, $k_3$, $\tilde{k}_1$, $g$, $g_1$, $\tilde{g}_1$, $h$, $\tilde{h}_1$ and antilinear form $\ell$ involved in \eqref{weakequilibriumfreqPML2D} and \eqref{weakequilibriumfreqPML3D}, we introduce the finite element (frequency-independent) real matrices $\Mbf$, $\Cbf$, $\Kbf$, $\Kbf_2$, $\Kbf_3$, $\widetilde{\Kbf}_1$, $\Gbf$, $\Gbf_1$, $\widetilde{\Gbf}_1$, $\Hbf$ and $\widetilde{\Hbf}_1$, and the finite element (frequency-dependent) complex vector $\hat{\Fbf}(\omega)$. Similarly, for the finite element discretization of sesquilinear forms $\tilde{m}$, $\tilde{c}$, $\tilde{k}$, $c_1$ and $k_1$ involved in \eqref{weakkinematicrelationfreqPML}, we introduce the finite element (frequency-independent) real matrices $\widetilde{\Mbf}$, $\widetilde{\Cbf}$, $\widetilde{\Kbf}$, $\Cbf_1$ and $\Kbf_1$.

In an algebraic setting, the finite element matrix system resulting from the weak formulation of time-harmonic PML equations \eqref{weakformfreqPML} reads
\begin{subequations}\label{algebraicformfreqPML}
\begin{align}
-\omega^2 \Mbf \hat{\Ubf} + \mathrm{i}\omega \Cbf \hat{\Ubf} + (\Kbf + \Kbf_2) \hat{\Ubf} + \widetilde{\Kbf}_1 \hat{\Ebf}& \notag\\
+ \frac{1}{\mathrm{i}\omega} \(\Gbf \hat{\Ubf} + \widetilde{\Gbf}_1 \hat{\Ebf}\) &= \hat{\Fbf} \quad \text{for } d=2,\label{algebraicequilibriumfreqPML2D}\\
-\omega^2 \Mbf \hat{\Ubf} + \mathrm{i}\omega \Cbf \hat{\Ubf} + (\Kbf + \Kbf_3) \hat{\Ubf} + \widetilde{\Kbf}_1 \hat{\Ebf}& \notag\\
 + \frac{1}{\mathrm{i}\omega} \((\Gbf + \Gbf_1) \hat{\Ubf} + \widetilde{\Gbf}_1 \hat{\Ebf}\) - \frac{1}{\omega^2} \(\Hbf \hat{\Ubf} + \widetilde{\Hbf}_1 \hat{\Ebf}\) &= \hat{\Fbf} \quad \text{for } d=3,\label{algebraicequilibriumfreqPML3D}\\
-\omega^2 \widetilde{\Mbf} \hat{\Ebf} + \mathrm{i}\omega \(\widetilde{\Cbf} \hat{\Ebf} + \Cbf_1 \hat{\Ubf}\) + \widetilde{\Kbf} \hat{\Ebf} + \Kbf_1 \hat{\Ubf} &= \zerob.\label{algebraickinematicrelationfreqPML}
\end{align}
\end{subequations}

Then, it comes down to solving the following system of linear algebraic equations for any angular frequency $\omega>0$:
\begin{subequations}\label{algebraicsysfreqPML}
\begin{align}
-\omega^2 \Mbb \hat{\Ubb} + \mathrm{i}\omega \Cbb \hat{\Ubb} + \Kbb \hat{\Ubb} + \frac{1}{\mathrm{i}\omega} \Gbb \hat{\Ubb} &= \hat{\Fbb} \quad \text{for } d=2,\label{algebraicsysfreqPML2D}\\
-\omega^2 \Mbb \hat{\Ubb} + \mathrm{i}\omega \Cbb \hat{\Ubb} + \Kbb \hat{\Ubb} + \frac{1}{\mathrm{i}\omega} \Gbb \hat{\Ubb} - \frac{1}{\omega^2} \Hbb \hat{\Ubb} &= \hat{\Fbb} \quad \text{for } d=3,\label{algebraicsysfreqPML3D}
\end{align}
\end{subequations}
%\begin{equation}\label{algebraicsysfreqPML}
%\(-\omega^2 \Mbb + \mathrm{i}\omega \Cbb + \Kbb + \frac{1}{\mathrm{i}\omega} \Gbb - \frac{1}{\omega^2} \Hbb\) \hat{\Ubb} = \hat{\Fbb},
%\end{equation}
where the (frequency-independent) real matrices $\Mbb$, $\Cbb$, $\Kbb$, $\Gbb$ and $\Hbb$, and the (frequency-dependent) complex vectors $\hat{\Fbb}(\omega)$ and $\hat{\Ubb}(\omega)$ are respectively defined by
\begin{align*}
\Mbb &= \begin{bmatrix}
\Mbf & \zerob\\
\zerob & \widetilde{\Mbf}
\end{bmatrix}, \quad
\Cbb = \begin{bmatrix}
\Cbf & \zerob\\
\Cbf_1 & \widetilde{\Cbf}
\end{bmatrix}, \quad
\Kbb = \begin{bmatrix}
\Kbf+\Kbf_d & \widetilde{\Kbf}_1\\
\Kbf_1 & \widetilde{\Kbf}
\end{bmatrix}, \\
\Gbb &= \begin{bmatrix}
\Gbf & \widetilde{\Gbf}_1\\
\zerob & \zerob
\end{bmatrix} \text{ for } d=2, \quad
\Gbb = \begin{bmatrix}
\Gbf+\Gbf_1 & \widetilde{\Gbf}_1\\
\zerob & \zerob
\end{bmatrix} \text{ for } d=3,% \quad 
%\Hbb = \zerob \text{ for } d=2, \quad
\\ 
\Hbb &= \begin{bmatrix}
\Hbf & \widetilde{\Hbf}_1\\
\zerob & \zerob
\end{bmatrix} \text{ only for } d=3, \quad
\hat{\Fbb} = \begin{bmatrix}
\hat{\Fbf}\\
\zerob
\end{bmatrix}
 \quad \text{and} \quad
\hat{\Ubb} = \begin{bmatrix}
\hat{\Ubf}\\
\hat{\Ebf}
\end{bmatrix}.
\end{align*}
Note that all the block-system matrices are frequency-independent square real matrices, but only the mass matrix $\Mbb$ is symmetric and has a block-diagonal structure, whereas the other matrices $\Kbb$, $\Cbb$, $\Gbb$ and $\Hbb$ are non-symmetric. Besides, the block-system \eqref{algebraicsysfreqPML} has a sparse structure at two levels, coming from the sparsity pattern inherent to each finite element matrix involved in \eqref{algebraicformfreqPML}, but also from the block-sparsity pattern of matrices $\widetilde{\Mbf}$, $\Cbf$, $\Cbf_1$, $\widetilde{\Cbf}$, $\Kbf_1$, $\Kbf_2$, $\Kbf_3$, $\widetilde{\Kbf}$, $\widetilde{\Kbf}_1$, $\Gbf$, $\Gbf_1$, $\widetilde{\Gbf}_1$, $\Hbf$ and $\widetilde{\Hbf}_1$ that are assembled only from their element-level constituent matrices defined over the PML region $\Omega_{\text{PML}}$. For practical implementation in the frequency domain, the FE matrix system \eqref{algebraicsysfreqPML} can be rewritten as
\begin{equation*}\label{algebraicsysfreqPMLbis}
\Kbb^{\text{dyn}}(\omega) \hat{\Ubb} = \hat{\Fbb},
\end{equation*}
where $\Kbb^{\text{dyn}}(\omega)$ is the complex dynamic (frequency-dependent) stiffness matrix defined by
\begin{equation*}\label{dynamicalstiffness}
\Kbb^{\text{dyn}}(\omega) = \begin{cases}
%-\omega^2 \Mbb + \mathrm{i}\omega \Cbb + \Kbb + \dfrac{1}{\mathrm{i}\omega} \Gbb = 
\(\Kbb-\omega^2 \Mbb\) + \mathrm{i}\(\omega \Cbb - \dfrac{1}{\omega} \Gbb\) & \text{for } d=2, \\
%-\omega^2 \Mbb + \mathrm{i}\omega \Cbb + \Kbb + \dfrac{1}{\mathrm{i}\omega} \Gbb - \dfrac{1}{\omega^2} \Hbb = 
\(\Kbb-\omega^2 \Mbb- \dfrac{1}{\omega^2} \Hbb\) + \mathrm{i}\(\omega \Cbb - \dfrac{1}{\omega} \Gbb\) & \text{for } d=3.
\end{cases}
\end{equation*}
%\begin{equation*}\label{dynamicalstiffness}
%\Kbb^{\text{dyn}}(\omega) = 
%%-\omega^2 \Mbb + \mathrm{i}\omega \Cbb + \Kbb + \dfrac{1}{\mathrm{i}\omega} \Gbb - \dfrac{1}{\omega^2} \Hbb = 
%\(\Kbb-\omega^2 \Mbb- \dfrac{1}{\omega^2} \Hbb\) + \mathrm{i}\(\omega \Cbb - \dfrac{1}{\omega} \Gbb\).
%\end{equation*}
Note that $\Kbb^{\text{dyn}}(\omega)$ is a non-symmetric complex matrix due to the non-symmetry of real matrices $\Kbb$, $\Cbb$, $\Gbb$ and $\Hbb$.

For practical implementation in the time domain, the FE matrix system \eqref{algebraicsysfreqPML3D} in the three-dimensional case can be recast as
\begin{equation}\label{newalgebraicsysfreqPML3D}
-\mathrm{i}\omega^3 \Mbb \hat{\Ubb} - \omega^2 \Cbb \hat{\Ubb} + \mathrm{i}\omega \Kbb \hat{\Ubb} + \Gbb \hat{\Ubb} + \frac{1}{\mathrm{i}\omega} \Hbb \hat{\Ubb} = \mathrm{i}\omega \hat{\Fbb} \quad \text{for } d=3.
\end{equation}

\subsubsection{Time-domain implementation}\label{sec:implementationtime}

The time-harmonic FE matrix system \eqref{algebraicsysfreqPML2D}-\eqref{newalgebraicsysfreqPML3D} can then be easily inverted back into the time domain for transient analysis due to its simple frequency-dependence. By applying the inverse Fourier transform in time to the time-harmonic FE matrix system \eqref{algebraicsysfreqPML2D}-\eqref{newalgebraicsysfreqPML3D}, we obtain the following second-order and third-order integro-differential equations (IDEs) in time
%\begin{subequations}\label{algebraicsystimePML}
%\begin{equation}\label{algebraicsystimePML2D}
%\Mbb \ddot{\Ubb} + \Cbb \dot{\Ubb} + \Kbb \Ubb + \Gbb \Ibb = \Fbb
%\end{equation}
%in the two-dimensional case (\ie{} for $d=2$), and
%\begin{equation}\label{algebraicsystimePML3D}
%\Mbb \ddot{\Ubb} + \Cbb \dot{\Ubb} + \Kbb \Ubb + \Gbb \Ibb + \Hbb \Jbb = \Fbb,
%\end{equation}
%in the three-dimensional case (\ie{} for $d=3$), 
%\end{subequations}
\begin{subequations}\label{algebraicsystimePML}
\begin{align}
\Mbb \ddot{\Ubb} + \Cbb \dot{\Ubb} + \Kbb \Ubb + \Gbb \Ibb &= \Fbb \quad \text{for } d=2,\label{algebraicsystimePML2D}\\
\Mbb \dddot{\Ubb} + \Cbb \ddot{\Ubb} + \Kbb \dot{\Ubb} + \Gbb \Ubb + \Hbb \Ibb &= \dot{\Fbb} \quad \text{for } d=3,\label{algebraicsystimePML3D}
\end{align}
\end{subequations}
%\begin{equation}\label{algebraicsystimePML}
%\Mbb \ddot{\Ubb} + \Cbb \dot{\Ubb} + \Kbb \Ubb + \Gbb \Ibb + \Hbb \Jbb = \Fbb,
%\end{equation}
for time $t>0$, in which the time-dependent real vectors $\Ubb(t)$, $\dot{\Ubb}(t)$, $\ddot{\Ubb}(t)$, $\dddot{\Ubb}(t)$, $\Ibb(t)$, %$\Jbb(t)$, 
$\Fbb(t)$ and $\dot{\Fbb}(t)$ are respectively defined by
\begin{align*}
\Ubb &= \begin{bmatrix}
\Ubf\\
\Ebf
\end{bmatrix},
\quad
\dot{\Ubb} = \begin{bmatrix}
\dot{\Ubf}\\
\dot{\Ebf}
\end{bmatrix},
\quad
\ddot{\Ubb} = \begin{bmatrix}
\ddot{\Ubf}\\
\ddot{\Ebf}
\end{bmatrix},
\quad
\dddot{\Ubb} = \begin{bmatrix}
\dddot{\Ubf}\\
\dddot{\Ebf}
\end{bmatrix},\\
\Ibb &= \int_{0}^t \Ubb \, d\tau,
%\quad \Jbb = \int_{0}^t \Ibb \, d\tau = \int_{0}^t \int_{0}^{\tau} \Ubb \, d\theta \, d\tau,
\quad
\Fbb = \begin{bmatrix}
\Fbf\\
\zerob
\end{bmatrix}
\quad \text{and} \quad
\dot{\Fbb} = \begin{bmatrix}
\dot{\Fbf}\\
\zerob
\end{bmatrix}.
\end{align*}
The IDEs in time \eqref{algebraicsystimePML} are supplemented with the following initial conditions at time $t=0$:
\begin{subequations}\label{discretizedinitialconditionsUdUddU}
\begin{equation}\label{discretizedinitialconditionsUdU}
\Ubb(0) = \Ubb^0 = \begin{bmatrix}
\Ubf^0\\
\zerob
\end{bmatrix}
\quad \text{and} \quad
\dot{\Ubb}(0) = \dot{\Ubb}^0 = \begin{bmatrix}
\Vbf^0\\
\zerob
\end{bmatrix},
\end{equation}
with
\begin{equation}\label{discretizedinitialconditionsddU}
\ddot{\Ubb}(0) = \ddot{\Ubb}^0 = \begin{bmatrix}
\Abf^0\\
\zerob
\end{bmatrix} \quad \text{only for } d=3,
\end{equation}
\end{subequations}
where $\Ubf^0$ and $\Vbf^0$ are the initial displacement and velocity FE vectors corresponding to the spatial discretization of the initial displacement and velocity fields $\ub_0$ and $\vb_0$ over $\Omega$, and $\Abf^0$ is the initial acceleration FE vector corresponding to the spatial discretization of the initial acceleration field $\ddot{\ub}(\cdot,0)$ over $\Omega$ and such that $\Mbf \Abf^0 = \Fbf^0 - \Cbf \Vbf^0 - \Kbf \Ubf^0$ with $\Fbf^0 = \Fbf(0)$ the initial FE vector of nodal forces corresponding to the spatial discretization of the initial body force field $\fb(\cdot,0)$ at time $t=0$. Note that, since $\fb$, $\ub_0$ and $\vb_0$ have bounded supports in $\Omega_{\text{BD}}$ (see the last paragraph of Section~\ref{sec:timeformulation}), $\Fbf$, $\Ubf^0$ and $\Vbf^0$ have non-zero components only for the degrees of freedom associated to the spatial discretization of $\fb$, $\ub_0$ and $\vb_0$ over the physical domain $\Omega_{\text{BD}}$.

In order to recover a second-order IDE in time in the three-dimensional case, one may analytically integrate \eqref{algebraicsystimePML3D} in time over $\intervalcc{0}{t}$ for any time $t>0$, using the initial conditions \eqref{discretizedinitialconditionsUdUddU}. We then obtain the following second-order IDE in time in the three-dimensional case
\begin{equation}\label{newalgebraicsystimePML3D}
\Mbb \ddot{\Ubb} + \Cbb \dot{\Ubb} + \Kbb \Ubb + \Gbb \Ibb + \Hbb \Jbb = \Fbb \quad \text{for } d=3,
\end{equation}
for time $t>0$, in which the time-dependent real vector $\Jbb(t)$ is defined by
\begin{equation*}
\Jbb = \int_{0}^t \Ibb \, d\tau = \int_{0}^t \int_{0}^{\tau} \Ubb \, d\theta \, d\tau.
\end{equation*}
The semi-discrete systems \eqref{algebraicsystimePML2D} in the two-dimensional case and \eqref{newalgebraicsystimePML3D} in the three-dimensional case are then both second-order in time, thus allowing the use of standard time integration schemes classically employed in computational structural dynamics. Also, the FE solution $\Ubb$ of the time-domain PML formulation obviously satisfies the causality principle.

\begin{remark}\label{rmrk:CALformulation}
In the numerical experiments presented in Section~\ref{sec:results}, the proposed PML formulation is compared to a CAL formulation with \textit{ad hoc} dissipative material properties introduced in the PML region $\Omega_{\text{PML}}$ but without a characteristic impedance matching the one of the physical domain $\Omega_{\text{BD}}$. In the frequency domain, the time-harmonic CAL formulation consists in solving the following system of linear algebraic equations at any angular frequency $\omega>0$:
%\begin{equation*}\label{algebraicsysfreqCAL}
%-\omega^2 \Mbf \hat{\Ubf} + \mathrm{i}\omega \Cbf \hat{\Ubf} + \Kbf \hat{\Ubf} = \hat{\Fbf},
%\end{equation*}
%or equivalently,
\begin{equation}\label{algebraicsysfreqCALbis}
\Kbf^{\text{dyn}}(\omega) \hat{\Ubf} = \hat{\Fbf},
\end{equation}
where $\Kbf^{\text{dyn}}(\omega)$ is the dynamic (frequency-dependent) stiffness matrix defined by
\begin{equation*}\label{dynamicalstiffnessCAL}
\Kbf^{\text{dyn}}(\omega) = 
%-\omega^2 \Mbf + \mathrm{i}\omega \Cbf + \Kbf = 
\Kbf-\omega^2 \Mbf + \mathrm{i} \omega \Cbf.
\end{equation*}
Note that $\Kbf^{\text{dyn}}(\omega)$ is a symmetric complex matrix due to the symmetry of real matrices $\Kbf$, $\Mbf$ and $\Cbf$.
In the time domain, the CAL formulation leads to the following classical second-order ordinary differential equation (ODE) in time:
\begin{equation}\label{algebraicsystimeCAL}
\Mbf \ddot{\Ubf} + \Cbf \dot{\Ubf} + \Kbf \Ubf = \Fbf
\end{equation}
for time $t>0$, with the initial conditions $\Ubf(0) = \Ubf^0$ and $\dot{\Ubf}(0) = \Vbf^0$ at time $t=0$.
\end{remark}

\subsection{Time sampling scheme}\label{sec:timescheme}

We use both a Newmark time scheme \cite{New59} and a finite difference time scheme based on the trapezoidal rule for solving the equations in the time domain for a given time sampling of $\Ubb$, $\dot{\Ubb}$, $\ddot{\Ubb}$, $\Ibb$ and $\Jbb$. Another possibility would consist in using an extended Newmark time scheme, as it was done in \cite{Kuc10,Ass12,Fat15a} for instance. For the sake of simplicity, we consider a uniform time step $\Delta t$ over a given time interval of interest $I = \intervalcc{0}{T}$, with $T=N\Delta t$ the final time and $N$ the number of time steps.

\subsubsection{Newmark scheme for the displacement vector $\Ubb$, velocity vector $\dot{\Ubb}$ and acceleration vector $\ddot{\Ubb}$}\label{sec:Newmarkscheme}

The Newmark scheme may be either explicit or implicit, unstable or (un)conditionally stable (depending on the values of the two parameters $\gamma$ and $\beta$) and it is based on the following recursive formulae expressed in acceleration format as
\begin{subequations}\label{Newmarktimescheme}
\begin{align}
\Ubb^n &= \Ubb^{n-1} + \Delta t \dot{\Ubb}^{n-1} + \dfrac{{\Delta t}^2}{2} \((1-2\beta) \ddot{\Ubb}^{n-1} + 2\beta \ddot{\Ubb}^n\),\label{NewmarktimeschemeU}\\
\dot{\Ubb}^n &= \dot{\Ubb}^{n-1} + \Delta t \((1-\gamma) \ddot{\Ubb}^{n-1} + \gamma \ddot{\Ubb}^n\),\label{NewmarktimeschemedU}
\end{align}
\end{subequations}
where $\Ubb^n$, $\dot{\Ubb}^n$ and $\ddot{\Ubb}^n$ denote respectively the displacement, velocity and acceleration FE vectors evaluated at time $t_n = n\Delta t$, with $\Delta t = t_n-t_{n-1}$ the discrete time step. Alternatively, assuming that $\beta\neq 0$, the Newmark time scheme \eqref{Newmarktimescheme} can also be rewritten in displacement format as
\begin{subequations}\label{Newmarktimeschemebis}
\begin{align}
\dot{\Ubb}^n &= \gamma_1 \(\Ubb^n-\Ubb^{n-1}\) + \(1+\gamma_2\) \dot{\Ubb}^{n-1} + \gamma_3 \ddot{\Ubb}^{n-1},\label{NewmarktimeschemedUbis}\\
\ddot{\Ubb}^n &= \beta_1 \(\Ubb^n-\Ubb^{n-1}\) + \beta_2 \dot{\Ubb}^{n-1} + \(1+\beta_3\) \ddot{\Ubb}^{n-1},\label{NewmarktimeschemeddU}
\end{align}
\end{subequations}
where
\begin{alignat*}{3}
\beta_1 &= \dfrac{1}{\beta {\Delta t}^2}, \quad &&\beta_2 = -\dfrac{1}{\beta \Delta t}, \quad &&\beta_3 = -\dfrac{1}{2\beta}, \\
\gamma_1 &= \dfrac{\gamma}{\beta \Delta t}, \quad &&\gamma_2 = -\dfrac{\gamma}{\beta}% \quad \text{and}
, \quad &&\gamma_3 = \(1-\dfrac{\gamma}{2\beta}\) \Delta t.
\end{alignat*}
The two parameters $\gamma$ and $\beta$ allow the degree of implicitness and the stability of the time scheme to be controlled \cite{Hug87}. Recall that the Newmark time scheme \eqref{Newmarktimescheme} (or \eqref{Newmarktimeschemebis}) is second-order accurate whatever the values of $\gamma$ and $\beta$. Also, when $\gamma<1/2$, it is unstable, whereas when $\gamma\geq 1/2$ and $2\beta\geq \gamma$, it is unconditionally stable for non-active and non-dispersive media \cite{Chi07,Mat11}, and when $\gamma\geq 1/2$ and $2\beta<\gamma$, it becomes conditionally stable, which means the choice of time step $\Delta t$ is limited and bounded by the Courant-Friedrichs-Lewy (CFL) stability condition \cite{Cou28,Cou67} for non-dispersive media \cite{Mat11}. Besides, when $\beta=0$, the time integration scheme \eqref{Newmarktimescheme} is explicit whatever the value of $\gamma$%, since the displacement FE vector $\Ubb^n$ at curent time $t_n$ only depends on the solution $(\Ubb^{n-1},\dot{\Ubb}^{n-1},\ddot{\Ubb}^{n-1})$ at previous time $t_{n-1}$ through formula \eqref{NewmarktimeschemeU}
. In particular, when $\gamma=1/2$ and $\beta=0$, one recovers the explicit conditionally stable central-difference scheme. Such an explicit time integration scheme has been recently combined with matrix lumping to avoid the computational burden associated with matrix inversion and applied in (un)split-field (CFS-)PML formulations to perform two- and three-dimensional transient elastodynamic analyses in unbounded (or very large) domains \cite{Ma06,Mar08b,Basu09,Mat11,Xie14,Zaf16}. In the numerical experiments shown in Section~\ref{sec:results}, we use the implicit unconditionally stable constant average acceleration scheme by taking $\gamma=1/2$ and $\beta=1/4$. Such an implicit time integration scheme requires the solution of a linear matrix system at every time step (which may be computationally expensive in very large three-dimensional transient elastodynamic analyses), but it does not suffer from the CFL stability condition in return, which means the choice of time step $\Delta t$ is not limited by the smallest element size (spatial resolution), which is of particular interest especially for highly irregular (unstructured) and/or fine spatial FE meshes. It has already been employed in unsplit-field (CFS-)PML formulations to perform transient analyses of second-order elastic wave equations in unbounded domains \cite{Basu04,Kang10,Kuc10,Kuc11,Kuc13,Mat11,Fat15a,Brun16,Zhou16}. Note that the unconditional stability is guaranteed only for non-active and non-dispersive media \cite{Chi07}, and therefore it cannot be ensured to hold in the PML region $\Omega_{\text{PML}}$ that is inherently an anisotropic dissipative (absorbing) medium. Nevertheless, several numerical experiments have shown that the proposed time stepping scheme is actually highly stable \cite{Mat11}.

\subsubsection{Finite difference scheme for the time-integral vectors $\Ibb$ and $\Jbb$}\label{sec:finitedifferencescheme}

The finite difference scheme may be either explicit or implicit, (un)conditionally stable (depending on the value of the parameter $\alpha$) and it is based on the following recursive formulae
\begin{subequations}\label{finitedifferencescheme}
\begin{align}
\Ibb^n &= \Ibb^{n-1} + \Delta t \((1-\alpha)\Ubb^{n-1} + \alpha\Ubb^n\),\label{finitedifferenceschemeiU}\\
\Jbb^n &= \Jbb^{n-1} + \Delta t \((1-\alpha)\Ibb^{n-1} + \alpha\Ibb^n\),\label{NewmarktimeschemeiiU}
\end{align}
\end{subequations}
where $\Ibb^n$ and $\Jbb^n$ denote respectively the time-integral $\Ibb$ of displacement FE vector $\Ubb$ and the time-integral $\Jbb$ of FE vector $\Ibb$ evaluated at time $t_n = n\Delta t$. Note that the time scheme \eqref{NewmarktimeschemeiiU} is introduced only in the three-dimensional case (\ie{} for $d=3$), since $\Jbb$ is involved in the left-hand side of \eqref{newalgebraicsystimePML3D} in the three-dimensional case (\ie{} for $d=3$), whereas it does not appear in the left-hand side of \eqref{algebraicsystimePML2D} in the two-dimensional case (\ie{} for $d=2$). Without loss of generality and for the sake of simplicity, we consider here the same parameter $\alpha$ for both time schemes \eqref{finitedifferenceschemeiU} and \eqref{NewmarktimeschemeiiU}. The parameter $\alpha$ allows the degree of implicitness and the stability of the time scheme to be controlled. Recall that when $\alpha\geq 1/2$, the finite difference scheme \eqref{finitedifferencescheme} is unconditionally stable, whereas when $\alpha< 1/2$, it becomes conditionally stable. Besides, when $\alpha=0$ (resp. $\alpha=1$), one recovers the explicit conditionally stable forward Euler time scheme (resp. the implicit unconditionally stable backward Euler time scheme), which is first-order accurate, whereas when $\alpha=1/2$, it leads to the implicit unconditionally stable Crank-Nicolson time scheme, which is second-order accurate. In the numerical experiments shown in Section~\ref{sec:results}, we use the Crank-Nicolson time scheme by taking $\alpha=1/2$ to preserve the intrinsic second-order accuracy and the unconditional stability of the Newmark time scheme \eqref{Newmarktimescheme} (or \eqref{Newmarktimeschemebis}) with $\gamma=1/2$ and $\beta=1/4$ (for non-active and non-dispersive media).

\subsubsection{Time sampling scheme for the PML formulation}\label{sec:timeschemePML}

Introducing the discretized version of the second-order IDEs in time \eqref{algebraicsystimePML2D} and \eqref{newalgebraicsystimePML3D} at time $t_n$ and using the above formulae \eqref{Newmarktimeschemebis} and \eqref{finitedifferencescheme}, we derive the time stepping scheme for the FE discretized wave equations \eqref{algebraicsystimePML2D} and \eqref{newalgebraicsystimePML3D} expressed in displacement format that reads as follows for $n\geq 1$,
%\begin{subequations}\label{timeschemePMLdisplacement}
%\begin{align}
%\dot{\Ubb}_{\gamma}^n &= -\gamma_1 \Ubb^{n-1} + \(1+\gamma_2\) \dot{\Ubb}^{n-1} + \gamma_3 \ddot{\Ubb}^{n-1},\label{predictordUbis}\\
%\ddot{\Ubb}_{\beta}^n &= -\beta_1 \Ubb^{n-1} + \beta_2 \dot{\Ubb}^{n-1} + \(1+\beta_3\) \ddot{\Ubb}^{n-1},\label{predictorddU}\\
%\Ibb_{\alpha}^n &= \Ibb^{n-1} + (1-\alpha) \Delta t \Ubb^{n-1},\label{predictoriUbis}\\
%\Jbb_{\alpha}^n &= \Jbb^{n-1} + (1-\alpha) \Delta t \Ibb^{n-1},\label{predictoriiUbis}\\
%\Kbb^{\text{eff}} \Ubb^n &= \Fbb^n - \Mbb \ddot{\Ubb}_{\beta}^n - \Cbb \dot{\Ubb}_{\gamma}^n - \(\Gbb + \alpha \Delta t \Hbb\) \Ibb_{\alpha}^n - \Hbb \Jbb_{\alpha}^n,\label{Ubis}\\
%\dot{\Ubb}^n &= \dot{\Ubb}_{\gamma}^n + \gamma_1 \Ubb^n,\label{dUbis}\\
%\ddot{\Ubb}^n &= \ddot{\Ubb}_{\beta}^n + \beta_1 \Ubb^n,\label{ddUbis}\\
%\Ibb^n &= \Ibb_{\alpha}^n + \alpha \Delta t \Ubb^n,\label{iUbis}\\
%\Jbb^n &= \Jbb_{\alpha}^n + \alpha \Delta t \Ibb^n,\label{iiUbis}
%\end{align}
%\end{subequations}
\begin{subequations}\label{timeschemePMLdisplacement}
\begin{alignat}{2}
\Kbb^{\text{eff}} \Ubb^n &= \Fbb^n - \Mbb \ddot{\Ubb}_{\beta}^n - \Cbb \dot{\Ubb}_{\gamma}^n - \Gbb \Ibb_{\alpha}^n &&\quad \text{for } d=2,\label{timeschemePMLdisplacement2D}\\
\Kbb^{\text{eff}} \Ubb^n &= \Fbb^n - \Mbb \ddot{\Ubb}_{\beta}^n - \Cbb \dot{\Ubb}_{\gamma}^n - \(\Gbb + \alpha \Delta t \Hbb\) \Ibb_{\alpha}^n - \Hbb \Jbb_{\alpha}^n &&\quad \text{for } d=3,\label{timeschemePMLdisplacement3D}
\end{alignat}
\end{subequations}
where $\dot{\Ubb}_{\gamma}^n = -\gamma_1 \Ubb^{n-1} + \(1+\gamma_2\) \dot{\Ubb}^{n-1} + \gamma_3 \ddot{\Ubb}^{n-1}$, $\ddot{\Ubb}_{\beta}^n = -\beta_1 \Ubb^{n-1} + \beta_2 \dot{\Ubb}^{n-1} + \(1+\beta_3\) \ddot{\Ubb}^{n-1}$, $\Ibb_{\alpha}^n = \Ibb^{n-1} + (1-\alpha) \Delta t \Ubb^{n-1}$ and $\Jbb_{\alpha}^n = \Jbb^{n-1} + (1-\alpha) \Delta t \Ibb^{n-1}$ are the respective predictors of unknown FE vectors $\dot{\Ubb}^n$, $\ddot{\Ubb}^n$, $\Ibb^n$ and $\Jbb^n$ at time $t_n$ such that $\dot{\Ubb}^n = \dot{\Ubb}_{\gamma}^n + \gamma_1 \Ubb^n$, $\ddot{\Ubb}^n = \ddot{\Ubb}_{\beta}^n + \beta_1 \Ubb^n$, $\Ibb^n = \Ibb_{\alpha}^n + \alpha \Delta t \Ubb^n$ and $\Jbb^n = \Jbb_{\alpha}^n + \alpha \Delta t \Ibb^n$, $\Fbb^n$ is the right-hand side FE vector $\Fbb$ evaluated at time $t_n$, and $\Kbb^{\text{eff}}$ is the effective stiffness matrix defined by
\begin{equation*}\label{effectivestiffness}
\Kbb^{\text{eff}} = \begin{cases}
\beta_1 \Mbb + \gamma_1 \Cbb + \Kbb + \alpha \Delta t \Gbb & \text{for } d=2, \\
\beta_1 \Mbb + \gamma_1 \Cbb + \Kbb + \alpha \Delta t \(\Gbb + \alpha \Delta t \Hbb\) & \text{for } d=3.
\end{cases}
\end{equation*}
%\begin{equation*}\label{effectivestiffness}
%\Kbb^{\text{eff}} = \beta_1 \Mbb + \gamma_1 \Cbb + \Kbb + \alpha \Delta t \(\Gbb + \alpha \Delta t \Hbb\).
%\end{equation*}
Note that $\Kbb^{\text{eff}}$ is a non-symmetric real matrix (due to the non-symmetry of real matrices $\Kbb$, $\Cbb$, $\Gbb$ and $\Hbb$) which depends on the two parameters $\gamma$ and $\beta$ involved in the Newmark time scheme \eqref{Newmarktimeschemebis} though $\gamma_1$ and $\beta_1$ and also on the parameter $\alpha$ involved the finite difference time scheme \eqref{finitedifferencescheme}. For practical purposes, introducing the predictors $\dot{\Ubb}_{\gamma}^n$, $\ddot{\Ubb}_{\beta}^n$, $\Ibb_{\alpha}^n$ and $\Jbb_{\alpha}^n$ into \eqref{timeschemePMLdisplacement} and using again the discretized version of the second-order IDEs in time \eqref{algebraicsystimePML2D} and \eqref{newalgebraicsystimePML3D} at previous time $t_{n-1}$, the linear matrix system \eqref{timeschemePMLdisplacement} to be solved at current time $t_n$ for $n\geq 1$ can be recast as
\begin{subequations}\label{matrixsystemdisplacement}
\begin{alignat}{3}
\Kbb^{\text{eff}} \Delta\Ubb^n &= \Fbb^n-\Fbb^{n-1} &&- \Delta t \Gbb \Ubb^{n-1} - \(\beta_2 \Mbb + \gamma_2 \Cbb\) \dot{\Ubb}^{n-1} && \notag\\
& &&- \(\beta_3 \Mbb + \gamma_3 \Cbb\) \ddot{\Ubb}^{n-1} &&%\quad 
\text{for } d=2,\label{matrixsystemdisplacement2D}\\
\Kbb^{\text{eff}} \Delta\Ubb^n &= \Fbb^n-\Fbb^{n-1} &&- \Delta t \(\Gbb + \alpha \Delta t \Hbb\) \Ubb^{n-1} - \(\beta_2 \Mbb + \gamma_2 \Cbb\) &&\dot{\Ubb}^{n-1} \notag\\
& &&- \(\beta_3 \Mbb + \gamma_3 \Cbb\) \ddot{\Ubb}^{n-1} - \Delta t \Hbb \Ibb^{n-1} &&%\quad 
\text{for } d=3,\label{matrixsystemdisplacement3D}
\end{alignat}
\end{subequations}
in which $\Delta\Ubb^n = \Ubb^n-\Ubb^{n-1}$ is the displacement increment FE vector. The time stepping scheme boils down to computing the displacement FE vector $\Ubb^n = \Ubb^{n-1} + \Delta\Ubb^n$ solution of the linear matrix system \eqref{matrixsystemdisplacement}, then updating the corresponding velocity and acceleration FE vectors $\dot{\Ubb}^n$ and $\ddot{\Ubb}^n$ as well as $\Ibb^n$ (only for $d=3$) through \eqref{NewmarktimeschemedUbis}, \eqref{NewmarktimeschemeddU} and \eqref{finitedifferenceschemeiU} at each time step $t_n$. Note that there is no need to compute and store the FE vectors $\Ibb^n$ in the two-dimensional case (\ie{} for $d=2$) and $\Jbb^n$ in the three-dimensional case (\ie{} for $d=3$). Even though the computation and storage of FE vector $\Ibb^n$ at each time $t_n$ are required in the three-dimensional case (\ie{} for $d=3$), it should be noted that $\Ibb^n$ has non-zero components only for the degrees of freedom associated to the spatial discretization of the time-histories $\Ub(\cdot,t_n) = \int_{0}^{t_n} \ub \, d\tau$ and $\Eb(\cdot,t_n) = \int_{0}^{t_n} \eb \, d\tau$ (evaluated at time $t_n$) of displacement field $\ub$ and auxiliary strain field $\eb$, respectively, over the PML region $\Omega_{\text{PML}}$.

\begin{remark}
Alternatively, considering the acceleration-based formula \eqref{Newmarktimescheme} of the Newmark time scheme, we derive an equivalent reformulation of the time stepping scheme \eqref{timeschemePMLdisplacement} expressed in acceleration format that reads as follows for $n\geq 1$,
%\begin{subequations}\label{timeschemePMLacceleration}
%\begin{align}
%\Ubb_{\beta}^n &= \Ubb^{n-1} + \Delta t \dot{\Ubb}^{n-1} + (1-2\beta) \dfrac{{\Delta t}^2}{2} \ddot{\Ubb}^{n-1},\label{predictorU}\\
%\dot{\Ubb}_{\gamma}^n &= \dot{\Ubb}^{n-1} + (1-\gamma) \Delta t \ddot{\Ubb}^{n-1},\label{predictordU}\\
%\Ibb_{\alpha}^n &= \Ibb^{n-1} + (1-\alpha) \Delta t \Ubb^{n-1},\label{predictoriU}\\
%\Jbb_{\alpha}^n &= \Jbb^{n-1} + (1-\alpha) \Delta t \Ibb^{n-1},\label{predictoriiU}\\
%\Mbb^{\text{eff}} \ddot{\Ubb}^n &= \Fbb^n - \Cbb \dot{\Ubb}_{\gamma}^n - \(\Kbb + \alpha \Delta t \(\Gbb + \alpha \Delta t \Hbb\)\) \Ubb_{\beta}^n - \(\Gbb + \alpha \Delta t \Hbb\) \Ibb_{\alpha}^n - \Hbb \Jbb_{\alpha}^n,\label{ddU}\\
%\Ubb^n &= \Ubb_{\beta}^n + \beta {\Delta t}^2 \ddot{\Ubb}^n,\label{U}\\
%\dot{\Ubb}^n &= \dot{\Ubb}_{\gamma}^n + \gamma \Delta t \ddot{\Ubb}^n,\label{dU}\\
%\Ibb^n &= \Ibb_{\alpha}^n + \alpha \Delta t \Ubb^n,\label{iU}\\
%\Jbb^n &= \Jbb_{\alpha}^n + \alpha \Delta t \Ibb^n,\label{iiU}
%\end{align}
%\end{subequations}
\begin{subequations}\label{timeschemePMLacceleration}
\begin{alignat}{3}
\Mbb^{\text{eff}} \ddot{\Ubb}^n &= \Fbb^n - \Cbb \dot{\Ubb}_{\gamma}^n &&- \(\Kbb + \alpha \Delta t \Gbb\) \Ubb_{\beta}^n - \Gbb \Ibb_{\alpha}^n &&\quad \text{for } d=2,\label{timeschemePMLacceleration2D}\\
\Mbb^{\text{eff}} \ddot{\Ubb}^n &= \Fbb^n - \Cbb \dot{\Ubb}_{\gamma}^n &&- \(\Kbb + \alpha \Delta t \(\Gbb + \alpha \Delta t \Hbb\)\) \Ubb_{\beta}^n \notag\\
& &&- \(\Gbb + \alpha \Delta t \Hbb\) \Ibb_{\alpha}^n - \Hbb \Jbb_{\alpha}^n &&\quad \text{for } d=3,\label{timeschemePMLacceleration3D}
\end{alignat}
\end{subequations}
where $\Ubb_{\beta}^n = \Ubb^{n-1} + \Delta t \dot{\Ubb}^{n-1} + (1-2\beta) \dfrac{{\Delta t}^2}{2} \ddot{\Ubb}^{n-1}$ and $\dot{\Ubb}_{\gamma}^n = \dot{\Ubb}^{n-1} + (1-\gamma) \Delta t \ddot{\Ubb}^{n-1}$ are the respective predictors of unknown FE vectors $\Ubb^n$ and $\dot{\Ubb}^n$ at time $t_n$ such that $\Ubb^n = \Ubb_{\beta}^n + \beta {\Delta t}^2 \ddot{\Ubb}^n$ and $\dot{\Ubb}^n = \dot{\Ubb}_{\gamma}^n + \gamma \Delta t \ddot{\Ubb}^n$, and $\Mbb^{\text{eff}}$ is the effective mass matrix defined by
\begin{equation*}
\Mbb^{\text{eff}} = \beta {\Delta t}^2 \Kbb^{\text{eff}} = \begin{cases}
\Mbb + \gamma \Delta t \Cbb + \beta {\Delta t}^2 \(\Kbb + \alpha \Delta t \Gbb\) & \text{for } d=2, \\
\Mbb + \gamma \Delta t \Cbb + \beta {\Delta t}^2 \(\Kbb + \alpha \Delta t \(\Gbb + \alpha \Delta t \Hbb\)\) & \text{for } d=3.
\end{cases}
\end{equation*}
%\begin{equation*}
%\Mbb^{\text{eff}} = \beta {\Delta t}^2 \Kbb^{\text{eff}} = \Mbb + \gamma \Delta t \Cbb + \beta {\Delta t}^2 \(\Kbb + \alpha \Delta t \(\Gbb + \alpha \Delta t \Hbb\)\).
%\end{equation*}
In the same way as $\Kbb^{\text{eff}}$, $\Mbb^{\text{eff}}$ is a non-symmetric real matrix that depends on both parameters $\gamma$ and $\beta$ and also on parameter $\alpha$. For practical purposes, introducing the predictors $\Ubb_{\beta}^n$, $\dot{\Ubb}_{\gamma}^n$, $\Ibb_{\alpha}^n$ and $\Jbb_{\alpha}^n$ into \eqref{timeschemePMLacceleration} and using the discretized version of the second-order IDEs in time \eqref{algebraicsystimePML2D} and \eqref{newalgebraicsystimePML3D} at previous time $t_{n-1}$, the linear matrix system \eqref{timeschemePMLacceleration} to be solved at current time $t_n$ for $n\geq 1$ can be recast as
\begin{subequations}\label{matrixsystemacceleration}
\begin{align}
\Mbb^{\text{eff}} \Delta\ddot{\Ubb}^n = \Fbb^n-\Fbb^{n-1} &- \Delta t \Gbb \Ubb^{n-1} - \Delta t \(\Kbb + \alpha \Delta t \Gbb\) \dot{\Ubb}^{n-1} \notag\\
&- \Delta t\(\Cbb + \dfrac{\Delta t}{2} \(\Kbb + \alpha \Delta t \Gbb\)\) \ddot{\Ubb}^{n-1} \label{matrixsystemacceleration2D}
\end{align}
in the two-dimensional case (\ie{} for $d=2$), and
\begin{align}
\Mbb^{\text{eff}} \Delta\ddot{\Ubb}^n = \Fbb^n&-\Fbb^{n-1} - \Delta t \(\Gbb + \alpha \Delta t \Hbb\) \Ubb^{n-1} - \Delta t \(\Kbb + \alpha \Delta t \(\Gbb + \alpha \Delta t \Hbb\)\) \dot{\Ubb}^{n-1} \notag\\
&- \Delta t\(\Cbb + \dfrac{\Delta t}{2} \(\Kbb + \alpha \Delta t \(\Gbb + \alpha \Delta t \Hbb\)\)\) \ddot{\Ubb}^{n-1} - \Delta t \Hbb \Ibb^{n-1} \label{matrixsystemacceleration3D}
\end{align}
in the three-dimensional case (\ie{} for $d=3$),
\end{subequations}
%\begin{equation}
%\begin{aligned}
%\Mbb^{\text{eff}} \Delta\ddot{\Ubb}^n = \Fbb^n-\Fbb^{n-1} &- \Delta t \(\Gbb + \alpha \Delta t \Hbb\) \Ubb^{n-1} - \Delta t \(\Kbb + \alpha \Delta t \(\Gbb + \alpha \Delta t \Hbb\)\) \dot{\Ubb}^{n-1}\\
%&- \Delta t\(\Cbb + \dfrac{\Delta t}{2} \(\Kbb + \alpha \Delta t \(\Gbb + \alpha \Delta t \Hbb\)\)\) \ddot{\Ubb}^{n-1} - \Delta t \Hbb \Ibb^{n-1},
%\end{aligned}
%\end{equation}
in which $\Delta\ddot{\Ubb}^n = \ddot{\Ubb}^n - \ddot{\Ubb}^{n-1}$ is the acceleration increment FE vector. The time stepping scheme boils down to computing the acceleration FE vector $\ddot{\Ubb}^n = \ddot{\Ubb}^{n-1} + \Delta\ddot{\Ubb}^n$ solution of the linear matrix system \eqref{matrixsystemacceleration}, then updating the displacement and velocity FE vectors $\Ubb^n$ and $\dot{\Ubb}^n$ as well as $\Ibb^n$ (only for $d=3$) through \eqref{NewmarktimeschemeU}, \eqref{NewmarktimeschemedU} and \eqref{finitedifferenceschemeiU} at each time step $t_n$. Note that once again there is no need to compute and store the FE vectors $\Ibb^n$ in the two-dimensional case (\ie{} for $d=2$) and $\Jbb^n$ in the three-dimensional case (\ie{} for $d=3$), whereas it is required to compute and store $\Ibb^n$ at each time $t_n$ in the three-dimensional case (\ie{} for $d=3$).
\end{remark}

%\begin{remark}
%In order to avoid the computation and storage of $\Ibb^n$ in the three-dimensional case (\ie{} for $d=3$), a possible alternative would consist in defining the PML region $\Omega_{\text{PML}}$ as the disjoint union of $d$ non-overlapping subregions $\set{\widetilde{\Omega}_i}_{i=1}^{d}$ such that $\widetilde{\Omega}_i = \setst{\xb\in \Omega_{\text{PML}}}{(\abs{x_i}-\ell_i)/h_i > (\abs{x_j}-\ell_j)/h_j \text{ for all } j \in \{1,\dots,d\} \text{ such that } j\neq i} \subset \Omega_i$. Then, only one damping function would be non-zero in each subregion of $\Omega_{\text{PML}}$. More precisely, in each PML subregion $\widetilde{\Omega}_i$, $d_i(x_i)\neq 0$ and $d_j(x_j) = 0$ for all $j \in \{1,\dots,d\}$ such that $j\neq i$, so that $\Db_3^{\Pi} = \zerob$, $\det(\Db) = 0$ and then the FE matrices $\Kbf_2=\Gbf_1=\zerob$, $\Kbf_3=\zerob$, $\Hbf=\zerob$, $\widetilde{\Hbf}_1=\zerob$, therefore $\Hbb=\zerob$ and the linear matrix system \eqref{matrixsystemdisplacement3D} (resp. \eqref{matrixsystemacceleration3D}) in the three-dimensional case (\ie{} for $d=3$) reduces to the one \eqref{matrixsystemdisplacement2D} (resp. \eqref{matrixsystemacceleration2D}) obtained in the two-dimensional case (\ie{} for $d=2$).
%\end{remark}

\begin{remark}
Following Remark~\ref{rmrk:CALformulation}, recall that the time-domain PML formulation is compared to a CAL formulation governed by the second-order ODE in time \eqref{algebraicsystimeCAL} in the numerical experiments shown in Section~\ref{sec:results}. In the time domain, such a CAL formulation comes down to solving the following linear matrix system at each time step $t_n$ for $n\geq 1$:
\begin{equation}\label{CALmatrixsystemdisplacement}
\Kbf^{\text{eff}} \Delta\Ubf^n = \Fbf^n-\Fbf^{n-1} - \(\beta_2 \Mbf + \gamma_2 \Cbf\) \dot{\Ubf}^{n-1} - \(\beta_3 \Mbf + \gamma_3 \Cbf\) \ddot{\Ubf}^{n-1},
\end{equation}
with the initial conditions given by $\Ubf^0$ and $\Vbf^0$, and where $\Delta\Ubf^n = \Ubf^n - \Ubf^{n-1}$ is the displacement increment FE vector and $\Kbf^{\text{eff}}$ is the effective stiffness matrix defined by
\begin{equation*}
\Kbf^{\text{eff}} = \beta_1 \Mbf^{\text{eff}} = \beta_1 \Mbf + \gamma_1 \Cbf + \Kbf,
\end{equation*}
with $\Mbf^{\text{eff}} = \Mbf + \gamma \Delta t \Cbf + \beta {\Delta t}^2 \Kbf$ the corresponding effective mass matrix.
\end{remark}

\section{Numerical results}\label{sec:results}

In order to assess the performances and the robustness of the proposed PML formulation, we present different numerical experiments for simulating the propagation and absorption of elastic waves in two-dimensional semi-infinite (unbounded) single- or multi-layer isotropic homogeneous elastic media (without any viscous damping) subjected to a directional force point-source. We study numerically the accuracy and stability of the proposed time-domain PML formulation governed by the second-order IDE in time \eqref{algebraicsystimePML2D} %for long-time simulations 
and compare it to the time-domain CAL formulation governed by the second-order ODE in time \eqref{algebraicsystimeCAL}.

\begin{figure}
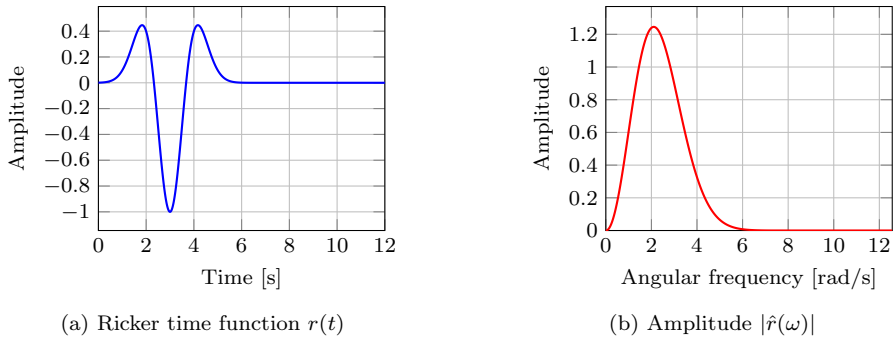

\centering
\begin{subfigure}[t]{0.45\textwidth}
	\centering
	\tikzsetnextfilename{Ricker_time}
	\input{Ricker_time_matlab}
	\caption{Ricker time function $r(t)$}\label{fig:Ricker_time}
\end{subfigure}\hfill
%\begin{subfigure}[t]{0.45\textwidth}
%	\centering
%	\tikzsetnextfilename{Ricker_freq}
%	%\input{Ricker_freq}
%	\input{Ricker_freq_matlab}
%	\caption{Amplitude $\abs{\hat{r}(f)}$}\label{fig:Ricker_freq}
%\end{subfigure}
\begin{subfigure}[t]{0.45\textwidth}
	\centering
	\tikzsetnextfilename{Ricker_omega}
	\input{Ricker_omega_matlab}
	\caption{Amplitude $\abs{\hat{r}(\omega)}$}\label{fig:Ricker_omega}
\end{subfigure}
%\caption{Evolutions of (\subref{fig:Ricker_time}) the Ricker wavelet $r(t)$ with respect to time $t$ over the time interval $\intervalcc{0}{12}$~s and (\subref{fig:Ricker_freq}) the amplitude $\abs{\hat{r}(f)}$ of its Fourier transform in time $\hat{r}(f)$ with respect to frequency $f$ over the the frequency band $\intervalcc{0}{2}$~Hz}\label{fig:Ricker}
\caption{Evolutions of (\subref{fig:Ricker_time}) the Ricker wavelet $r(t)$ with respect to time $t$ over the time interval $\intervalcc{0}{12}$~s and (\subref{fig:Ricker_omega}) the amplitude $\abs{\hat{r}(\omega)}$ of its Fourier transform in time $\hat{r}(\omega)$ with respect to angular frequency $\omega$ over the angular frequency band $\intervalcc{0}{4\pi}$~rad/s (corresponding to the frequency band $\intervalcc{0}{2}$~Hz)}\label{fig:Ricker}
\end{figure}

In all the numerical experiments, an explosive source term $\fb(\xb,t)$, acting as a downward vertical force, is applied to a source point $S$ located in the physical domain $\Omega_{\text{BD}}$ at the position $\xb^S$ and defined as
\begin{equation*}
\fb(\xb,t) = r(t) g(\xb) \ebf_2,
\end{equation*}
where the source spatial function $g(\xb) = \delta(\xb-\xb^S)$ is the Dirac delta function $\delta$ (also known as the impulse function) on $\Rbb^d$ at the origin (\ie{} at point $\zerob$) shifted by the source position $\xb^S$, and the source time function $r(t)$ is a second-order Ricker wavelet\footnote{The Ricker wavelet $r(t)$ defined in \eqref{Rickerwavelet} is such that $r(t) = \dfrac{d^2p}{d t^2}(t)$, where $p(t) = \dfrac{1}{2(\pi f_d)^2} \exp(-(\pi f_d)^2(t - t_d)^2)$ corresponds to a modified second-order derivative of the Gaussian probability density function $t\mapsto \dfrac{1}{\sigma \sqrt{2\pi}} \exp\(-\dfrac{(t - t_d)^2}{2\sigma^2}\)$ with mean value $t_d$ and standard deviation $\sigma=\sqrt{2}/\omega_d$ up to a multiplicative constant $4\sqrt{\pi}/\omega_d^3$ with $\omega_d=2\pi f_d$ the central (or dominant) angular frequency.} (sometimes called a Mexican hat wavelet) given by \cite{Beca03,App06b,Meza08,Meza10,Meza12,Mat11,Brun16,Zaf16,Li18,He19}
\begin{equation}\label{Rickerwavelet}
%r(t) = \(2(\pi f_d)^2(t-t_d)^2 - 1\) e^{-(\pi f_d)^2(t - t_d)^2},
r(t) = \(2(\pi f_d)^2(t-t_d)^2 - 1\) \exp(-(\pi f_d)^2(t - t_d)^2),
\end{equation}
with a central (or dominant) frequency $f_d = 1/3$~Hz, and a source time delay $t_d = 1/f_d = 3$~s. The time-Fourier transform $\hat{r}(\omega)$ of the wavelet function $r(t)$ is written as
\begin{equation*}
%\hat{r}(\omega) = -4\sqrt{\pi} \dfrac{\omega^2}{\omega_d^3} e^{-\omega^2/\omega_d^2} e^{-\mathrm{i} \omega t_d}
\hat{r}(\omega) = -4\sqrt{\pi} \dfrac{\omega^2}{\omega_d^3} \exp\(-\dfrac{\omega^2}{\omega_d^2}\) \exp(-\mathrm{i} \omega t_d)
\end{equation*}
with a central (or dominant) angular frequency $\omega_d = 2\pi f_d = 2\pi/3 \approx 2.0944$~rad/s. Figure~\ref{fig:Ricker} shows the evolutions of the Ricker wavelet $r(t)$ with respect to time $t$ and of the amplitude $\abs{\hat{r}(\omega)}$ of its time-Fourier transform $\hat{r}(\omega)$ with respect to angular frequency $\omega=2\pi f$.

\def \a {1}
\def \fd {1/3}
\def \omegad {2*pi*\fd}
\def \td {3}

For the time integration scheme used for solving the time-domain equations, a constant time step $\Delta t = 0.025$~s is chosen to correctly reproduce the source time variation and adequately captured the numerical solutions for all the numerical tests. The final sampled time of the numerical simulations is $T=20$~s, thus requiring $N=800$ time steps. Besides, the elastic medium is assumed to be at rest at initial time $t=0$, so that zero initial conditions $\ub(\xb,0) = \ub_0(\xb) = \zerob$ and $\dot{\ub}(\xb,0) = \vb_0(\xb) = \zerob$ hold for any $\xb$ in $\Rbb^d$. As regards the spatial discretization, all the finite element meshes considered in the numerical experiments are generated with Gmsh \cite{Geu09}. Finally, all material parameters and physical quantities are expressed in terms of the International System of Units (universally abbreviated SI units), such as times in [s], frequencies in [Hz], angular frequencies in [rad.s$^{-1}$], lengths and displacements in [m], velocities in [m.s$^{-1}]$, mass densities in [kg.m$^{-3}]$, elastic moduli in [Pa], energies in [J], \etc, and we further drop units for the sake of readability.

\subsection{Lamb's problem: directional force point-source applied onto the free surface of a two-dimensional semi-infinite isotropic homogeneous elastic medium}\label{sec:Brun2016_Lamb}

We first consider the well-known Lamb's problem \cite{Lamb04}, which consists in modeling the propagation of elastic waves in an isotropic homogeneous %elastic 
half-space with free-surface boundary conditions and a directional force point-source applied onto the top free surface of the semi-infinite (unbounded) domain.

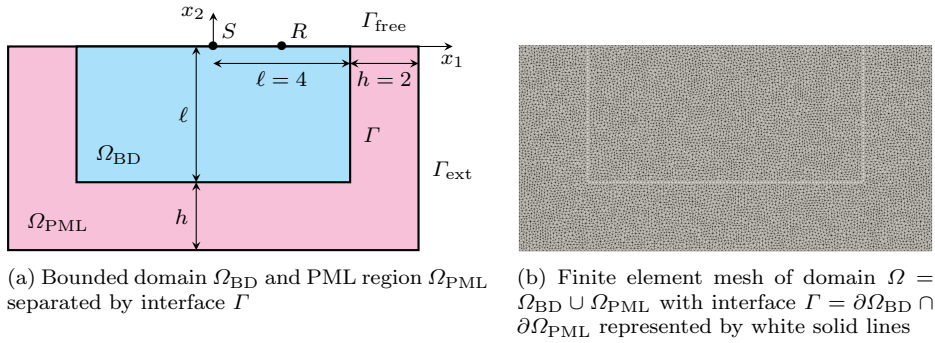
\begin{figure}
\centering
\begin{subfigure}[t]{0.52\textwidth}
	\centering
	\tikzsetnextfilename{Brun2016_Lamb_domain}
	\begin{tikzpicture}[scale=0.45]

\def \lx {4.0}
\def \ly {4.0}
\def \hx {2.0}
\def \hy {2.0}
%\draw[help lines,step=.5] (-\lx-\hx-1,-\ly-\hy-1) grid (\lx+\hx+1,2);

% Origin-Source
\coordinate (O) at (0,0);

% Receiver
\coordinate (R) at (\lx/2,0);

% Domain coordinates
\coordinate (A) at (-\lx,-\ly);
\coordinate (B) at (\lx,0);
\coordinate (C) at (-\lx,0);
\coordinate (D) at (\lx,-\ly);

% PML coordinates
\coordinate (E) at (-\lx-\hx,-\ly-\hy);
\coordinate (F) at (\lx+\hx,0);
\coordinate (G) at (-\lx-\hx,0);
\coordinate (H) at (\lx+\hx,-\ly-\hy);
\coordinate (Interface) at (\lx,-\ly/2);
\coordinate (Boundary) at (\lx+\hx,-\ly/2-\hy/2);
\coordinate (FreeBoundary) at (\lx+\hx/2,-0);

% PML
\draw[black,thick,fill=magenta!30] (E) rectangle (F);
\node[above right=0.2] at (E) {$\Omega_{\text{PML}}$};
\draw[solid,black,stealth-stealth] (\lx,-0.5) -- (\lx+\hx,-0.5) node[midway,below]{$h = 2$};
\draw[solid,black,stealth-stealth] (-0.5,-\ly) -- (-0.5,-\ly-\hy) node[midway,left]{$h$};

% Domain
\draw[black,thick,fill=cyan!30] (A) rectangle (B);
\draw (O) node[above right] {$S$} node{$\bullet$};
\draw (R) node[above right] {$R$} node{$\bullet$};
\node[above right=0.2] at (A) {$\Omega_{\text{BD}}$};
\draw[solid,black,stealth-stealth] (0,-0.5) -- (\lx,-0.5) node[midway,below]{$\ell = 4$};
\draw[solid,black,stealth-stealth] (-0.5,0) -- (-0.5,-\ly) node[midway,left]{$\ell$};

% Interface
\node[below right=0.1] at (Interface) {$\Gamma$};
\node[below right=0.1] at (Boundary) {$\Gamma_{\text{ext}}$};
\node[above=0.1] at (FreeBoundary) {$\Gamma_{\text{free}}$};

% Axis
\draw[solid,black,-stealth] (0,0) -- (\lx+\hx+1,0) node[below]{$x_1$};
\draw[solid,black,-stealth] (0,0) -- (0,1) node[left]{$x_2$};

\end{tikzpicture}
	\caption{Bounded domain $\Omega_{\text{BD}}$ and PML region $\Omega_{\text{PML}}$ separated by interface $\Gamma$}\label{fig:Brun2016_Lamb_domain}
\end{subfigure}\hfill
\begin{subfigure}[t]{0.45\textwidth}
	\centering
	\includegraphics[height=0.92\figureheight]{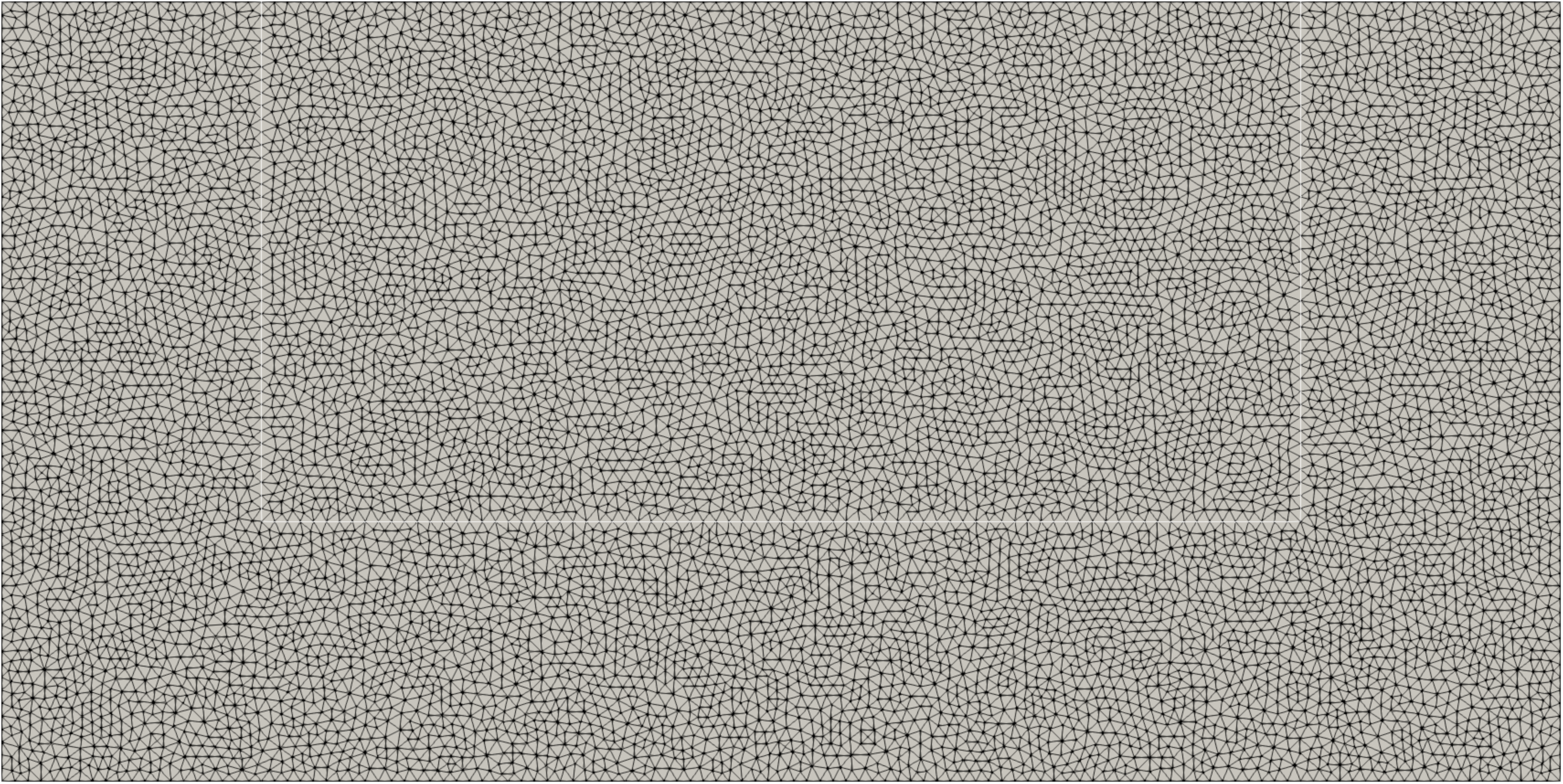}
	\caption{Finite element mesh of domain $\Omega = \Omega_{\text{BD}} \cup \Omega_{\text{PML}}$ with interface $\Gamma = \partial\Omega_{\text{BD}} \cap \partial\Omega_{\text{PML}}$ represented by white solid lines}\label{fig:Brun2016_Lamb_mesh}
\end{subfigure}
\caption{Lamb's problem: (\subref{fig:Brun2016_Lamb_domain}) two-dimensional semi-infinite domain (half-space) modeled by a physical bounded domain $\Omega_{\text{BD}}$ surrounded by an artificial PML region $\Omega_{\text{PML}}$, subjected to a downward vertical force point-source $S$, with a receiver $R$, both located at the top free surface $\Gamma_{\text{free}}$ of $\Omega_{\text{BD}}$, and (\subref{fig:Brun2016_Lamb_mesh}) associated finite element mesh}\label{fig:Brun2016_Lamb}
\end{figure}

We aims at numerically simulating the propagation of elastic waves in the half-plane $\intervaloo{-\infty}{+\infty} \times \intervaloc{-\infty}{0} \subset \Rbb^2$ (with a top free surface $\Gamma_{\text{free}}$ located at $x_2=0$) generated by a vertical force point-source $S$ located at the center of the top free surface $\Gamma_{\text{free}}$ of the semi-infinite domain, \ie{} at the position $\xb^S = (0, 0)$ of the frame origin (see Figure~\ref{fig:Brun2016_Lamb_domain}). The material properties of the isotropic homogeneous elastic medium are characterized by a mass density $\rho=1$, a Young's modulus $E=2.5$ and a Poisson's ratio $\nu=0.25$ under the 2D plane strain assumption, so that the Lam\'e's coefficients are $\lambda = \mu = 1$, the longitudinal $P$-wave velocity (resp. wavelength) is $c_p\approx 1.7321$ (resp. $\lambda_p = c_p/f_d \approx 5.1962$) and the transverse $S$-wave velocity (resp. wavelength) is $c_s=1$ (resp. $\lambda_s = c_s/f_d = 3$). This academic test case has already been studied in \cite{Brun16,Zaf16}. The exact analytical solution involves bulk $P$- and $S$-waves as well as Rayleigh waves generated by the point-source.

We introduce a rectangular (two-dimensional) bounded domain $\Omega_{\text{BD}} = \intervalcc{-\ell}{\ell} \times \intervalcc{-\ell}{0}$ with length $2\ell = 8$ and width $\ell = 4$, surrounded by a PML region $\Omega_{\text{PML}} = \setst{\xb \in \Rbb^2}{\ell< \abs{x_1}\leq \ell+h \text{ and } -\ell-h\leq x_2< -\ell}$ corresponding to three overlapping absorbing layers (on the right, left and bottom sides of the physical domain of interest $\Omega_{\text{BD}}$) with the same finite thickness $h=\ell/2=2$ for each absorbing layer (see Figure~\ref{fig:Brun2016_Lamb_domain}). The entire computational domain $\Omega = \Omega_{\text{BD}}\cup \Omega_{\text{PML}} = \intervalcc{-\ell-h}{\ell+h} \times \intervalcc{-\ell-h}{0}$ is then a rectangular bounded domain with length $2(\ell+h) = 12$ and width $\ell+h = 6$ over which the calculations are performed. As already mentioned in Section~\ref{sec:absorptionproperties}, due to the expected absorption properties in the PML region $\Omega_{\text{PML}}$, we apply homogeneous Dirichlet boundary conditions $\ub = \zerob$ on the exterior boundary $\Gamma_{\text{ext}}$ of $\Omega_{\text{PML}}$, that is at $\abs{x_1}=\ell+h=6$ and at $x_2=-\ell-h=-6$, and we consider stress-free boundary conditions $\scalprod{\sigmab}{\nb} = \zerob$ at the top surface $\Gamma_{\text{free}}$ of the entire domain $\Omega = \Omega_{\text{BD}} \cup \Omega_{\text{PML}}$, that is at $x_2=0$. For the damping profiles in the PML region $\Omega_{\text{PML}}$, we use the empirical damping function $d_i(x_i)$ defined in \eqref{powerlawfunction} with a polynomial degree $p=2$ and a damping coefficient $d_i^{\mathrm{max}}$ given by \eqref{dampingcoefficientgeneral} with a phase velocity $v_p = c_p$ (being equal to the longitudinal $P$-wave speed) and an amplitude $A_i = 6$ corresponding to a small theoretical reflection coefficient $R_i = 10^{-4}$ in \eqref{dampingcoefficient}. Note that the PML thickness $h=2$ is slightly lower than a half ($1/2$) of the longitudinal $P$-wave wavelength $\lambda_p \approx 5.1962$ and equal to two thirds ($2/3$) of the shear $S$-wave wavelength $\lambda_s = 3$.

\begin{figure}
	\centering
	\includegraphics[width=\textwidth]{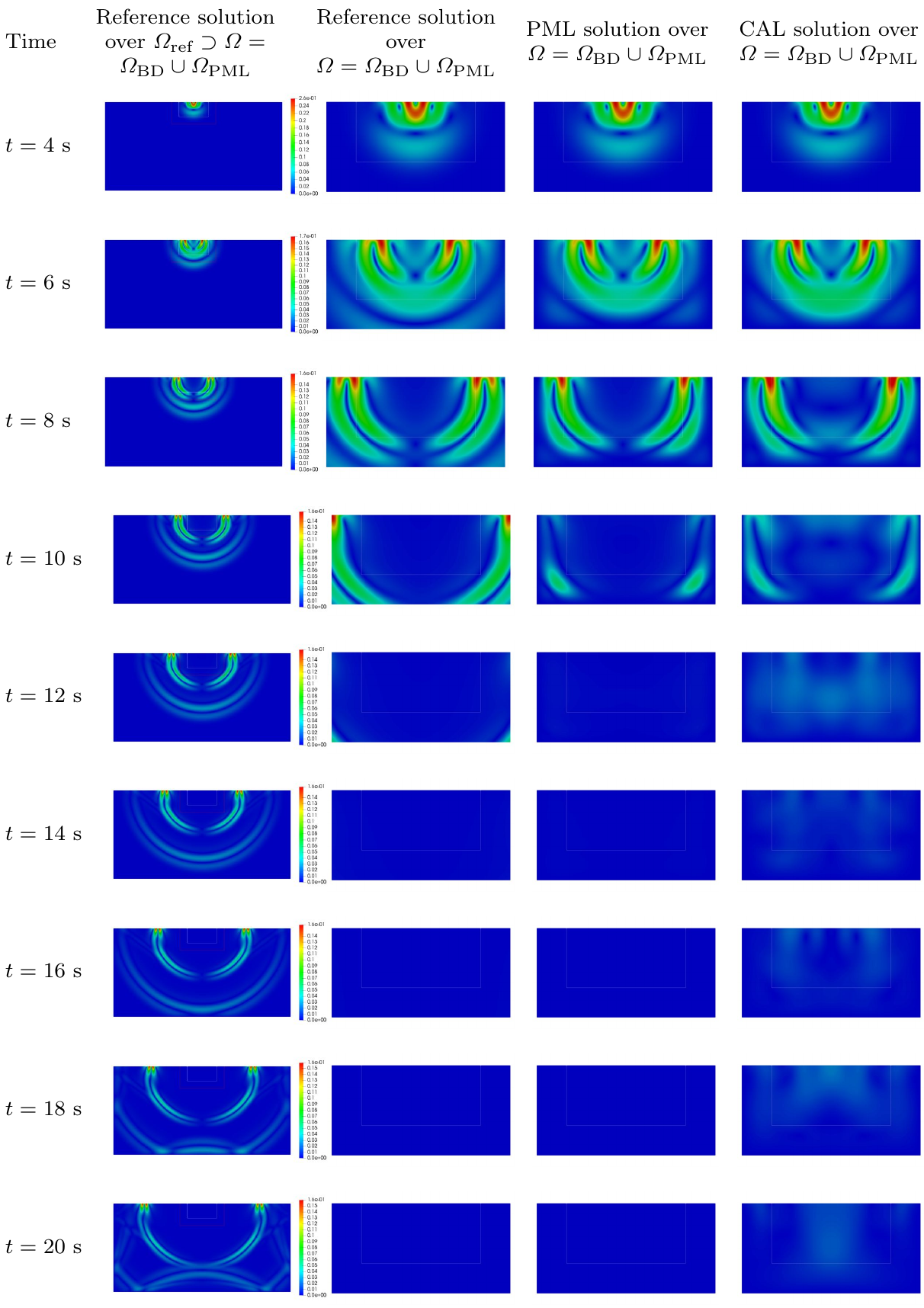}
	\caption{Lamb's problem: snapshots of the displacement field magnitude $\xb \mapsto \norm{\ub(\xb,t)}$ taken at different times $t=4,6,8,10,12,14,16,18,20$~s (from top to bottom) for the reference solution over an extended computational domain $\Omega_{\text{ref}} \supset \Omega = \Omega_{\text{BD}} \cup \Omega_{\text{PML}}$ (first column) with interface $\Gamma = \partial\Omega_{\text{BD}} \cap \partial\Omega_{\text{PML}}$ indicated by white solid lines and the exterior boundary $\Gamma_{\text{ext}}$ of $\Omega$ indicated by red solid lines, the reference solution over the computational domain $\Omega = \Omega_{\text{BD}} \cup \Omega_{\text{PML}}$ (second column), the PML solution over $\Omega$ (third column) and the CAL solution over $\Omega$ (fourth column)}\label{fig:Brun2016_Lamb_displacement}
\end{figure}

For the spatial discretization, the finite element mesh is an unstructured triangulation of the entire domain $\Omega$ that is composed of $6$-nodes quadratic triangular elements with a characteristic element size $\Delta x = 0.1$ such that $\Delta x = \lambda_s/30 \approx \lambda_p/50$ (see Figure~\ref{fig:Brun2016_Lamb_mesh}). It thus contains $38\,209$ nodes (\ie{} $76\,418$ degrees of freedom) and $18\,924$ elements. We use quadratic finite elements for the spatial discretization of both displacement field $\ub$ and auxiliary strain field $\eb$, since they usually outperform linear finite elements in terms of accuracy \cite{Bin05}. Also, standard isoparametric quadratic finite elements have already been used and found to be numerically stable in other PML formulations \cite{Kang10,Kuc10,Kuc11,Ass12,Kuc13,Fat15a,Fat15b,Assi15}. We validate the proposed PML approach by comparing the numerical PML solution $\ub$ to a reference solution $\ub^{\text{ref}}$ computed by using a standard purely displacement-based formulation on the same time interval $I=\intervalcc{0}{T}$ with the same time step $\Delta t = 0.025$~s, but on an extended (or enlarged) computational domain $\Omega_{\text{ref}} = \intervalcc{-L}{L} \times \intervalcc{-L}{0}$ (containing the physical domain of interest $\Omega_{\text{BD}}$) with a top free boundary and left, right and bottom fixed exterior boundaries, and whose side length $L = 6\ell = 24$ is equal to six times the size $\ell$ of the physical domain $\Omega_{\text{BD}}$, so that any outgoing waves (exiting $\Omega_{\text{BD}}$) remain within $\Omega_{\text{ref}}\setminus \Omega_{\text{BD}}$ and do not travel back and do not interfere with the wave solution in $\Omega_{\text{BD}}$ during the time interval of interest $I=\intervalcc{0}{T}$ (see the first column in Figure~\ref{fig:Brun2016_Lamb_displacement}). This extended domain is discretized using the same characteristic element size $\Delta x = 0.1$ as the one considered for the spatial discretization of domain $\Omega$. The reference finite element mesh then comprises $610\,113$ nodes (\ie{} $1\,220\,226$ degrees of freedom) and $304\,336$ quadratic triangular elements. The extended computational domain $\Omega_{\text{ref}}$ is naturally made of the same isotropic homogeneous linear elastic material as the physical domain $\Omega_{\text{BD}}$. Note that the reference solution $\ub^{\text{ref}}$ computed within $\Omega_{\text{ref}}$ is completely independent from the PML solution $\ub$ computed within $\Omega$. Both reference and PML solutions are compared only within the physical domain $\Omega_{\text{BD}}$.

A receiver $R$ is located on the top free surface $\Gamma_{\text{free}}$, at a distance of $2$ from the source $S$ in the $x_1$-direction, \ie{} at the position $\xb^R = (2, 0)$ (see Figure~\ref{fig:Brun2016_Lamb_domain}), and records the two components of the displacement vector $\ub=(u_1,u_2)$ over the time period $I=\intervalcc{0}{20}$~s. Figure~\ref{fig:Brun2016_Lamb_displacement_R} represents the time evolutions of the horizontal and vertical displacements $u_1(\xb^R,t)$ and $u_2(\xb^R,t)$ at receiver $R$ for both PML and CAL solutions compared to the reference solution. For a more quantitative analysis, the corresponding absolute errors with respect to the reference solution are plotted in a logarithmic scale. The PML transient dynamical response is very close to the reference one, whereas the CAL transient dynamical response is clearly not acceptable from approximately time $t=7$~s due to multiple reflections occurring at the interface $\Gamma$ between the elastic physical domain $\Omega_{\text{BD}}$ and the absorbing unphysical layers constituting $\Omega_{\text{PML}}$.

\begin{figure}
\centering
\begin{subfigure}[t]{0.45\textwidth}
	\centering
	\tikzsetnextfilename{Brun2016_Lamb_displacement_Ux}
	\input{Brun2016_Lamb_displacement_Ux_R}
	\caption{%Time history of horizontal displacement
	Horizontal displacement
	at receiver $R$
	}\label{fig:Brun2016_Lamb_displacement_Ux_R}
\end{subfigure}\hfill
\begin{subfigure}[t]{0.45\textwidth}
	\centering
	\tikzsetnextfilename{Brun2016_Lamb_error_displacement_Ux_R}
	\input{Brun2016_Lamb_error_displacement_Ux_R}
	\caption{%Time history of the error on horizontal displacement
	Error on horizontal displacement
	at receiver $R$
	}\label{fig:Brun2016_Lamb_error_displacement_Ux_R}
\end{subfigure}\\
\begin{subfigure}[t]{0.45\textwidth}
	\centering
	\tikzsetnextfilename{Brun2016_Lamb_displacement_Uy_R}
	\input{Brun2016_Lamb_displacement_Uy_R}
	\caption{%Time history of vertical displacement
	Vertical displacement
	at receiver $R$
	}\label{fig:Brun2016_Lamb_displacement_Uy_R}
\end{subfigure}\hfill
\begin{subfigure}[t]{0.45\textwidth}
	\centering
	\tikzsetnextfilename{Brun2016_Lamb_error_displacement_Uy_R}
	\input{Brun2016_Lamb_error_displacement_Uy_R}
	\caption{%Time history of the error on vertical displacement
	Error on vertical displacement
	at receiver $R$
	}\label{fig:Brun2016_Lamb_error_displacement_Uy_R}
\end{subfigure}
\caption{Lamb's problem: evolutions of (\subref{fig:Brun2016_Lamb_displacement_Ux_R}) the horizontal displacement $u_1$, (\subref{fig:Brun2016_Lamb_error_displacement_Ux_R}) the absolute error $\abs{u_1 - u_1^{\text{ref}}}$ on the horizontal displacement, (\subref{fig:Brun2016_Lamb_displacement_Uy_R}) the vertical displacement $u_2$ and (\subref{fig:Brun2016_Lamb_error_displacement_Uy_R}) the absolute error $\abs{u_2 - u_2^{\text{ref}}}$ on the vertical displacement at receiver $R$ with respect to time $t$ for the PML and CAL solutions compared to the reference solution}\label{fig:Brun2016_Lamb_displacement_R}
\end{figure}

\begin{figure}
	\centering
	\includegraphics[width=\textwidth]{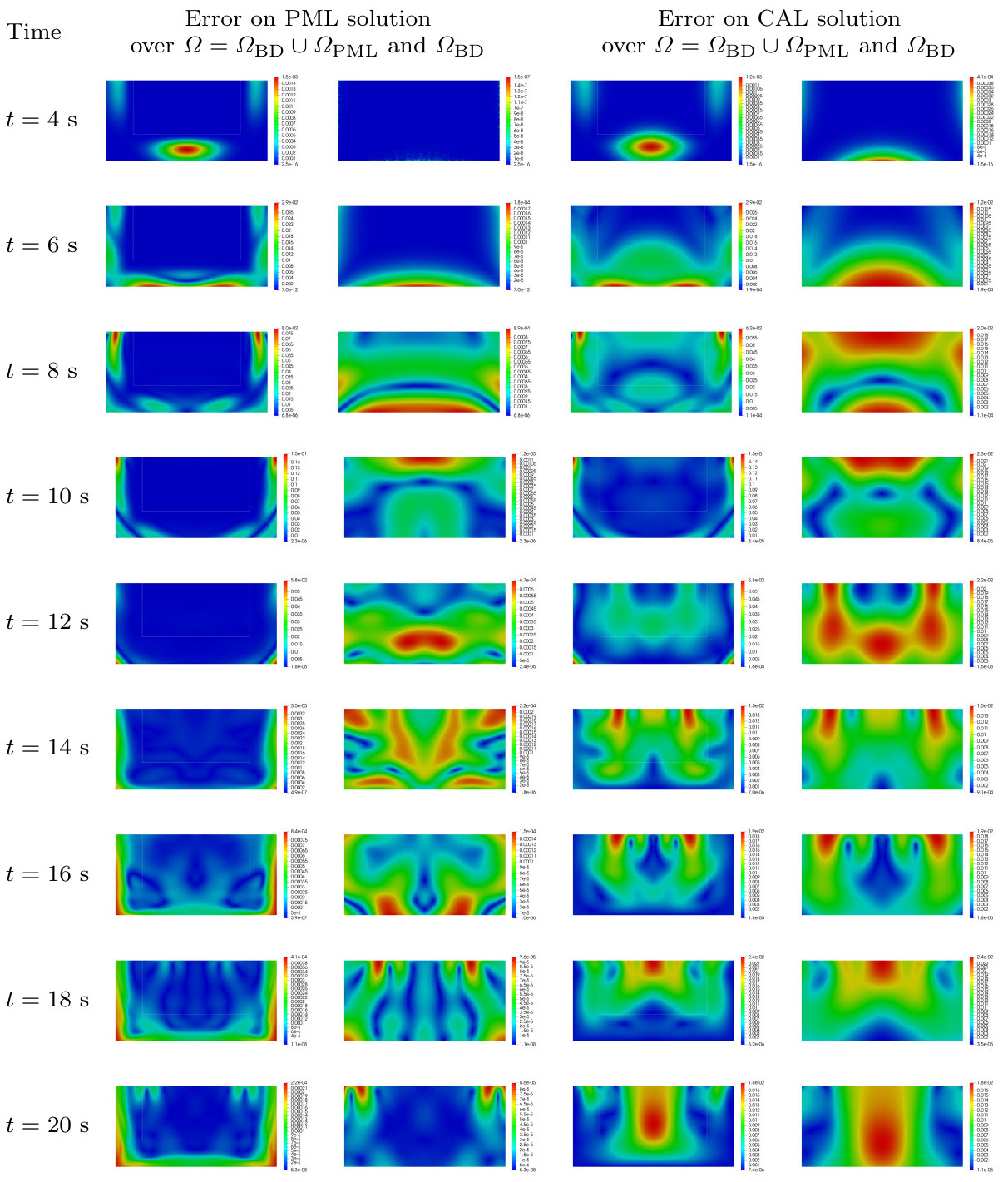}
	\caption{Lamb's problem: snapshots of the absolute error on the displacement field magnitude $\xb \mapsto \abs{\norm{\ub(\xb,t)} - \norm{\ub^{\text{ref}}(\xb,t)}}$ taken at different times $t=4,6,8,10,12,14,16,18,20$~s (from top to bottom) for the PML solution over the entire computational domain $\Omega = \Omega_{\text{BD}} \cup \Omega_{\text{PML}}$ (first column) and its restriction to the physical bounded domain $\Omega_{\text{BD}}$ (second column) and the CAL solution over $\Omega$ and $\Omega_{\text{BD}}$ (third and fourth columns)}\label{fig:Brun2016_Lamb_error_displacement}
\end{figure}

Snapshots of the displacement field magnitude $\xb \mapsto \norm{\ub(\xb,t)} = \sqrt{\ub(\xb,t)^T \ub(\xb,t)} = \sqrt{u_1(\xb,t)^2+u_2(\xb,t)^2}$ taken at different times are displayed in Figure~\ref{fig:Brun2016_Lamb_displacement} for the reference solution computed over the extended computational domain $\Omega_{\text{ref}}$ and truncated over the computational domain $\Omega = \Omega_{\text{BD}} \cup \Omega_{\text{PML}}$ as well as for the PML and CAL solutions over $\Omega$. We observe that the vertical force point-source has first generated two weak bulk waves, a longitudinal $P$-wave followed by a transverse $S$-wave (the former being faster with a lower amplitude than the latter) and then produced two strong Rayleigh waves propagating at the free surface $\Gamma_{\text{free}}$ with higher amplitudes than the bulk waves. The PML solution is in very good agreement with the reference solution inside the physical bounded domain $\Omega_{\text{BD}}$ without any visible reflections, unlike the CAL solution that suffers from spurious reflections at the interface $\Gamma$ between $\Omega_{\text{BD}}$ and $\Omega_{\text{PML}}$. For comparison purposes, Figure~\ref{fig:Brun2016_Lamb_error_displacement} shows snapshots of the absolute error on the displacement field magnitude $\xb \mapsto \abs{\norm{\ub(\xb,t)} - \norm{\ub^{\text{ref}}(\xb,t)}}$ at different times for the PML and CAL solutions compared to the reference solution $\ub^{\text{ref}}$ over both the entire computational domain $\Omega = \Omega_{\text{BD}} \cup \Omega_{\text{PML}}$ and the physical domain $\Omega_{\text{BD}}$. The errors obtained for the PML solution remain very small, between one and three orders of magnitude lower than that obtained for the CAL solution. The PML method is clearly more efficient than the CAL method for absorbing both bulk and Rayleigh waves.

To further illustrate the accuracy and efficiency of the proposed PML method, we compute the total mechanical energy $E(t)$ stored in the physical bounded domain $\Omega_{\text{BD}}$ as the sum of kinetic and internal energies at each time $t$
\begin{equation*}
E(t) = E_k(t) + E_i(t),
\end{equation*}
where $E_k(t)$ is the kinetic energy and $E_i(t)$ is the internal energy, respectively defined by
\begin{alignat*}{2}
E_k(t) &= \frac{1}{2} \int_{\Omega_{\text{BD}}} \rho \norm{\dot{\ub}(\xb,t)}^2 \, d\Omega &&= \frac{1}{2} \dot{\Ubf}_{\text{BD}}(t)^T \Mbf_{\text{BD}} \dot{\Ubf}_{\text{BD}}(t),\\
E_i(t) &= \frac{1}{2} \int_{\Omega_{\text{BD}}} \sigmab(\xb,t): \epsilonb(\ub(\xb,t)) \, d\Omega &&= \frac{1}{2} \Ubf_{\text{BD}}(t)^T \Kbf_{\text{BD}} \Ubf_{\text{BD}}(t),
\end{alignat*}
where $\norm{\dot{\ub}(\xb,t)} = \sqrt{\dot{\ub}(\xb,t)^T \dot{\ub}(\xb,t)} = \sqrt{\dot{u}_1(\xb,t)^2+\dot{u}_2(\xb,t)^2}$ is the amplitude of the vector-valued velocity field $\dot{\ub}(\xb,t)$ at time $t\geq 0$ and position $\xb \in \Omega_{\text{BD}}$, $\Ubf_{\text{BD}}(t)$ and $\dot{\Ubf}_{\text{BD}}(t)$ are the finite element real vectors associated to displacement and velocity fields $\ub(\cdot,t)$ and $\dot{\ub}(\cdot,t)$ over the physical domain $\Omega_{\text{BD}}$, and $\Kbf_{\text{BD}}$ and $\Mbf_{\text{BD}}$ are the corresponding standard finite element stiffness and mass matrices over $\Omega_{\text{BD}}$. Figure~\ref{fig:Brun2016_Lamb_energies} shows the evolutions of the kinetic energy $E_k(t)$, internal energy $E_i(t)$ and total energy $E(t)$ (contained in the physical domain $\Omega_{\text{BD}}$) with respect to time $t$, as well as the ones of the respective absolute errors $\abs{E_k(t) - E_k^{\text{ref}}(t)}$, $\abs{E_i(t) - E_i^{\text{ref}}(t)}$ and $\abs{E(t) - E^{\text{ref}}(t)}$ plotted in a logarithmic scale, where the superscript ${}^{\text{ref}}$ over a variable denotes its reference counterpart computed using the reference solution $\ub^{\text{ref}}$ over $\Omega_{\text{BD}}$. We observe that the vertical force point-source injects energy in the physical domain $\Omega_{\text{BD}}$ from initial time $t=0$~s to approximately time $t=4.5$~s. The energy is then carried by bulk ($P$- and $S$-waves) and Rayleigh waves until they progressively leave the physical domain $\Omega_{\text{BD}}$ between approximately time $t=4.5$~s for the first $P$-wave and time $t=9$~s for the last Rayleigh wave (see Figure~\ref{fig:Brun2016_Lamb_displacement}). The errors on kinetic, internal and total energies computed using the CAL solution reach much higher values than the ones calculated using the PML solution due to the amount of energy carried by the spurious waves reflected at the interface $\Gamma$ and coming back into the physical domain of interest $\Omega_{\text{BD}}$. The energy decay of the PML solution within the physical domain $\Omega_{\text{BD}}$ is gradual and closely follows that of the reference solution without any distinguishable discrepancy, thus indicating the efficiency of the proposed PML method.

\begin{figure}
\centering
\begin{subfigure}[t]{0.45\textwidth}
	\centering
	\tikzsetnextfilename{Brun2016_Lamb_kinetic_energy}
	\input{Brun2016_Lamb_kinetic_energy}
	\caption{%Time history of kinetic energy
	Kinetic energy
	}\label{fig:Brun2016_Lamb_kinetic_energy}
\end{subfigure}\hfill
\begin{subfigure}[t]{0.45\textwidth}
	\centering
	\tikzsetnextfilename{Brun2016_Lamb_error_kinetic_energy}
	\input{Brun2016_Lamb_error_kinetic_energy}
	\caption{%Time history of the error on kinetic energy
	Error on kinetic energy
	}\label{fig:Brun2016_Lamb_error_kinetic_energy}
\end{subfigure}\\
\begin{subfigure}[t]{0.45\textwidth}
	\centering
	\tikzsetnextfilename{Brun2016_Lamb_internal_energy}
	\input{Brun2016_Lamb_internal_energy}
	\caption{%Time history of internal energy
	Internal energy
	}\label{fig:Brun2016_Lamb_internal_energy}
\end{subfigure}\hfill
\begin{subfigure}[t]{0.45\textwidth}
	\centering
	\tikzsetnextfilename{Brun2016_Lamb_error_internal_energy}
	\input{Brun2016_Lamb_error_internal_energy}
	\caption{%Time history of the error on internal energy
	Error on internal energy
	}\label{fig:Brun2016_Lamb_error_internal_energy}
\end{subfigure}\\
\begin{subfigure}[t]{0.45\textwidth}
	\centering
	\tikzsetnextfilename{Brun2016_Lamb_total_energy}
	\input{Brun2016_Lamb_total_energy}
	\caption{%Time history of total energy
	Total energy
	}\label{fig:Brun2016_Lamb_total_energy}
\end{subfigure}\hfill
\begin{subfigure}[t]{0.45\textwidth}
	\centering
	\tikzsetnextfilename{Brun2016_Lamb_error_total_energy}
	\input{Brun2016_Lamb_error_total_energy}
	\caption{%Time history of the error on total energy
	Error on total energy
	}\label{fig:Brun2016_Lamb_error_total_energy}
\end{subfigure}
\caption{Lamb's problem: evolutions of (\subref{fig:Brun2016_Lamb_kinetic_energy}) the kinetic energy $E_k(t)$, (\subref{fig:Brun2016_Lamb_error_kinetic_energy}) the absolute error $\abs{E_k(t) - E_k^{\text{ref}}(t)}$ on the kinetic energy, (\subref{fig:Brun2016_Lamb_internal_energy}) the internal energy $E_i(t)$, (\subref{fig:Brun2016_Lamb_error_internal_energy}) the absolute error $\abs{E_i(t) - E_i^{\text{ref}}(t)}$ on the internal energy, (\subref{fig:Brun2016_Lamb_total_energy}) the total energy $E(t)$ and (\subref{fig:Brun2016_Lamb_error_total_energy}) the absolute error $\abs{E(t) - E^{\text{ref}}(t)}$ on the total energy stored in the physical domain $\Omega_{\text{BD}}$ with respect to time $t$ for the PML and CAL solutions compared to the reference solution}\label{fig:Brun2016_Lamb_energies}
\end{figure}

\subsection{Directional force point-source buried within a two-dimensional semi-infinite isotropic homogeneous elastic medium}\label{sec:Matzen2011_exp1}
 
We again study the propagation of elastic waves in an isotropic homogeneous % elastic 
half-space with free-surface boundary conditions but the directional force point-source is now buried within the semi-infinite (unbounded) domain at a given depth.

We perform the numerical simulation of elastic waves propagating in the half-plane $\intervaloo{-\infty}{+\infty} \times \intervaloc{-\infty}{\ell} \subset \Rbb^2$ (with a top free surface $\Gamma_{\text{free}}$ located at $x_2=\ell=4$) submitted to a vertical force point-source buried within the semi-infinite domain at a depth $\ell$, \ie{} at the position $\xb^S = (0, 0)$ of the frame origin (see Figure~\ref{fig:Matzen2011_exp1_domain}). We keep the same isotropic homogeneous elastic material properties as the ones considered in the previous numerical example. A similar test case has already been investigated in \cite{Mat11} with different geometric features and material properties. As for the previous test case presented in Section~\ref{sec:Brun2016_Lamb}, the exact analytical solution brings bulk $P$- and $S$-waves into play as well as Rayleigh waves induced by the vertical force point-source. 

We consider a squared (two-dimensional) physical domain $\Omega_{\text{BD}} = \intervalcc{-\ell}{\ell} \times \intervalcc{-\ell}{\ell}$ with side length $2\ell = 8$, bounded by a PML region $\Omega_{\text{PML}} = \setst{\xb \in \Rbb^2}{\ell< \abs{x_1}\leq \ell+h \text{ and } -\ell-h\leq x_2< -\ell}$ gathering three overlapping absorbing layers (on the right, left and bottom sides of the physical domain of interest $\Omega_{\text{BD}}$) with the same finite thickness $h=2$ as in the first numerical experiment (see Figure~\ref{fig:Matzen2011_exp1_domain}). The entire computational domain $\Omega = \Omega_{\text{BD}}\cup \Omega_{\text{PML}} = \intervalcc{-\ell-h}{\ell+h} \times \intervalcc{-\ell-h}{\ell}$ is then a rectangular domain with length $2(\ell+h) = 12$ and width $2\ell+h = 10$ over which the numerical simulations are run. Recall that the point-source $S$ is located at the center of the physical domain $\Omega_{\text{BD}}$. Therefore, bulk and Rayleigh waves are not expected to be formed instantly as in the first numerical example, as already mentioned in \cite{Mat11}. We also expect the reflection and conversion of waves at the top free surface. For the same reasons as in the first numerical experiment shown in Section~\ref{sec:Brun2016_Lamb}, we apply homogeneous Dirichlet boundary conditions $\ub = \zerob$ on the exterior boundary $\Gamma_{\text{ext}}$ of $\Omega_{\text{PML}}$, that is at $\abs{x_1}=\ell+h=6$ and at $x_2=-\ell-h=-6$. We also impose stress-free boundary conditions $\scalprod{\sigmab}{\nb} = \zerob$ at the top surface $\Gamma_{\text{free}}$ of the entire domain $\Omega = \Omega_{\text{BD}} \cup \Omega_{\text{PML}}$, that is at $x_2=\ell$. We use again a quadratic damping function $d_i(x_i)$ in the PML region $\Omega_{\text{PML}}$ but with a higher amplitude $A_i = 12$ corresponding to a very small theoretical reflection coefficient $R_i = 10^{-8}$ according to \eqref{dampingcoefficient}.

\setlength\figureheight{0.15\textheight}
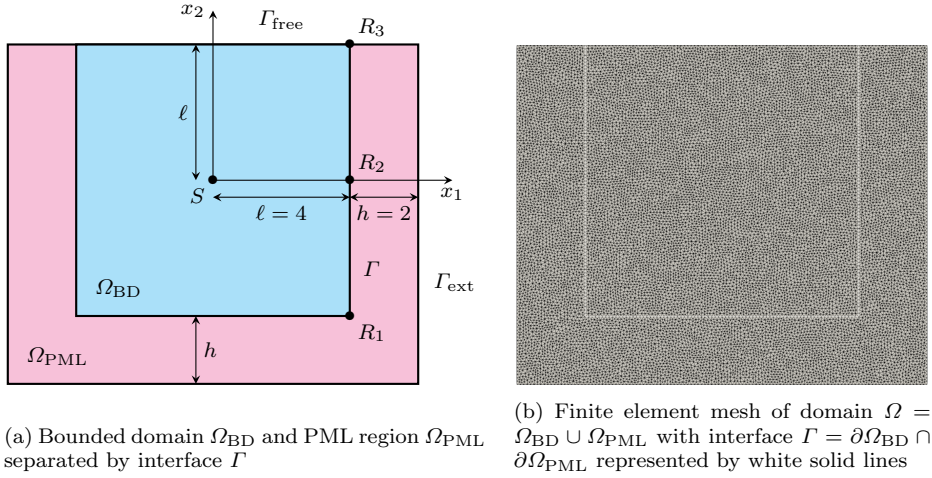
\begin{figure}
\centering
\begin{subfigure}[t]{0.52\textwidth}
	\centering
	\tikzsetnextfilename{Matzen2011_exp1_domain}
	\begin{tikzpicture}[scale=0.45]

\def \lx {4.0}
\def \ly {4.0}
\def \hx {2.0}
\def \hy {2.0}
%\draw[help lines,step=.5] (-\lx-\hx-1,-\ly-\hy-1) grid (\lx+\hx+1,\ly+1);

% Origin-Source
\coordinate (O) at (0,0);

% Receiver
\coordinate (R3) at (\lx,\ly);
\coordinate (R2) at (\lx,0);
\coordinate (R1) at (\lx,-\ly);

% Domain coordinates
\coordinate (A) at (-\lx,-\ly);
\coordinate (B) at (\lx,\ly);
\coordinate (C) at (-\lx,\ly);
\coordinate (D) at (\lx,-\ly);

% PML coordinates
\coordinate (E) at (-\lx-\hx,-\ly-\hy);
\coordinate (F) at (\lx+\hx,\ly);
\coordinate (G) at (-\lx-\hx,\ly);
\coordinate (H) at (\lx+\hx,-\ly-\hy);
\coordinate (Interface) at (\lx,-\ly/2);
\coordinate (Boundary) at (\lx+\hx,-\ly/2-\hy/2);
\coordinate (FreeBoundary) at (\lx/2,\ly);

% PML
\draw[black,thick,fill=magenta!30] (E) rectangle (F);
\node[above right=0.2] at (E) {$\Omega_{\text{PML}}$};
\draw[solid,black,stealth-stealth] (\lx,-0.5) -- (\lx+\hx,-0.5) node[midway,below]{$h = 2$};
\draw[solid,black,stealth-stealth] (-0.5,-\ly) -- (-0.5,-\ly-\hy) node[midway,right]{$h$};

% Domain
\draw[black,thick,fill=cyan!30] (A) rectangle (B);
\draw (O) node[below left] {$S$} node{$\bullet$};
\draw (R1) node[below right] {$R_1$} node{$\bullet$};
\draw (R2) node[above right] {$R_2$} node{$\bullet$};
\draw (R3) node[above right] {$R_3$} node{$\bullet$};
\node[above right=0.2] at (A) {$\Omega_{\text{BD}}$};
\draw[solid,black,stealth-stealth] (0,-0.5) -- (\lx,-0.5) node[midway,below]{$\ell = 4$};
\draw[solid,black,stealth-stealth] (-0.5,0) -- (-0.5,\ly) node[midway,left]{$\ell$};

% Interface
\node[below right=0.1] at (Interface) {$\Gamma$};
\node[right=0.1] at (Boundary) {$\Gamma_{\text{ext}}$};
\node[above=0.1] at (FreeBoundary) {$\Gamma_{\text{free}}$};

% Axis
\draw[solid,black,-stealth] (0,0) -- (\lx+\hx+1,0) node[below]{$x_1$};
\draw[solid,black,-stealth] (0,0) -- (0,\ly+1) node[left]{$x_2$};

\end{tikzpicture}
	\caption{Bounded domain $\Omega_{\text{BD}}$ and PML region $\Omega_{\text{PML}}$ separated by interface $\Gamma$}\label{fig:Matzen2011_exp1_domain}
\end{subfigure}\hfill
\begin{subfigure}[t]{0.45\textwidth}
	\centering
	\includegraphics[height=1.52\figureheight]{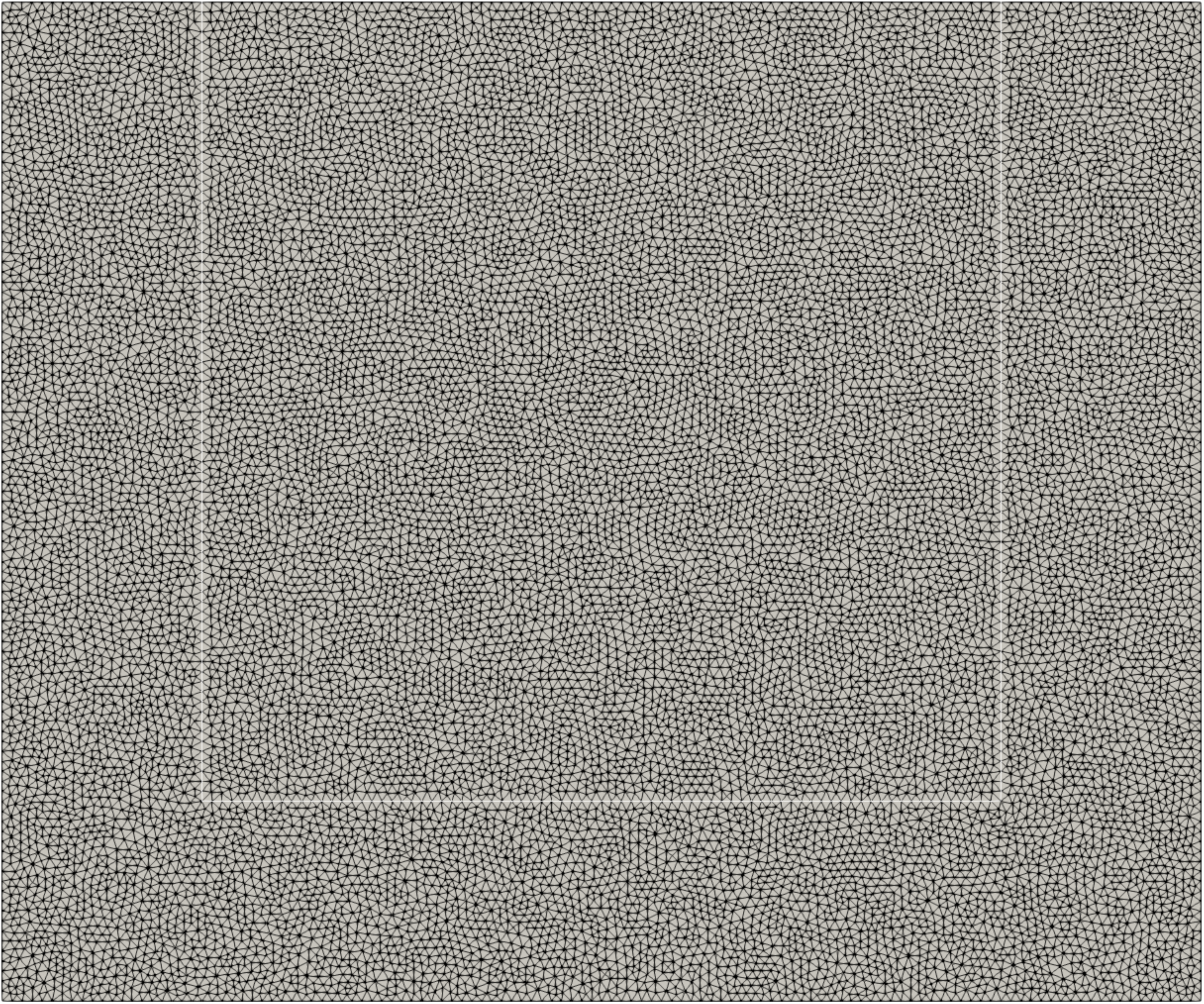}
	\caption{Finite element mesh of domain $\Omega = \Omega_{\text{BD}} \cup \Omega_{\text{PML}}$ with interface $\Gamma = \partial\Omega_{\text{BD}} \cap \partial\Omega_{\text{PML}}$ represented by white solid lines}\label{fig:Matzen2011_exp1_mesh}
\end{subfigure}
\caption{Buried source problem: (\subref{fig:Matzen2011_exp1_domain}) two-dimensional semi-infinite domain (half-space) modeled by a physical bounded domain $\Omega_{\text{BD}}$ surrounded by an artificial PML region $\Omega_{\text{PML}}$, subjected to a downward vertical force point-source $S$ buried in $\Omega_{\text{BD}}$, with three receivers $R_1$, $R_2$ and $R_3$ located along the interface $\Gamma$ between $\Omega_{\text{BD}}$ and $\Omega_{\text{PML}}$, and (\subref{fig:Matzen2011_exp1_mesh}) associated finite element mesh}\label{fig:Matzen2011_exp1}
\end{figure}

At the spatial level, the entire computational domain $\Omega$ is discretized using an unstructured finite element mesh made of $6$-nodes quadratic triangular elements with the same mesh density $\Delta x = 0.1$ as in the first test case (see Figure~\ref{fig:Matzen2011_exp1_mesh}). It then consists of $63\,525$ nodes (\ie{} $127\,050$ degrees of freedom) and $31\,862$ elements. The extended rectangular computational domain $\Omega_{\text{ref}} = \intervalcc{-L}{L} \times \intervalcc{-L}{\ell}$ (with $L = 6\ell = 24$) used to compute a reference solution $\ub^{\text{ref}}$ is six times larger than the physical domain of interest $\Omega_{\text{BD}}$ in order to mimic the exact solution inside the physical domain $\Omega_{\text{BD}}$ over the specified time interval $I=\intervalcc{0}{T}$. The reference finite element mesh is composed of $711\,445$ nodes (\ie{} $1\,422\,890$ degrees of freedom) and $354\,962$ quadratic triangular elements.

The time histories of the horizontal component $u_1$ and vertical component $u_2$ of the displacement vector $\ub$ are recorded over the time interval $I=\intervalcc{0}{20}$~s at three receivers $R_1$, $R_2$ and $R_3$ along the interface $\Gamma$, respectively located at the bottom right corner position $\xb^{R_1} = (4, -4)$, at the middle right position $\xb^{R_2} = (4, -2)$ and at the top right corner position $\xb^{R_3} = (4, 0)$ of the physical domain $\Omega_{\text{BD}}$ (see Figure~\ref{fig:Matzen2011_exp1_domain}). The monitored time evolutions of the two components of the displacement vector for both PML and CAL solutions and their respective absolute errors with respect to the reference solution are displayed in Figures~\ref{fig:Matzen2011_exp1_displacement_Ux_R123} and \ref{fig:Matzen2011_exp1_displacement_Uy_R123}. The PML solution closely matches the reference solution for each of the three receivers $R_1$, $R_2$ and $R_3$ placed at the interface $\Gamma$ between the physical domain $\Omega_{\text{BD}}$ and the PML region $\Omega_{\text{PML}}$, while the CAL solution is highly distorted and exhibits large spurious oscillations from approximately time $t=6$~s for receivers $R_1$ and $R_3$ and time $t=8$~s for receiver $R_2$ due to significant spurious reflections at the interface $\Gamma$.

\begin{figure}
\centering
\begin{subfigure}[t]{0.45\textwidth}
	\centering
	\tikzsetnextfilename{Matzen2011_exp1_displacement_Ux_R1}
	\input{Matzen2011_exp1_displacement_Ux_R1}
	\caption{%Time history of horizontal displacement 
	Horizontal displacement
	at receiver $R_1$
	}\label{fig:Matzen2011_exp1_displacement_Ux_R1}
\end{subfigure}\hfill
\begin{subfigure}[t]{0.45\textwidth}
	\centering
	\tikzsetnextfilename{Matzen2011_exp1_error_displacement_Ux_R1}
	\input{Matzen2011_exp1_error_displacement_Ux_R1}
	\caption{%Time history of the error on horizontal displacement 
	Error on horizontal displacement
	at receiver $R_1$
	}\label{fig:Matzen2011_exp1_error_displacement_Ux_R1}
\end{subfigure}\\
\begin{subfigure}[t]{0.45\textwidth}
	\centering
	\tikzsetnextfilename{Matzen2011_exp1_displacement_Ux_R2}
	\input{Matzen2011_exp1_displacement_Ux_R2}
	\caption{%Time history of horizontal displacement 
	Horizontal displacement
	at receiver $R_2$
	}\label{fig:Matzen2011_exp1_displacement_Ux_R2}
\end{subfigure}\hfill
\begin{subfigure}[t]{0.45\textwidth}
	\centering
	\tikzsetnextfilename{Matzen2011_exp1_error_displacement_Ux_R2}
	\input{Matzen2011_exp1_error_displacement_Ux_R2}
	\caption{%Time history of the error on horizontal displacement 
	Error on horizontal displacement
	at receiver $R_2$
	}\label{fig:Matzen2011_exp1_error_displacement_Ux_R2}
\end{subfigure}\\
\begin{subfigure}[t]{0.45\textwidth}
	\centering
	\tikzsetnextfilename{Matzen2011_exp1_displacement_Ux_R3}
	\input{Matzen2011_exp1_displacement_Ux_R3}
	\caption{%Time history of horizontal displacement 
	Horizontal displacement
	at receiver $R_3$
	}\label{fig:Matzen2011_exp1_displacement_Ux_R3}
\end{subfigure}\hfill
\begin{subfigure}[t]{0.45\textwidth}
	\centering
	\tikzsetnextfilename{Matzen2011_exp1_error_displacement_Ux_R3}
	\input{Matzen2011_exp1_error_displacement_Ux_R3}
	\caption{%Time history of the error on horizontal displacement 
	Error on horizontal displacement
	at receiver $R_3$
	}\label{fig:Matzen2011_exp1_error_displacement_Ux_R3}
\end{subfigure}
\caption{Buried source problem: evolutions of (\subref{fig:Matzen2011_exp1_displacement_Uy_R1}-\subref{fig:Matzen2011_exp1_displacement_Uy_R2}-\subref{fig:Matzen2011_exp1_displacement_Uy_R3}) the horizontal displacement $u_1$ and (\subref{fig:Matzen2011_exp1_error_displacement_Uy_R1}-\subref{fig:Matzen2011_exp1_error_displacement_Uy_R2}-\subref{fig:Matzen2011_exp1_error_displacement_Uy_R3}) the corresponding absolute error $\abs{u_1 - u_1^{\text{ref}}}$ at three receivers $R_1$ (first row), $R_2$ (second row) and $R_3$ (third row) with respect to time $t$ for the PML and CAL solutions compared to the reference solution}\label{fig:Matzen2011_exp1_displacement_Ux_R123}
\end{figure}

\begin{figure}
\centering
\begin{subfigure}[t]{0.45\textwidth}
	\centering
	\tikzsetnextfilename{Matzen2011_exp1_displacement_Uy_R1}
	\input{Matzen2011_exp1_displacement_Uy_R1}
	\caption{%Time history of vertical displacement 
	Vertical displacement
	at receiver $R_1$
	}\label{fig:Matzen2011_exp1_displacement_Uy_R1}
\end{subfigure}\hfill
\begin{subfigure}[t]{0.45\textwidth}
	\centering
	\tikzsetnextfilename{Matzen2011_exp1_error_displacement_Uy_R1}
	\input{Matzen2011_exp1_error_displacement_Uy_R1}
	\caption{%Time history of the error on vertical displacement 
	Error on vertical displacement
	at receiver $R_1$
	}\label{fig:Matzen2011_exp1_error_displacement_Uy_R1}
\end{subfigure}\\
\begin{subfigure}[t]{0.45\textwidth}
	\centering
	\tikzsetnextfilename{Matzen2011_exp1_displacement_Uy_R2}
	\input{Matzen2011_exp1_displacement_Uy_R2}
	\caption{%Time history of vertical displacement 
	Vertical displacement
	at receiver $R_2$
	}\label{fig:Matzen2011_exp1_displacement_Uy_R2}
\end{subfigure}\hfill
\begin{subfigure}[t]{0.45\textwidth}
	\centering
	\tikzsetnextfilename{Matzen2011_exp1_error_displacement_Uy_R2}
	\input{Matzen2011_exp1_error_displacement_Uy_R2}
	\caption{%Time history of the error on vertical displacement 
	Error on vertical displacement
	at receiver $R_2$
	}\label{fig:Matzen2011_exp1_error_displacement_Uy_R2}
\end{subfigure}\\
\begin{subfigure}[t]{0.45\textwidth}
	\centering
	\tikzsetnextfilename{Matzen2011_exp1_displacement_Uy_R3}
	\input{Matzen2011_exp1_displacement_Uy_R3}
	\caption{%Time history of vertical displacement 
	Vertical displacement
	at receiver $R_3$
	}\label{fig:Matzen2011_exp1_displacement_Uy_R3}
\end{subfigure}\hfill
\begin{subfigure}[t]{0.45\textwidth}
	\centering
	\tikzsetnextfilename{Matzen2011_exp1_error_displacement_Uy_R3}
	\input{Matzen2011_exp1_error_displacement_Uy_R3}
	\caption{%Time history of the error on vertical displacement 
	Error on vertical displacement
	at receiver $R_3$
	}\label{fig:Matzen2011_exp1_error_displacement_Uy_R3}
\end{subfigure}
\caption{Buried source problem: evolutions of (\subref{fig:Matzen2011_exp1_displacement_Ux_R1}-\subref{fig:Matzen2011_exp1_displacement_Ux_R2}-\subref{fig:Matzen2011_exp1_displacement_Ux_R3}) the vertical displacement $u_2$ and (\subref{fig:Matzen2011_exp1_error_displacement_Ux_R1}-\subref{fig:Matzen2011_exp1_error_displacement_Ux_R2}-\subref{fig:Matzen2011_exp1_error_displacement_Ux_R3}) the corresponding absolute error $\abs{u_2 - u_2^{\text{ref}}}$ at three receivers $R_1$ (first row), $R_2$ (second row) and $R_3$ (third row) with respect to time $t$ for the PML and CAL solutions compared to the reference solution}\label{fig:Matzen2011_exp1_displacement_Uy_R123}
\end{figure}

\begin{figure}
	\centering
	\includegraphics[width=\textwidth]{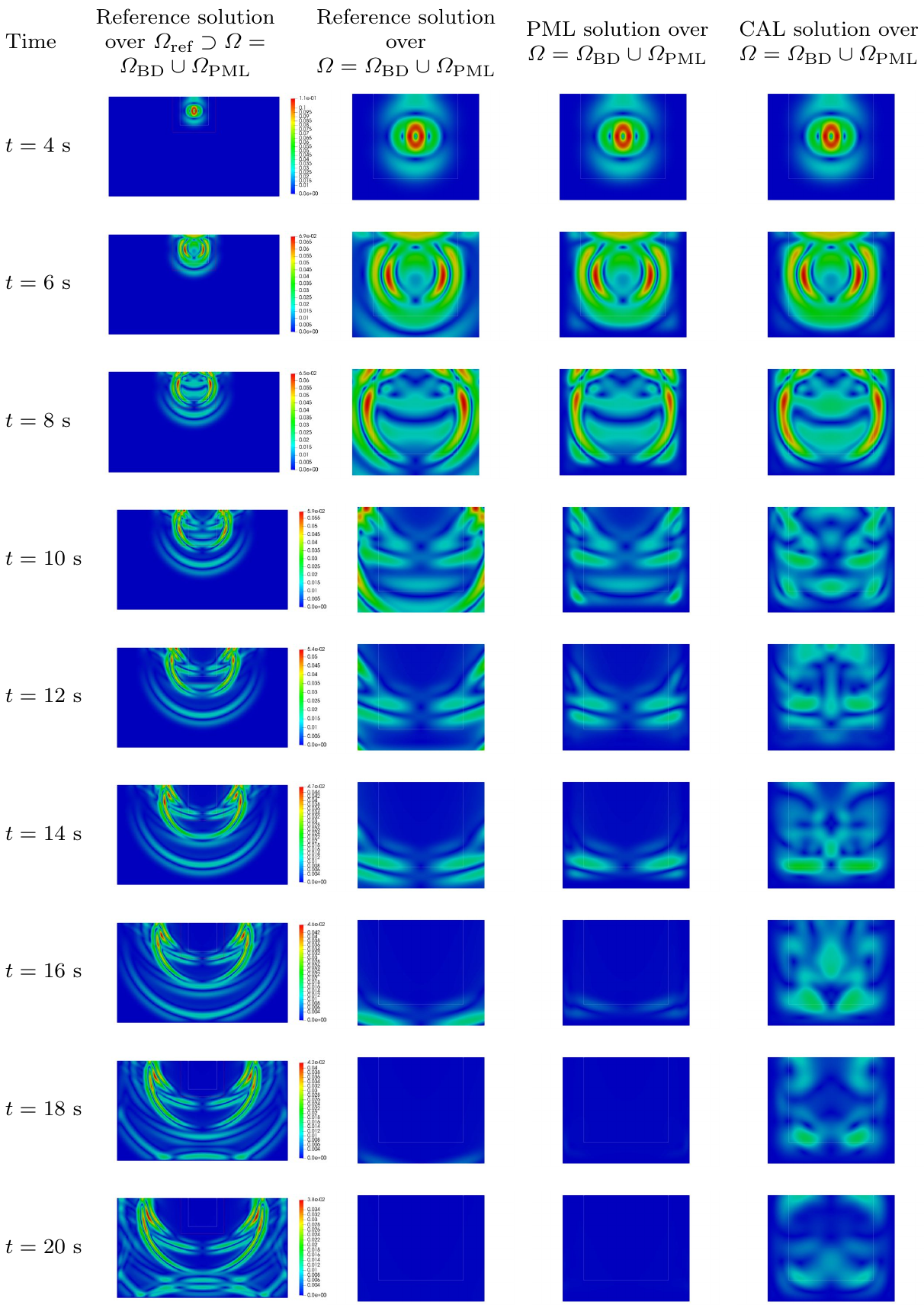}
	\caption{Buried source problem: snapshots of the displacement field magnitude $\xb \mapsto \norm{\ub(\xb,t)}$ taken at different times $t=4,6,8,10,12,14,16,18,20$~s (from top to bottom) for the reference solution over an extended computational domain $\Omega_{\text{ref}} \supset \Omega = \Omega_{\text{BD}} \cup \Omega_{\text{PML}}$ (first column) with interface $\Gamma = \partial\Omega_{\text{BD}} \cap \partial\Omega_{\text{PML}}$ indicated by white solid lines and the exterior boundary $\Gamma_{\text{ext}}$ of $\Omega$ indicated by red solid lines, the reference solution over the computational domain $\Omega = \Omega_{\text{BD}} \cup \Omega_{\text{PML}}$ (second column), the PML solution over $\Omega$ (third column) and the CAL solution over $\Omega$ (fourth column)}\label{fig:Matzen2011_exp1_displacement}
\end{figure}

\begin{figure}
	\centering
	\includegraphics[width=\textwidth]{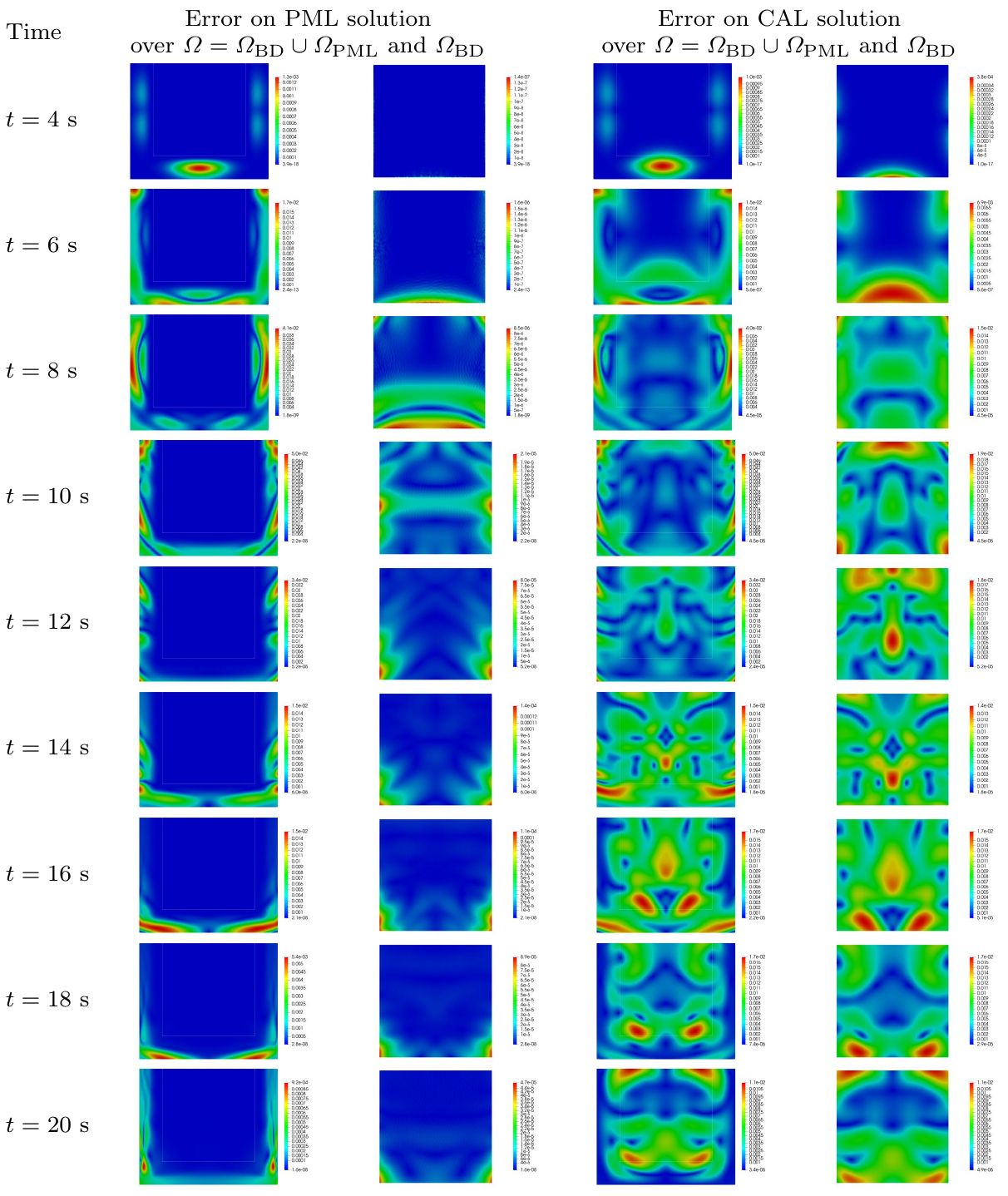}
	\caption{Buried source problem: snapshots of the error on the displacement field magnitude $\xb \mapsto \abs{\norm{\ub(\xb,t)} - \norm{\ub^{\text{ref}}(\xb,t)}}$ taken at different times $t=4,6,8,10,12,14,16,18,20$~s (from top to bottom) for the PML solution over the entire computational domain $\Omega = \Omega_{\text{BD}} \cup \Omega_{\text{PML}}$ (first column) and its restriction to the physical bounded domain $\Omega_{\text{BD}}$ (second column) and the CAL solution over $\Omega$ and $\Omega_{\text{BD}}$ (third and fourth columns)}\label{fig:Matzen2011_exp1_error_displacement}
\end{figure}

Snapshots of the displacement field magnitude $\xb \mapsto \norm{\ub(\xb,t)}$ taken at different times from time $t=4$~s to final time $t=T=20$~s are shown in Figure~\ref{fig:Matzen2011_exp1_displacement} for the three performed simulations, namely for the reference solution computed over the extended computational domain $\Omega_{\text{ref}}$ and restricted to the computational domain $\Omega = \Omega_{\text{BD}} \cup \Omega_{\text{PML}}$ as well as for the PML and CAL solutions over $\Omega$. It can be observed that two bulk waves have been formed by the driving buried source, a primary faster longitudinal $P$-wave followed by a second larger shear $S$-wave. Both bulk waves propagate towards the boundaries of the physical domain $\Omega_{\text{BD}}$ and enter into the PML region $\Omega_{\text{PML}}$ at normal incidence. When reaching and impinging upon the top free surface $\Gamma_{\text{free}}$, the direct incident bulk waves give rise to reflected bulk and Rayleigh waves that interfere together. The large Rayleigh waves can be easily distinguished based upon their elliptical polarizations and large amplitudes in comparison with the bulk waves characterized by circular polarizations and smaller amplitudes. The large-amplitude Rayleigh waves can be observed on the transient dynamical responses of the vertical displacement $u_2$ recorded at receivers $R_2$ and $R_3$ around time $t=6$~s and $t=8$~s, respectively (see Figures~\ref{fig:Matzen2011_exp1_displacement_Uy_R2} and \ref{fig:Matzen2011_exp1_displacement_Uy_R3}). We observe that no reflected waves are visible for the PML solution (neither any spurious reflections from the interface $\Gamma$, nor any residual reflections from the fixed exterior boundary $\Gamma_{\text{ext}}$), whereas multiple reflections coming from the interface $\Gamma$ arise for the CAL solution. For further analysis, Figure~\ref{fig:Matzen2011_exp1_error_displacement} shows some snapshots of the absolute errors on the displacement field magnitude $\xb \mapsto \abs{\norm{\ub(\xb,t)} - \norm{\ub^{\text{ref}}(\xb,t)}}$ obtained by comparing the PML and CAL solutions to the reference one. It confirms that the PML method is clearly superior to the CAL method for the absorption of both bulk and Rayleigh waves in this numerical experiment.

Figure~\ref{fig:Matzen2011_exp1_energies} displays the time evolutions of the kinetic energy $E_k(t)$, internal energy $E_i(t)$ and total energy $E(t)$ stored in the physical
domain $\Omega_{\text{BD}}$, as well as that of the corresponding absolute errors $\abs{E_k(t) - E_k^{\text{ref}}(t)}$, $\abs{E_i(t) - E_i^{\text{ref}}(t)}$ and $\abs{E(t) -
 E^{\text{ref}}(t)}$. We can see that some energy is injected by the point-source into the physical domain $\Omega_{\text{BD}}$ between initial time $t=0$~s and approximately time $t=4$~s. Then, the total energy transported by bulk ($P$- and $S$-waves) and Rayleigh waves decreases gradually as the outgoing waves leave the physical domain $\Omega_{\text{BD}}$ and are progressively absorbed inside the PML region $\Omega_{\text{PML}}$ starting from around time $t=4.5$~s for the first $P$-wave to time $t=16$~s for the last reflected bulk wave (see Figure~\ref{fig:Matzen2011_exp1_displacement}). The discrepancies between PML and CAL solutions in terms of energy decay are caused by the spurious waves that send energy back into the physical domain $\Omega_{\text{BD}}$. The numerical results highlight the poor performances of the CAL solution compared to the PML solution that is in almost perfect agreement with the reference solution.

\begin{figure}
\centering
\begin{subfigure}[t]{0.45\textwidth}
	\centering
	\tikzsetnextfilename{Matzen2011_exp1_kinetic_energy}
	\input{Matzen2011_exp1_kinetic_energy}
	\caption{%Time history of kinetic energy
	Kinetic energy
	}\label{fig:Matzen2011_exp1_kinetic_energy}
\end{subfigure}\hfill
\begin{subfigure}[t]{0.45\textwidth}
	\centering
	\tikzsetnextfilename{Matzen2011_exp1_error_kinetic_energy}
	\input{Matzen2011_exp1_error_kinetic_energy}
	\caption{%Time history of the error on kinetic energy
	Error on kinetic energy
	}\label{fig:Matzen2011_exp1_error_kinetic_energy}
\end{subfigure}\\
\begin{subfigure}[t]{0.45\textwidth}
	\centering
	\tikzsetnextfilename{Matzen2011_exp1_internal_energy}
	\input{Matzen2011_exp1_internal_energy}
	\caption{%Time history of internal energy
	Internal energy
	}\label{fig:Matzen2011_exp1_internal_energy}
\end{subfigure}\hfill
\begin{subfigure}[t]{0.45\textwidth}
	\centering
	\tikzsetnextfilename{Matzen2011_exp1_error_internal_energy}
	\input{Matzen2011_exp1_error_internal_energy}
	\caption{%Time history of the error on internal energy
	Error on internal energy
	}\label{fig:Matzen2011_exp1_error_internal_energy}
\end{subfigure}\\
\begin{subfigure}[t]{0.45\textwidth}
	\centering
	\tikzsetnextfilename{Matzen2011_exp1_total_energy}
	\input{Matzen2011_exp1_total_energy}
	\caption{%Time history of total energy
	Total energy
	}\label{fig:Matzen2011_exp1_total_energy}
\end{subfigure}\hfill
\begin{subfigure}[t]{0.45\textwidth}
	\centering
	\tikzsetnextfilename{Matzen2011_exp1_error_total_energy}
	\input{Matzen2011_exp1_error_total_energy}
	\caption{%Time history of the error on total energy
	Error on total energy
	}\label{fig:Matzen2011_exp1_error_total_energy}
\end{subfigure}
\caption{Buried source problem: evolutions of (\subref{fig:Matzen2011_exp1_kinetic_energy}) the kinetic energy $E_k(t)$, (\subref{fig:Matzen2011_exp1_error_kinetic_energy}) the absolute error $\abs{E_k(t) - E_k^{\text{ref}}(t)}$ on the kinetic energy, (\subref{fig:Matzen2011_exp1_internal_energy}) the internal energy $E_i(t)$, (\subref{fig:Matzen2011_exp1_error_internal_energy}) the absolute error $\abs{E_i(t) - E_i^{\text{ref}}(t)}$ on the internal energy, (\subref{fig:Matzen2011_exp1_total_energy}) the total energy $E(t)$ and (\subref{fig:Matzen2011_exp1_error_total_energy}) the absolute error $\abs{E(t) - E^{\text{ref}}(t)}$ on the total energy stored in the physical domain $\Omega_{\text{BD}}$ with respect to time $t$ for the PML and CAL solutions compared to the reference solution}\label{fig:Matzen2011_exp1_energies}
\end{figure}

\subsection{Directional force point-source applied onto the free surface of a two-dimensional semi-infinite multi-layer isotropic homogeneous elastic medium}\label{sec:Brun2016_multilayer}

We finally investigate the propagation of elastic waves in a two-layer isotropic homogeneous %elastic 
half-space excited by a directional force point-source placed at the top free surface in order to assess the accuracy and efficiency of the PML method in the presence of an interface separating two homogeneous media with different elastic material properties.

\begin{figure}
\centering
\begin{subfigure}[t]{\textwidth}
	\centering
	\tikzsetnextfilename{Brun2016_multilayer_domain}
	\begin{tikzpicture}[scale=0.45]

\def \lx {4.0}
\def \ly {4.0}
\def \hx {6.0}
\def \hy {6.0}
%\draw[help lines,step=.5] (-\lx-\hx-1,-\ly-\hy-1) grid (\lx+\hx+1,2);

% Origin-Source
\coordinate (O) at (0,0);

% Receiver
\coordinate (R) at (\lx/2,0);

% Domain coordinates
\coordinate (A) at (-\lx,-\ly);
\coordinate (B) at (\lx,0);
\coordinate (C) at (-\lx,0);
\coordinate (D) at (\lx,-\ly);

% PML coordinates
\coordinate (E) at (-\lx-\hx,-\ly-\hy);
\coordinate (F) at (\lx+\hx,0);
\coordinate (G) at (-\lx-\hx,0);
\coordinate (H) at (\lx+\hx,-\ly-\hy);
\coordinate (InterfaceR) at (\lx,-\ly/2);
\coordinate (BoundaryR) at (\lx+\hx,-\ly/2);
\coordinate (InterfaceL) at (-\lx,-\ly/2);
\coordinate (BoundaryL) at (-\lx-\hx,-\ly/2);
\coordinate (Boundary) at (\lx+\hx,-\ly/2-\hy/2);
\coordinate (FreeBoundary) at (\lx+\hx/2,-0);

% PML
\draw[black,thick,fill=magenta!30] (E) rectangle (F);
\node[above right=0.2] at (E) {\footnotesize $\Omega_{\text{PML}}$};
\draw[solid,black,stealth-stealth] (\lx,-0.5) -- (\lx+\hx,-0.5) node[midway,below]{\footnotesize $h = 6$};
\draw[solid,black,stealth-stealth] (-0.5,-\ly) -- (-0.5,-\ly-\hy) node[midway,left]{\footnotesize $h$};

% Domain
\draw[black,thick,fill=cyan!30] (A) rectangle (B);
\draw (O) node[above right] {\footnotesize $S$} node{\footnotesize $\bullet$};
\draw (R) node[above right] {\footnotesize $R$} node{\footnotesize $\bullet$};
%\node[above right=0.2] at (A) {\footnotesize $\Omega_{\text{BD}}$};
\node[above left=0.2] at (D) {\footnotesize $\Omega_{\text{BD}}$};
\draw[solid,black,stealth-stealth] (0,-0.5) -- (\lx,-0.5) node[midway,below]{\footnotesize $\ell = 4$};
\draw[solid,black,stealth-stealth] (-0.5,0) -- (-0.5,-\ly/2) node[midway,left]{\footnotesize $\ell/2$};
\draw[solid,black,stealth-stealth] (-0.5,-\ly/2) -- (-0.5,-\ly) node[midway,left]{\footnotesize $\ell/2$};
%\node[above left=0.1] at (-0.5,-2) {\footnotesize $\ell$};

% Interface
%\node[below right=0.1] at (InterfaceR) {\footnotesize $\Gamma$};
%\node[below right=0.1] at (BoundaryR) {\footnotesize $\Gamma_{\text{ext}}$};
\node[above right=0.1] at (D) {\footnotesize $\Gamma$};
\node[above right=0.1] at (Boundary) {\footnotesize $\Gamma_{\text{ext}}$};
\node[above=0.1] at (FreeBoundary) {\footnotesize $\Gamma_{\text{free}}$};

% Axis
\draw[solid,black,-stealth] (0,0) -- (\lx+\hx+1,0) node[below]{\footnotesize $x_1$};
\draw[solid,black,-stealth] (0,0) -- (0,1) node[left]{\footnotesize $x_2$};

% Layer
\draw[dashed,black] (BoundaryL) -- (InterfaceL);% node[near start,above right]{\footnotesize Layer 1 $(\rho_1,E_1,\nu_1)$} node[near start,below right]{\footnotesize Layer 2 ($\rho_2,E_2,\nu_2$)};
\draw[dashed,black] (InterfaceL) -- (InterfaceR);
\draw[dashed,black] (InterfaceR) -- (BoundaryR);
\node[above right=0.1] at (BoundaryL) {\footnotesize Layer 1 $(\rho_1,E_1,\nu_1)$};
\node[below right=0.1] at (BoundaryL) {\footnotesize Layer 2 $(\rho_2,E_2,\nu_2)$};
\end{tikzpicture}
	\caption{Bounded domain $\Omega_{\text{BD}}$ and PML region $\Omega_{\text{PML}}$ separated by interface $\Gamma$ represented by solid lines and the horizontal material interface separating the two elastic layers represented by a dashed line}\label{fig:Brun2016_multilayer_domain}
\end{subfigure}\\
\begin{subfigure}[t]{\textwidth}
	\centering
	\includegraphics[height=1.5\figureheight]{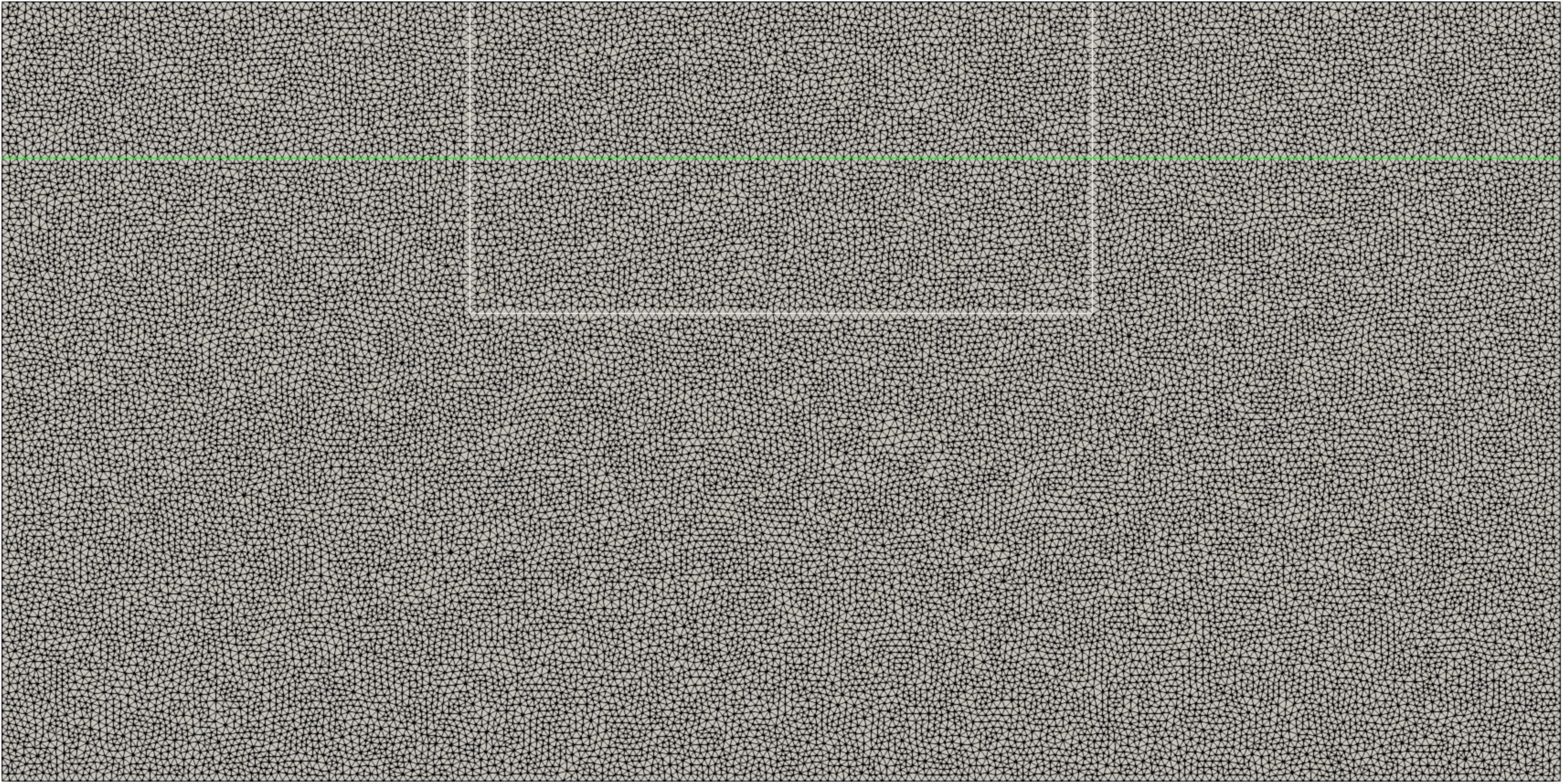}
	\caption{Finite element mesh of domain $\Omega = \Omega_{\text{BD}} \cup \Omega_{\text{PML}}$ with interface $\Gamma = \partial\Omega_{\text{BD}} \cap \partial\Omega_{\text{PML}}$ represented by white solid lines and the horizontal material interface separating the two elastic layers represented by a green solid line}\label{fig:Brun2016_multilayer_mesh}
\end{subfigure}
\caption{Multi-layer problem: (\subref{fig:Brun2016_multilayer_domain}) two-dimensional (two-)layered semi-infinite domain (half-space) modeled by a physical bounded domain $\Omega_{\text{BD}}$ surrounded by an artificial PML region $\Omega_{\text{PML}}$, subjected to a downward vertical force point-source $S$ with a receiver $R$, both located at the top free surface $\Gamma_{\text{free}}$ of $\Omega_{\text{BD}}$, and (\subref{fig:Brun2016_multilayer_mesh}) associated finite element mesh}\label{fig:Brun2016_multilayer}
\end{figure}

The semi-infinite (unbounded) domain corresponding to the half-plane $\intervaloo{-\infty}{+\infty} \times \intervaloc{-\infty}{0} \subset \Rbb^2$ (with a top free surface $\Gamma_{\text{free}}$ located at $x_2=0$ as in the first test case) is composed of two elastic layers (with different mechanical properties) separated by an horizontal interface located at a depth of $2$ below the top free surface, \ie{} at $x_2 = -2$ (see Figure~\ref{fig:Brun2016_multilayer_domain}). The linear elastic material is assumed to be isotropic and homogeneous in each of the two horizontal layers. The top layer (denoted Layer 1 in Figure~\ref{fig:Brun2016_multilayer_domain}) has the same material properties as in the two previous test cases, namely a mass density $\rho_1=1$, a Young's modulus $E_1=2.5$ and a Poisson's ratio $\nu_1=0.25$ in a state of 2D plane strain, leading to the following Lam\'e's coefficients $\lambda_1 = \mu_1 = 1$, $P$- and $S$-wave velocities $c_{p,1}\approx 1.7321$ and $c_{s,1}=1$, and corresponding wavelengths $\lambda_{p,1} = c_{p,1}/f_d \approx 5.1962$ and $\lambda_{s,1} = c_{s,1}/f_d = 3$. The point-source $S$ is located at the center of the free surface $\Gamma_{\text{free}}$ of the top layer (Layer 1), \ie{} at the position $\xb^S = (0, 0)$ of the frame origin. A receiver $R$ is also placed at $\Gamma_{\text{free}}$ away from the source $S$, the distance between the source $S$ and the receiver $R$ being equal to $2$. The bottom layer (denoted Layer 2 in Figure~\ref{fig:Brun2016_multilayer_domain}) is characterized by the following material properties: mass density $\rho_2=\rho_1=1$, Young's modulus $E_2=5E_1=12.5$, Poisson's ratio $\nu_2=\nu_1=0.25$, Lam\'e's coefficients $\lambda_2 = 5\lambda_1 = 5$ and $\mu_2 = 5\mu_1 = 5$, $P$- and $S$-wave velocities $c_{p,2} = \sqrt{5}c_{p,1}\approx 3.8730$ and $c_{s,2}= \sqrt{5}c_{s,1} = \sqrt{5}\approx 2.2361$, and corresponding wavelengths $\lambda_{p,2} = \sqrt{5}\lambda_{p,1}\approx 11.619$ and $\lambda_{s,2}= \sqrt{5}\lambda_{s,1} = 3\sqrt{5}\approx 6.7082$. Thus, the Young's and shear moduli (resp. the wave velocities and associated wavelengths) are five times (resp. more than two times) higher in the bottom layer (Layer 2) than in the top layer (Layer 1). As a consequence, the two-layer elastic system is a dispersive medium and the propagative waves are expected to undergo multiple reflections and refractions at the material interface between the two elastic layers. As already pointed out in \cite{Brun16}, this wave propagation problem is similar to an open elastic waveguide system in which the elastic waves are guided and tightly confined within the top layer (Layer 1) with possible radiations in the bottom layer (Layer 2), due to the contrast in the material properties between the two elastic layers. In addition to bulk waves (such as compressional $P$-waves and shear $S$-waves propagating within the two elastic media) and Rayleigh waves (propagating along the free surface), Stoneley waves (propagating along the material interface between the two elastic media) can also occur.

We consider the same physical domain $\Omega_{\text{BD}} = \intervalcc{-\ell}{\ell} \times \intervalcc{-\ell}{0}$ (with $\ell = 4$) as in the first test case, bounded at the top by a free surface $\Gamma_{\text{free}}$ and on the other three remaining sides by a three-overlapping-layer PML region $\Omega_{\text{PML}} = \setst{\xb \in \Rbb^2}{\ell< \abs{x_1}\leq \ell+h \text{ and } -\ell-h\leq x_2< -\ell}$ with a thickness $h=3\ell/2=6$ three times higher than in the two previous test cases (see Figure~\ref{fig:Brun2016_multilayer_domain}). As a consequence, the PML thickness $h=6$ is slightly lower than a half ($1/2$) of the largest longitudinal $P$-wave wavelength $\lambda_{p,2} \approx 11.619$ and close to the shear $S$-wave wavelength $\lambda_{s,2} \approx 6.7082$ in the underneath layer (Layer 2) corresponding to a medium with higher velocity. Note that both the physical domain $\Omega_{\text{BD}}$ and the nearby PML region $\Omega_{\text{PML}}$ feature different material properties on either side of the horizontal interface between the two elastic layers in order to fulfill the perfect matching property at the interface $\Gamma$ between $\Omega_{\text{BD}}$ and $\Omega_{\text{PML}}$. The computations are performed over the entire rectangular computational domain $\Omega = \Omega_{\text{BD}}\cup \Omega_{\text{PML}} = \intervalcc{-\ell-h}{\ell+h} \times \intervalcc{-\ell-h}{0}$ with length $2(\ell+h) = 20$ and width $\ell+h = 10$. We apply the same boundary conditions as in the two previous test cases, namely clamped conditions $\ub=\zerob$ on the right, left and bottom exterior boundaries $\Gamma_{\text{ext}}$ of $\Omega$ (\ie{} at $\abs{x_1}=\ell+h=10$ and at $x_2=-\ell-h=-10$) and stress-free conditions $\scalprod{\sigmab}{\nb} = \zerob$ at the top surface $\Gamma_{\text{free}}$ (\ie{} at $x_2=0$). The damping functions $d_i(x_i)$ allowing the propagative waves to be absorbed in the PML region $\Omega_{\text{PML}}$ are identical to those used in the previous (second) numerical experiment with a quadratic polynomial degree $p=2$ and a rather high amplitude $A_i=12$.

We use again a standard (displacement-based) finite element method based on piecewise quadratic basis functions on a triangulation of the entire domain $\Omega$ with the same characteristic element size $\Delta x=0.1$ as in the two previous examples. The unstructured finite element mesh of the entire domain $\Omega = \Omega_{\text{BD}} \cup \Omega_{\text{PML}}$ is represented in Figure~\ref{fig:Brun2016_multilayer_mesh}. It contains $105\,881$ nodes (\ie{} $211\,762$ degrees of freedom) and $52\,640$ elements. A very large extended computational domain $\Omega_{\text{ref}} = \intervalcc{-L}{L} \times \intervalcc{-L}{0}$ with $L=40$ equal to $10$ times the size $\ell=4$ of the physical domain $\Omega_{\text{BD}}$ is used to provide a reference solution $\ub^{\text{ref}}$ in $\Omega_{\text{BD}}$ with no parasite (spurious) waves being reflected at the clamped external boundaries of $\Omega_{\text{ref}}$ (\ie{} at $\abs{x_1}=L=40$ and at $x_2=-L=-40$) and sent back into the physical domain $\Omega_{\text{BD}}$ over the considered time interval $I = \intervalcc{0}{T}$. The associated reference finite element mesh is made of $845\,762$ quadratic finite elements with a total of $1\,693\,925$ nodes (\ie{} $3\,387\,850$ degrees of freedom).

In Figure~\ref{fig:Brun2016_multilayer_displacement_R}, we plot the time evolutions of the horizontal and vertical components $u_1(\xb^R,t)$ and $u_2(\xb^R,t)$ of the displacement vector $\ub(\xb^R,t)$ at the position $\xb^R$ of the receiver $R$. Despite the numerous reflections and refractions arising from the material interface between the two elastic layers, the PML solution fits the reference solution very well at receiver $R$ during the whole time interval $\intervalcc{0}{20}$~s, whereas the CAL solution exhibits some discrepancies. According to the time evolutions of the corresponding errors shown in Figures~\ref{fig:Brun2016_multilayer_error_displacement_Ux_R} and \ref{fig:Brun2016_multilayer_error_displacement_Uy_R}, the PML solution presents higher match with the reference solution than the CAL solution at receiver $R$.

\begin{figure}
\centering
\begin{subfigure}[t]{0.45\textwidth}
	\centering
	\tikzsetnextfilename{Brun2016_multilayer_displacement_Ux}
	\input{Brun2016_multilayer_displacement_Ux_R}
	\caption{%Time history of horizontal displacement
	Horizontal displacement
	at receiver $R$
	}\label{fig:Brun2016_multilayer_displacement_Ux_R}
\end{subfigure}\hfill
\begin{subfigure}[t]{0.45\textwidth}
	\centering
	\tikzsetnextfilename{Brun2016_multilayer_error_displacement_Ux_R}
	\input{Brun2016_multilayer_error_displacement_Ux_R}
	\caption{%Time history of the error on horizontal displacement
	Error on horizontal displacement
	at receiver $R$
	}\label{fig:Brun2016_multilayer_error_displacement_Ux_R}
\end{subfigure}\\
\begin{subfigure}[t]{0.45\textwidth}
	\centering
	\tikzsetnextfilename{Brun2016_multilayer_displacement_Uy_R}
	\input{Brun2016_multilayer_displacement_Uy_R}
	\caption{%Time history of vertical displacement
	Vertical displacement
	at receiver $R$
	}\label{fig:Brun2016_multilayer_displacement_Uy_R}
\end{subfigure}\hfill
\begin{subfigure}[t]{0.45\textwidth}
	\centering
	\tikzsetnextfilename{Brun2016_multilayer_error_displacement_Uy_R}
	\input{Brun2016_multilayer_error_displacement_Uy_R}
	\caption{%Time history of the error on vertical displacement
	Error on vertical displacement
	at receiver $R$
	}\label{fig:Brun2016_multilayer_error_displacement_Uy_R}
\end{subfigure}
\caption{Multi-layer problem: evolutions of (\subref{fig:Brun2016_multilayer_displacement_Ux_R}) the horizontal displacement $u_1$, (\subref{fig:Brun2016_multilayer_error_displacement_Ux_R}) the absolute error $\abs{u_1 - u_1^{\text{ref}}}$ on the horizontal displacement, (\subref{fig:Brun2016_multilayer_displacement_Uy_R}) the vertical displacement $u_2$ and (\subref{fig:Brun2016_multilayer_error_displacement_Uy_R}) the absolute error $\abs{u_2 - u_2^{\text{ref}}}$ on the vertical displacement at receiver $R$ with respect to time $t$ for the PML and CAL solutions compared to the reference solution}\label{fig:Brun2016_multilayer_displacement_R}
\end{figure}

\begin{figure}
	\centering
	\includegraphics[width=\textwidth]{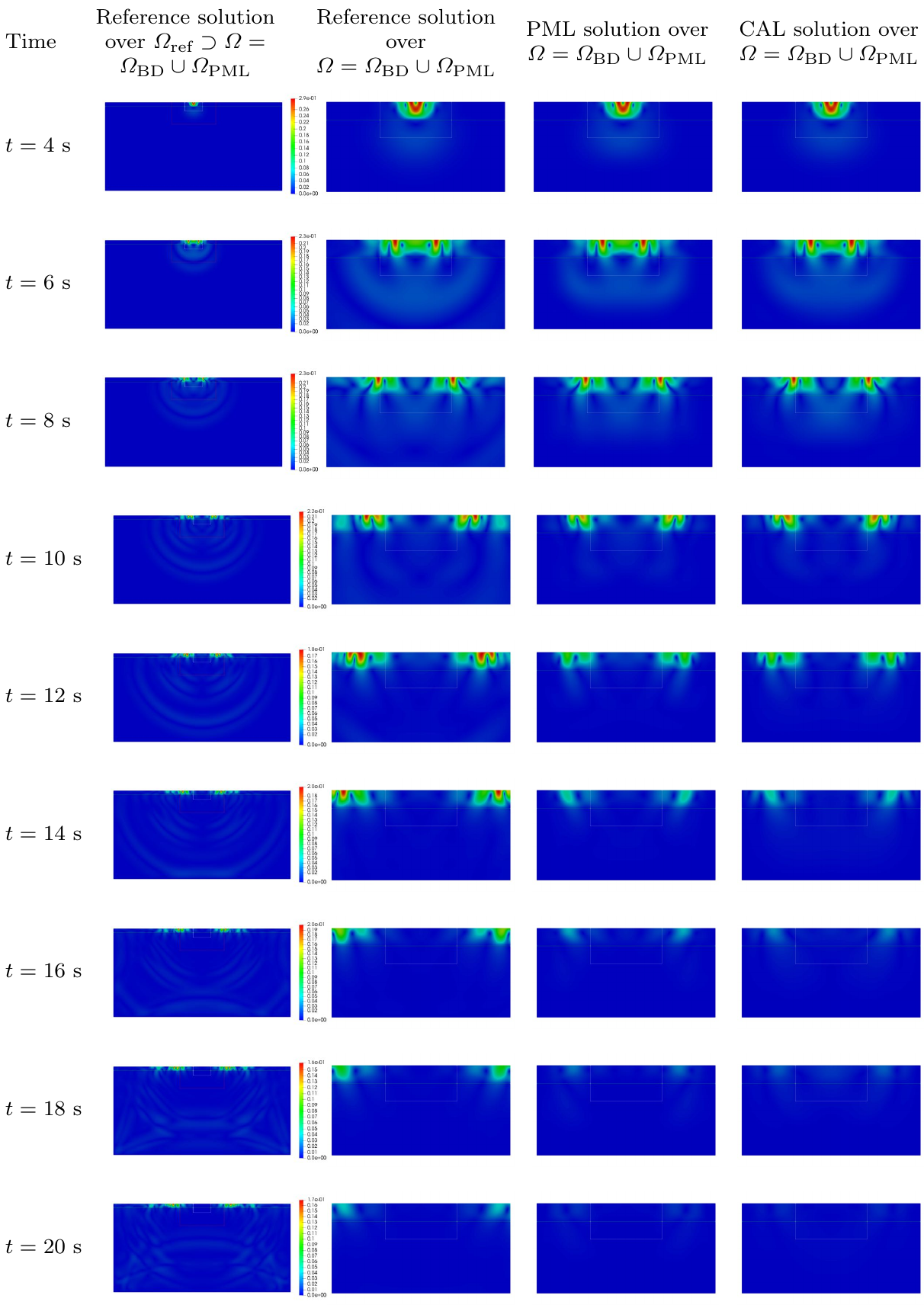}
	\caption{Multi-layer problem: snapshots of the displacement field magnitude $\xb \mapsto \norm{\ub(\xb,t)}$ taken at different times $t=4,6,8,10,12,14,16,18,20$~s (from top to bottom) for the reference solution over an extended computational domain $\Omega_{\text{ref}} \supset \Omega = \Omega_{\text{BD}} \cup \Omega_{\text{PML}}$ (first column) with interface $\Gamma = \partial\Omega_{\text{BD}} \cap \partial\Omega_{\text{PML}}$ indicated by white solid lines and the exterior boundary $\Gamma_{\text{ext}}$ of $\Omega$ indicated by red solid lines, the reference solution over the computational domain $\Omega = \Omega_{\text{BD}} \cup \Omega_{\text{PML}}$ (second column), the PML solution over $\Omega$ (third column) and the CAL solution over $\Omega$ (fourth column)}\label{fig:Brun2016_multilayer_displacement}
\end{figure}

\begin{figure}
	\centering
	\includegraphics[width=\textwidth]{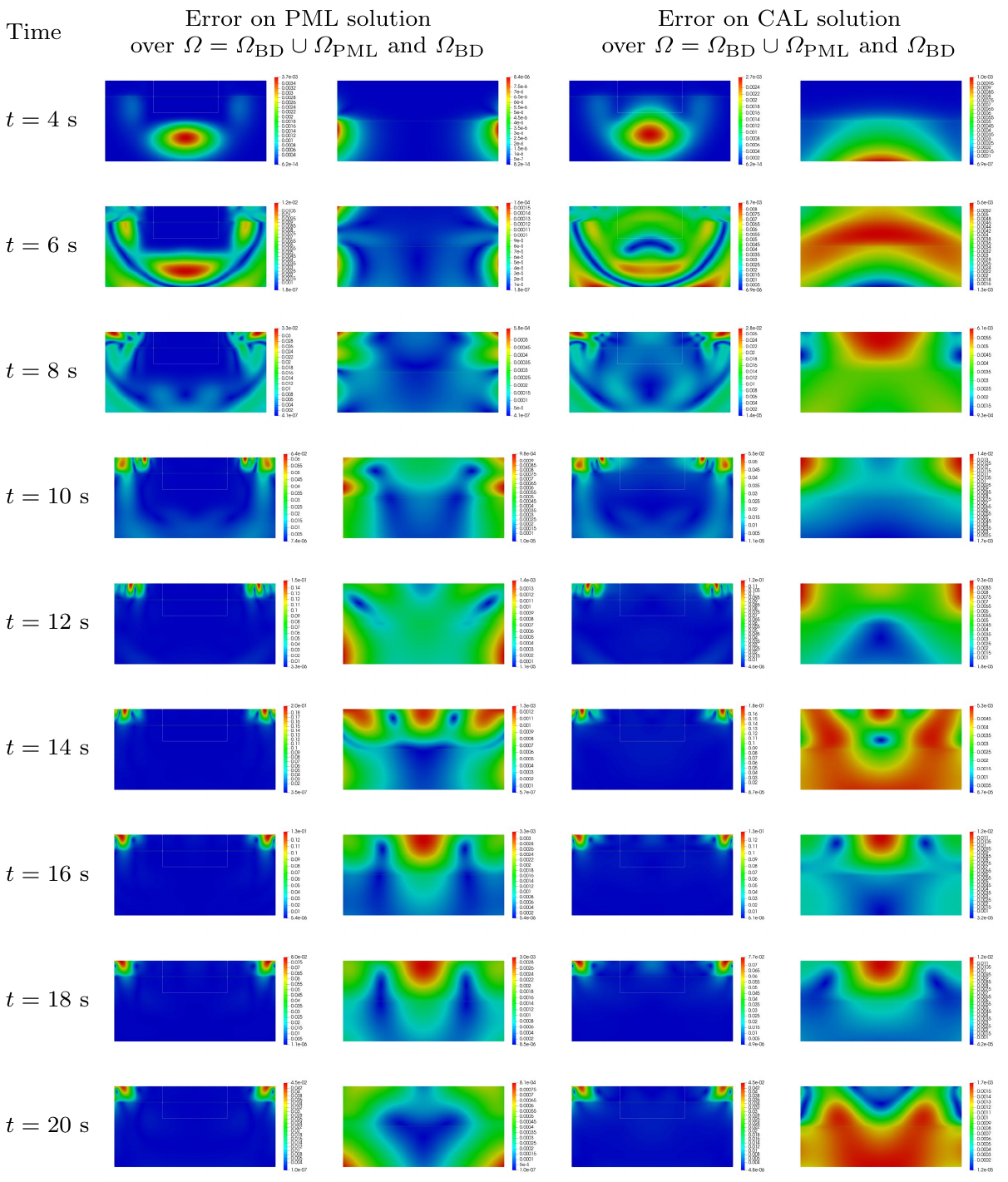}
	\caption{Multi-layer problem: snapshots of the error on the displacement field magnitude $\xb \mapsto \abs{\norm{\ub(\xb,t)} - \norm{\ub^{\text{ref}}(\xb,t)}}$ taken at different times $t=4,6,8,10,12,14,16,18,20$~s (from top to bottom) for the PML solution over the entire computational domain $\Omega = \Omega_{\text{BD}} \cup \Omega_{\text{PML}}$ (first column) and its restriction to the physical bounded domain $\Omega_{\text{BD}}$ (second column) and the CAL solution over $\Omega$ and $\Omega_{\text{BD}}$ (third and fourth columns)}\label{fig:Brun2016_multilayer_error_displacement}
\end{figure}

Figure~\ref{fig:Brun2016_multilayer_displacement} displays snapshots of the displacement field magnitude $\xb \mapsto \norm{\ub(\xb,t)}$ at several times for all the three numerical simulations performed up to $T=20$~s over the extended computational domain $\Omega_{\text{ext}}$ (for the reference solution) and the computational domain $\Omega$ (for the PML and CAL solutions). In both PML and CAL numerical simulations, the propagative waves leave the physical domain then damp out with an exponential decay inside the absorbing layers so as to mimic the propagation inside a semi-infinite two-layered medium. Both direct bulk and Rayleigh waves can be observed as well as multiple reflected and transmitted waves at the material interface between the two elastic media. Note that a fast bulk $P$-wave with a low amplitude has already entered the (high velocity) bottom layer (Layer 2) and left the physical domain $\Omega_{\text{BD}}$ after only $t=4$~s. As expected, most of the waves are guided in the top layer (Layer 1) with minor radiations in the bottom layer (Layer 2). Figure~\ref{fig:Brun2016_multilayer_error_displacement} depicts some snapshots of the absolute errors on the displacement field magnitude obtained by comparing the PML and CAL solutions to the reference one. Despite no noticeable spurious reflections can be clearly seen from the snapshots in Figure~\ref{fig:Brun2016_multilayer_displacement}, we observe that the errors computed on the PML solution inside the physical domain $\Omega_{\text{BD}}$ are at least around one order of magnitude lower than that computed on the CAL solution. All the outgoing waves are then slightly better absorbed in the PML region $\Omega_{\text{PML}}$ using the PML approach than using the CAL approach.

In order to provide further insight into the performances of the PML approach, Figure~\ref{fig:Brun2016_multilayer_energies} shows the evolutions of the kinetic, internal and total energies, namely $E_k(t)$, $E_i(t)$ and $E(t)$, accumulated in the physical domain $\Omega_{\text{BD}}$ as functions of time $t$ for the PML, CAL and reference solutions. Although all energy curves are almost perfectly superimposed due to the relatively large thickness $h=6$ of the PML region $\Omega_{\text{PML}}$, the PML solution is more accurate than the CAL solution anyway. Indeed, the errors committed on the kinetic energy $E_k(t)$, internal energy $E_i(t)$ and total energy $E(t)$ are obviously lower for the PML solution than the CAL solution, especially before all the outgoing waves have left the physical domain $\Omega_{\text{BD}}$, that is up to approximately time $t=14$~s. Overall, the performances of the proposed PML method in terms of absorptive capacity are very good, with no discernible spurious reflections or numerical instabilities, even in the presence of material heterogeneities.

\begin{figure}
\centering
\begin{subfigure}[t]{0.45\textwidth}
	\centering
	\tikzsetnextfilename{Brun2016_multilayer_kinetic_energy}
	\input{Brun2016_multilayer_kinetic_energy}
	\caption{%Time history of kinetic energy
	Kinetic energy
	}\label{fig:Brun2016_multilayer_kinetic_energy}
\end{subfigure}\hfill
\begin{subfigure}[t]{0.45\textwidth}
	\centering
	\tikzsetnextfilename{Brun2016_multilayer_error_kinetic_energy}
	\input{Brun2016_multilayer_error_kinetic_energy}
	\caption{%Time history of the error on kinetic energy
	Error on kinetic energy
	}\label{fig:Brun2016_multilayer_error_kinetic_energy}
\end{subfigure}\\
\begin{subfigure}[t]{0.45\textwidth}
	\centering
	\tikzsetnextfilename{Brun2016_multilayer_internal_energy}
	\input{Brun2016_multilayer_internal_energy}
	\caption{%Time history of internal energy
	Internal energy
	}\label{fig:Brun2016_multilayer_internal_energy}
\end{subfigure}\hfill
\begin{subfigure}[t]{0.45\textwidth}
	\centering
	\tikzsetnextfilename{Brun2016_multilayer_error_internal_energy}
	\input{Brun2016_multilayer_error_internal_energy}
	\caption{%Time history of the error on internal energy
	Error on internal energy
	}\label{fig:Brun2016_multilayer_error_internal_energy}
\end{subfigure}\\
\begin{subfigure}[t]{0.45\textwidth}
	\centering
	\tikzsetnextfilename{Brun2016_multilayer_total_energy}
	\input{Brun2016_multilayer_total_energy}
	\caption{%Time history of total energy
	Total energy
	}\label{fig:Brun2016_multilayer_total_energy}
\end{subfigure}\hfill
\begin{subfigure}[t]{0.45\textwidth}
	\centering
	\tikzsetnextfilename{Brun2016_multilayer_error_total_energy}
	\input{Brun2016_multilayer_error_total_energy}
	\caption{%Time history of the error on total energy
	Error on total energy
	}\label{fig:Brun2016_multilayer_error_total_energy}
\end{subfigure}
\caption{Multi-layer problem: evolutions of (\subref{fig:Brun2016_multilayer_kinetic_energy}) the kinetic energy $E_k(t)$, (\subref{fig:Brun2016_multilayer_error_kinetic_energy}) the absolute error $\abs{E_k(t) - E_k^{\text{ref}}(t)}$ on the kinetic energy, (\subref{fig:Brun2016_multilayer_internal_energy}) the internal energy $E_i(t)$, (\subref{fig:Brun2016_multilayer_error_internal_energy}) the absolute error $\abs{E_i(t) - E_i^{\text{ref}}(t)}$ on the internal energy, (\subref{fig:Brun2016_multilayer_total_energy}) the total energy $E(t)$ and (\subref{fig:Brun2016_multilayer_error_total_energy}) the absolute error $\abs{E(t) - E^{\text{ref}}(t)}$ on the total energy stored in the physical domain $\Omega_{\text{BD}}$ with respect to time $t$ for the PML and CAL solutions compared to the reference solution}\label{fig:Brun2016_multilayer_energies}
\end{figure}

% Because of the expected absorbing character of the PML region $\Omega_{\text{PML}}$, we apply a homogeneous (\ie{} zero) Dirichlet boundary condition $\ub = \zerob$ on its exterior boundary.
% The PML region $\Omega_{\text{PML}}$ is placed sufficiently far away from all sources of excitation so that the outgoing propagative waves are expected not to be incident at near-grazing angles on the interface $\Gamma$ and would be mostly damped.

\section{Conclusion}\label{sec:concl}

A review of the current state-of-the-art of the perfectly matched layer (PML) method and its variants (leading to different PML formulations) developed over the past 25 years has been first presented with a focus on the elastic wave propagation in unbounded media. An efficient PML formulation has then been proposed for the numerical simulation and modeling of second-order linear elastodynamic equations in two- and three-dimensional unbounded domains. Both time-harmonic (frequency-domain) and time-dependent (time-domain) PML formulations have been addressed. The frequency-domain PML formulation is based on a classical complex coordinate stretching and involves a specific auxiliary strain field. The time-domain PML formulation is then obtained by applying an inverse Fourier transform in time. The resulting unsplit-field PML weak formulation is discretized in space using a standard Galerkin finite element method and integrated in time through classical second-order accurate implicit time schemes, namely a Newmark time scheme combined with a Crack-Nicolson (finite difference) time scheme. Such a PML formulation turns out to be intrusive but it can be easily implemented into existing finite element codes with minor modifications. Furthermore, it does not require any field splitting, any nonlinear solver, any mass lumping, any limitation in the choice of time step size, and it does not involve any complicated (or costly) convolution operations in time or any high-order derivatives with respect to space and time. Besides, it preserves the intrinsic second-order accuracy of the second-order linear elastodynamic equations.

The performances of the proposed PML method have been illustrated on two-dimensional linear elastic wave propagation problems defined in semi-infinite (unbounded) domains and stated in Cartesian coordinates. Numerical examples include (horizontally-layered) isotropic homogeneous elastic half-spaces subjected to a directional force point-source applied onto the surface (Lamb's problem) or buried within the volume. The accuracy of the PML approach is compared to a classical absorbing layer (CAL) approach with respect to a reference solution computed on an extended computational domain. Numerical results highlight that the PML acts as a highly efficient absorbing boundary condition (compared to the CAL) used to attenuate the amplitude of the waves traveling outside the physical domain of interest. Eventually, the proposed PML method could be applied to three-dimensional wave propagation problems and readily extended to other coordinate systems, such as polar, cylindrical or spherical coordinates. Also, it would be interested to extend the PML formulation to fluid-solid (or other multiphysics) interaction problems. Another ambitious perspective would consist in investigating the intrinsic stability issues arising in certain kinds of anisotropic (visco)elastic materials and exhibited by other PML formulations \cite{Beca03,App06b,Kom07,Mar08b,Meza08,Li10,Dmi12,Duru12a,Kre13,Zha14,Ping14a,Ping16,Assi15,Beca15,Gao17a,Li17,Li18}. To date, the construction of a stable PML formulation for the linear elastodynamic equations in arbitrary anisotropic elastic media (whatever the level of material anisotropy, \ie{} for all the material symmetry classes ranging from isotropy to pure anisotropy) remains an open problem.

\begin{acknowledgements}
The authors gratefully acknowledge and thank Christian Soize, Professor at Universit\'e Gustave Eiffel, Laboratoire MSME, for very helpful discussions, constructive remarks and valuable comments.
\end{acknowledgements}

\section*{Conflict of interest}

The authors declare that they have no conflict of interest regarding the publication of this article.

\bibliographystyle{spmpsci}      % mathematics and physical sciences
\bibliography{Biblio}   % name your BibTeX data base

%\begin{thebibliography}{}
%%
%% and use \bibitem to create references. Consult the Instructions
%% for authors for reference list style.
%\end{thebibliography}

\end{document}